\documentclass[a4paper,11pt]{article}
\pdfoutput=1
\emergencystretch=\maxdimen
\usepackage{jcappub} 
\usepackage{gensymb}
\usepackage{graphicx}	
\usepackage{amsmath}	
\usepackage{amsfonts}
\usepackage{float}
\usepackage{subcaption} 
\usepackage{xurl}
\usepackage{siunitx}    
\usepackage{wrapfig}
\usepackage{mathrsfs,mathtools}
\usepackage{tensor}
\usepackage[numbers]{natbib}
\title{\boldmath Observational Predictions of LQG Motivated Polymerized Black Holes and Constraints From Sgr A* and M87*}
\author[a]{Rahul Kumar Walia}
\affiliation[a]{Astrophysics Research Centre, School of Mathematics, Statistics and Computer Science, University of KwaZulu-Natal, Private Bag 54001, Durban 4000, South Africa}
\emailAdd{rahul.phy3@gmail.com}
\date{\today}

\abstract{Loop quantum gravity inspired partial polymer quantization in four-dimensional spacetime leads to a globally regular black hole with a single horizon. The polymerized black hole metric is characterized by the minimum length parameter $k$, and mimics the Schwarzschild black hole in the weak-field limit. We present an analytic and numerical investigation of the strong gravitational lensing and shadow morphology to determine the observational impacts of quantum effects. Interestingly, the light deflection angle, the angular separation between the outermost relativistic image, and magnification are significantly larger than those for the Schwarzschild black hole. Using the ray-tracing technique, we simulate the black hole shadows under three distinct optically thin accretion models: static spherical accretion, radially infalling spherical accretion, and the thin accretion disk model. Polymerized black holes' shadow morphology strongly depends on $k$. We derive constraints on $k$ from the M87* and Sgr A* black hole shadow observations from the Event Horizon Telescope.}
	
\begin{document}
\maketitle
\section{Introduction}\label{Sec-1}
The presence of spacetime singularities, manifested as curvature scalar divergence or geodesic incompleteness, is an undesirable but apparently inevitable feature of general relativity and many other alternate classical theories of gravity \cite{Oppenheimer:1939ue,Penrose:1969pc}. Several independent attempts through quantum gravity models have been made to regularize the curvature invariant and to evade the ultra-violet incompleteness of general relativity. 
In this direction, Loop Quantum Gravity (LQG) turned out to be one of the few successful attempts to understand the quantum nature of gravity \cite{Rovelli:1994ge}. LQG is one of the non-perturbative and background-independent approaches to quantum gravity. It has been shown that the underlying idea of quantization of area and volume observables is the fundamental ingredient in the LQG models to solve the black hole singularity problem \cite{Rovelli:1997yv,Modesto:2004xx,Modesto:2005zm,Ashtekar:2005qt}. In particular,  the classical singularity is replaced by a spherical quantum bounce $S^2$ with the non-zero minimum area which connects the collapsing phase with a re-expanding branch in the LQG black hole models. However, because of the inherent complexity involved in the complete quantum treatment of LQG, it is easier to work with an effective-field theory approach. In this line of research, the phase space quantization or semi-classical polymerization that preserves the LQG's idea of spacetime discreteness turns out to be very interesting and fruitful \cite{Ashtekar:2002sn,Boehmer:2007ket}. Peltola and Kunstatter \cite{Peltola:2008pa,Peltola:2009jm} have reported that polymerization of the generalized area variable alone leads to a complete, regular, single-horizon spacetime, in which the classical singularity is replaced by a bounce. This four-dimensional quantum-corrected polymerized black hole spacetime has fascinating properties and advantages over other regular black hole candidates, including: (i) single horizon, which eliminates the problem of mass inflation at the inner horizon, (ii) globally hyperbolic spacetime, (iii) the global spacetime structure is quite different from other regular black holes, in that the areal radius decreases to a minimum value in the black hole interior and then re-expands into a Kantowski-Sachs universe (iv) geodesically complete spacetime. Interestingly, Daghigh \textit{et al.} \cite{Daghigh:2020mog} have proven that these black holes are stable against small massless scalar perturbations, contrary to other regular black holes \cite{Carballo-Rubio:2018pmi}. Bronnikov \textit{et al.} \cite{Bronnikov:2006fu} called the black bounce black hole with single horizon as the ``black universe".

Exploring the consequences of a minimal length scale is one of the best motivated avenues to make contact with the phenomenology of quantum gravity.  One can therefore ask whether the resulting LQG-motivated corrections around $r=0$ can propagate to the black hole exterior and leave imprints on the black hole's observational features. To put it another way, is it possible to use the black hole observational features, particularly in the electromagnetic spectrum, to extract information about the quantum gravity signatures in the black hole spacetime? How do a singular black hole of general relativity and a regular black hole of the same mass arising in LQG differ in their observational features? We address these questions in this paper. We investigate the gravitational lensing of light around the polymerized black hole, and calculate the direct and relativistic image positions, magnification, and time delay in the formation of primary and secondary images. We compute the effects of the polymer parameter on the lensing observables and compare them with those for the Schwarzschild black hole. One of the salient features of strong gravitational lensing by a black hole is the logarithmic divergence of the deflection angle in the impact parameter \cite{Bozza:2001xd,Bozza:2002zj}. This divergence accounts for the formation of a photon sphere around the black hole and the existence of a photon ring enclosing the ``black hole shadow" \cite{Bardeen:1973tla}. Following that, we construct the polymerized black hole shadows under various accretion flow models to get a better insight into the causal and observational features of emission arising near a polymerized black hole. While the photon ring solely depends on the spacetime metric, in astrophysically realistic scenarios, the optical appearance of a black hole is highly dependent on the details of the accretion models and the emission process. Consequently, by comparing the intensity distribution in synthetic shadows, one can assess the differences/similarities between the polymerized and Schwarzschild black hole shadows. In a magnificent achievement, EHT made the first horizon-scale radio observations of the supermassive black holes Sgr A* and M87* and unraveled their characteristic shadows enclosed with a bright photon emission ring \cite{Akiyama:2019cqa,EventHorizonTelescope:2019pgp,Akiyama:2019eap,EventHorizonTelescope:2022xnr,EventHorizonTelescope:2022wok,EventHorizonTelescope:2022urf,EventHorizonTelescope:2022xqj}. We constrain the polymer parameter using the EHT shadow results. Therefore, the present study is not only crucial to discovering the impact of quantum gravity corrections on the observational aspects of the black hole, it is also relevant in light of the EHT observations. 
Of course one can expects the quantum gravity effects to be relevant for microscopic black holes, but it is important to study the effects of such corrections both quantitatively and qualitatively for the astrophysical black holes. This is because there are proposals that, for the black holes with large entropy, the length scale of quantum gravitational effects need not be $\ell _p$ but could be far below the Planck scale $N^n\ell_p$; $N$ is large number of black hole microstates and $n>0$. With this assertion, a new possible window for observing quantum gravitational  effects in astrophysical black hole spacetime has been recently pointed out by Rovelli and Vidotto \cite{Rovelli:2014cta}. As a result, various studies, concentrated on finding the quantum gravity observational signatures in black hole spacetime,  have been actively addressed in the literature  \cite{Wei:2015dua,Held:2019xde,Kumar:2019ohr,Liu:2020ola,Zhu:2020tcf,Brahma:2020eos,Fu:2021fxn,Zeng:2021dlj,Sahu:2015dea}.

The study of gravitational lensing was led by Einstein \cite{Einstein:1936llh} and Darwin \cite{Darwin} in the weak-field limit, and later the theory was developed systematically in the strong-field limits as well \cite{Synge:1966okc,Refsdal:1964yk,Liebes:1964zz,1992grle.book.....S,Frittelli:1999yf,Gralla:2019drh}. However, Virbhadra \cite{Virbhadra:1999nm,Virbhadra:2002ju,Virbhadra:2007kw}, Bozza et al. \cite{Bozza:2001xd,Bozza:2002zj, Bozza:2004kq, Bozza:2007gt,Bozza:2008ev}, and Tsukamoto \cite{Tsukamoto:2016jzh,Tsukamoto:2016oca,Tsukamoto:2020iez,Tsukamoto:2021fsz} brought a significant interest into the field and provided an analytical framework to investigate the strong gravitational lensing effect for a generic static spherically symmetric metric. A black hole's shadow is a manifestation of the strong gravitational lensing around it. Black holes embedded in the optically thin accreting region are expected to reveal a dark ``shadow" caused by the photon capture and strong gravitational lensing \cite{Bardeen:1973tla,CT}. Luminet \cite{Luminet:1979nyg} presented a visual appearance of a thin emitting accretion disk around the Schwarzschild black hole. Similar hot accretion flows are found around many supermassive black holes in the universe, and they are natural candidates to reveal shadows in their images \cite{Narayan:2019imo}.  The black hole shadow theory has evolved over the past decades and has resulted in a flourish of studies (see \cite{Perlick:2021aok,Cunha:2018acu} for review). Lately, the strong gravitational lensing \cite{Bhadra:2003zs,Eiroa:2004gh,Keeton:2005jd,Chen:2009eu,Reyes:2010tr, Shaikh:2019itn,Javed:2019qyg,Reji:2019brv,Islam:2020xmy,Kumar:2020sag,Virbhadra:2022iiy,Virbhadra:2022ybp} and black hole shadow \cite{Falcke:1999pj,Takahashi:2004xh,Huang:2007us,Doeleman:2008qh, Hioki:2009na,Psaltis:2014mca,Abdujabbarov:2015xqa,Mishra:2019trb,Shaikh:2019fpu,Held:2019xde,Gralla:2020nwp,Konoplya:2019goy,Kumar:2019pjp,Kumar:2020hgm,Kumar:2020owy,Roy:2019esk,Shaikh:2018lcc,Bronzwaer:2021lzo,Mizuno:2018lxz,Kumar:2020owy,Afrin:2021imp,Ghosh:2020spb,Kumar:2020ltt,Kumar:2020yem,Kumar:2018ple,Nampalliwar:2021oqr,Nampalliwar:2021tyz,Gralla:2020srx,Devi:2021ctm,Roy:2021uye} have been extensively used in testing theories of gravity at the horizon-scale regime, black hole parameter estimation, and deducing the nature of any matter distributions in the black hole background. Therefore, strong gravitational lensing features and shadows can be used as an effective way to study polymerized black holes and can provide us with valuable information about the underlying quantum gravitational corrections. 

The rest of the paper is organized as follows. In Sect.~\ref{Sec-2}, we discuss the geometric properties of the static spherically symmetric polymerized black hole and demonstrate that this metric appears as an exact solution of the Einstein field equations minimally coupled with the phantom scalar field and the nonlinear electrodynamics field (NED) associated with the magnetic field. In Sect.~\ref{Sec-3}, we present the study of gravitational lensing in the weak and strong deflection limits. The image positions, magnifications, Einstein ring, strong-lensing observables, and numerical estimations of deflection angle are presented in Sect.~\ref{Sec-4}. In Sect.~\ref{Sec-5}, lensing by supermassive black holes Sgr A* and M87* is discussed. The polymerized black hole shadows under different accretion flow models are reported in Sect.~\ref{Sec-6}, and constraints on the polymer parameter are deduced using the Sgr A* and M87* black hole shadow observational data in Sect.~\ref{Sec-8}. Finally, we summarize our main findings in Sect.~\ref{Sec-8}.\\

\section{LQG motivated $4D$ polymerized black hole}\label{Sec-2}
The dynamical field equations in the four-dimensional partially polymerized theory admit a static and spherically symmetric black hole solution \cite{Peltola:2008pa,Peltola:2009jm}
\begin{equation}\label{metric1}
	ds^{2}=-A(r)dt^{2}+B(r)dr^{2}+C(r)(d\theta^{2}+\sin^{2}\theta d\phi^{2}),
\end{equation}
with 
\begin{eqnarray}
	A(r)&=&\left(\sqrt{1-\frac{k^2}{r^2}}-\frac{2M}{r}\right),\nonumber\\
	B(r)&=&\Bigg(\left(\sqrt{1-\frac{k^2}{r^2}}-\frac{2M}{r}\right)\left(1-\frac{k^2}{r^2}\right)\Bigg)^{-1},\; C(r)=r^2.\nonumber
\end{eqnarray}
The black hole metric depends on two parameters: black hole mass $M$ and polymer parameter $k$. The solution is asymptotically flat at $r\to \infty$, and the Ricci and Kretschmann scalars read as follows
\begin{eqnarray}
	R^{\mu\nu}R_{\mu\nu}&=&\frac{1}{2 r^{10}}\Big(-9 k^6+k^4 \left(36 M (M-\Sigma  r)+r^2\right)\nonumber\\
	&&+4 k^2 r^3 (-2 \Sigma  (2 M+r)+4 M+r)-8 (\Sigma -1) r^6\Big),\nonumber\\
	R^{\mu\nu\sigma\rho}R_{\mu\nu\sigma\rho}&=&\frac{1}{r^{10}}\Big(-33 k^6+k^4 \left(132 M (M-\Sigma  r)+41 r^2\right)\nonumber\\
	&&+4 k^2 r^2 \left(-36 M^2+2 M (11 \Sigma -2) r+(2 \Sigma -3) r^2\right)\nonumber\\
	&&+8 r^4 \left(6 M^2-\Sigma  r (2 M+r)+2 M r+r^2\right)\Big),
\end{eqnarray}
with $\Sigma=\sqrt{1-\frac{k^2}{r^2}}$. Curvature scalars are everywhere finite, and vanish rapidly at far distances from the black hole. In the limit $r\to k$, the scalars take simplified form
\begin{eqnarray}
	R^{\mu\nu}R_{\mu\nu}&=&\frac{2 \left(k^2+4 k M+9 M^2\right)}{k^6},\\
	R^{\mu\nu\sigma\rho}R_{\mu\nu\sigma\rho}&=&\frac{4 (k^2 + 9 M^2)}{k^6}.
\end{eqnarray}
Clearly, the curvature scalars are bounded from above at the black hole center and in the limit $k\to 0$, the bounce radius goes to 0 and the curvature scalar diverge.
As a result, the polymerized black hole metric describes a globally regular spacetime. One of the most striking features of this four-dimensional quantum-corrected black hole metric (\ref{metric1}) is that it has a single horizon at $r\equiv r_+=\sqrt{4M^2+k^2}$. From the metric function, it is clear that $r\geq k$; the radial coordinate $r$ admits a minimum value at $r=k$ in a nonstatic ($A(r) < 0$) spacetime region and called as ``black bounce''. Notably, the curvature singularity at $r=0$ is now replaced by a spacelike 2-sphere of radius $r=k$ bouncing into an infinitely expanding Kantowski-Sachs spacetime. The parameter $k$ defines the minimum radius of the bounce, such that for $k\to 0$ metric (\ref{metric1}) recovers the Schwarzschild black hole metric. It immediately follows that the energy conditions are violated at the center. Bronnikov \textit{et al.}, in a series of paper, \cite{Bronnikov:2005gm,Bronnikov:2006fu,Bronnikov:2015kea,Bronnikov:2018vbs} have shown that geometries containing a black bounce are described by solutions to the Einstein equations with phantom scalar fields. Nevertheless, the coordinate singularity at $r=k$, can be transformed away by using the transformation $r=\sqrt{k^2+y^2}$; the transformed metric reads as
\begin{eqnarray}
	ds^{2}&=&-\left(\frac{y-2M}{\sqrt{y^2+k^2}}\right)dt^{2}+\frac{1}{\left(\frac{y-2M}{\sqrt{y^2+k^2}}\right)}dy^{2}+(y^2+k^2)(d\theta^{2}+\sin^{2}\theta d\phi^{2}).\label{metric2}
\end{eqnarray}
Here, the radial coordinate $y$ assume the full range $0\leq y\leq \infty$. It is important to note that most of the regular black holes that are closely connected to a potential theory of quantum gravity, can not be viewed as derived from quantum-corrected field equations, but rather \emph{inspired} by quantum gravity principles. As such, they constitute useful phenomenological models but making their physical justification less straightforward \cite{Ashtekar:2004eh}. 
Here, we present the source for this polymerized black hole. Considering the validity of the Einstein field equations $G_{\mu\nu}=T_{\mu\nu}$ (with $8\pi G=1, c=1$), we calculate the corresponding energy-momentum tensor (EMT) for the metric (\ref{metric2}):
\begin{align}
	T^t_t\equiv G^t_t&=  \frac{-(k^2+y^2)^{3/2}+y^3 -k^2 \left(4 M-3y\right)} {\left(k^2+y^2\right)^{5/2}},\label{EMT1}\\
	T^y_y \equiv G^y_y&=  \frac{y}{\left(k^2+y^2\right)^{3/2}}-\frac{1}{k^2+y^2},\label{EMT2}\\
	T^{\theta}_{\theta}\equiv G^{\theta}_{\theta}&= \frac{k^2 (y-2 M)}{2 \left(k^2+y^2\right)^{5/2}} \label{EMT3}.
\end{align}
We start with the Einstein-Hilbert action and a minimally coupled uncharged scalar field $\Phi(y)$ and the NED field with Lagrangian density $\mathcal{L(F)}$ \cite{Bronnikov:2021uta}
\begin{equation}
	S=\int\sqrt{-g}d^4y\left(\mathcal{R} + 2\epsilon g^{\mu\nu} \nabla_{\mu}\Phi\nabla_{\nu}\Phi - 2V(\Phi) -\mathcal{L(F)} \right),\label{Action}
\end{equation}
where $\epsilon=+1 (-1)$ characterizes the canonical (phantom) scalar field with positive (negative) kinetic energy, which will be fixed later. $\mathcal{L(F)}$ is the Lorentz invariant NED field Lagrangian density with $\mathcal{F} \equiv F_{\mu\nu}F^{\mu\nu}$ as the Faraday invariant defined in terms of the NED field tensor $F_{\mu\nu}$. On varying the action (\ref{Action}) with the metric tensor field $g_{\mu\nu}$ leads to the following gravitational field equations
\begin{align}
	G_{\mu\nu}=T_{\mu\nu} \equiv T^{(\Phi)}_{\mu\nu} + T^{(\text{EM})}_{\mu\nu}, \label{FieldEq}
\end{align}
with $T^{(\Phi)}_{\mu\nu}$ and $T^{(\text{EM})}_{\mu\nu}$, respectively, being the EMT for the scalar field and the NED field as follows \cite{Bronnikov:2021uta}
\begin{eqnarray}
	T^{(\Phi)}_{\mu\nu}&=&2\epsilon \nabla_{\mu}\Phi\nabla_{\nu}\Phi-g_{\mu\nu}\left(\epsilon g^{\alpha\beta}\nabla_{\alpha}\Phi\nabla_{\beta}\Phi +V(\Phi)\right),\label{TT1}\\
	T^{\text{(EM)}}_{\mu\nu}&=&2\left( \frac{\partial \mathcal{L(F)}}{\partial \mathcal{F}} F_{\mu\sigma} \tensor{F}{_\nu^\sigma}-\frac{1}{4}g_{\mu\nu}\mathcal{L(F)} \right).\label{TT2}
\end{eqnarray}
Energy density is positive (negative) for canonical (phantom) scalar field.
Whereas on varying the action with scalar field $\Phi$ and $F_{\mu\nu}$, the corresponding dynamical equations read as
\begin{align}
	&2\epsilon g^{\mu\nu}\nabla_{\mu}\nabla_{\nu}\Phi-\frac{d V(\Phi)}{d \Phi}=0,\\
	& \nabla_{\mu}\Big( \frac{\partial \mathcal{L(F)}}{\partial \mathcal{F}} F^{\mu\nu}\Big)=0.
\end{align}
With the time translational and spherical symmetry of the metric (\ref{metric1}), we can make some assumptions about the scalar and NED fields. In particular, we assume that the scalar field $\Phi$ is time-independent and function of only spatial coordinates, whereas the NED field Faraday tensor has only two possible nonzero components: $F_{ty}= -F_{yt}$ (a radial electric field) and $F_{\theta\phi} =-F_{\phi\theta}$ (a radial magnetic field). Because of the intrinsic issues with the electric field NED describing a regular black hole spacetime \cite{Bronnikov:2000vy}, we will take the case of only the magnetic field, such that $F_{\theta\phi} = p \sin\theta$, where $p$ is the magnetic monopole charge. The Faraday invariant takes the form $\mathcal{F} = \frac{2 p^2}{\left(k^2+y^2\right)^2}$. Under these assumptions, the EMT becomes
\begin{align}
	T_{\mu}^{\nu(\Phi)} &= \epsilon \frac{y-2 M}{\sqrt{k^2+y^2}} \Phi'^2 \text{diag} (-1, +1, -1, -1) - \delta_{\mu}^{\nu}  V(\Phi), \label{EMT-S}\\
	T_{\mu}^{\nu\text{(EM)}} &=  \frac{-\mathcal{L(F)}}{2}\text{diag} \Big(1,\, 1,\,  1 -\frac{4 p^2 \mathcal{L' (F)}}{\mathcal{L(F)}(k^2 + y^2)^2}, 1 -\frac{4 p^2 \mathcal{L' (F)}}{\mathcal{L(F)}(k^2 + y^2)^2}\Big)\label{EMT-NED}. 
\end{align}
Importantly, $T^t_t=-\rho$, the energy density, and $T^y_y=P_y$, the radial pressure, is valid only outside the event horizon. Inside the horizon, $y$ is a temporal coordinate and $t$ is a spatial one, therefore, $T^t_t=P_y$, $T^y_y=-\rho$. It is worthwhile to note that neither the scalar field nor the NED field can independently be the source for the LQG-motivated polymerized black hole.
The EMT for a scalar field (\ref{EMT-S}) admits $T_t^{t(\Phi)}=T_{\theta}^{\theta(\Phi)}$ whereas for the NED field, $T_t^{t(\text{EM})}=T_{y}^{y(\text{EM})}$, but these are not satisfied individually for the polymerized black hole as given in Eqs.~(\ref{EMT1})-(\ref{EMT3}). Adopting the approach proposed in Ref.~\cite{Bronnikov:2021uta} helps us assess the necessity of including both the NED and scalar fields in the gravitational action. In what follows, we show that a linear combination of a suitable scalar field and the NED field generates a polymerized black hole.

From Eqs.~(\ref{EMT-S}) and (\ref{EMT-NED}), the difference of Eqs.~(\ref{EMT1}) and (\ref{EMT2}) is independent of the NED field, and thus integrating it leads to the scalar field solution
\begin{eqnarray}
	\Phi(y)=c_1\pm \frac{1}{\sqrt{-\epsilon }}\arctan\left(\frac{y}{k}\right),\label{Sol-Sc}
\end{eqnarray}
where $c_1$ is an integration constant. Clearly, $\epsilon=1$ leads to an nonphysical scalar field solution, thus only the phantom scalar field is an acceptable source. This was to be expected because, in the absence of curvature singularity, the energy conditions are violated around the black hole center. Similarly, the difference between the Eqs.~(\ref{EMT1}) and (\ref{EMT3}) has a contribution only from the NED field, and integrating it yields the NED field Lagrangian density solution
\begin{equation}
	\mathcal{L(F)}= 2\Bigg(\frac{25 k^2 \left( \sqrt{\frac{2p^2}{\mathcal{F}}}-k^2\right)^{3/2}+16 \left( \sqrt{\frac{2p^2}{\mathcal{F}}} -k^2\right)^{5/2}}{15 k^2 \left(\frac{2p^2}{\mathcal{F}}\right)^{5/4}} +\frac{6 M k^2}{5}\left(\frac{\mathcal{F}}{2p^2}\right)^{5/4}+\sqrt{\frac{\mathcal{F}}{2p^2}}-\frac{16}{15 k^2}\Bigg).\label{Sol-NED}
\end{equation}
Using the expressions for scalar field and the NED field Lagrangian density, one can use the field equation to determine the scalar field potential
\begin{equation}
	V(y)=\frac{2 \left(6 k^4 M-15 k^4 y-20 k^2 y^3+8 \left(k^2+y^2\right)^{5/2}-8 y^5\right)}{15 k^2 \left(k^2+y^2\right)^{5/2}}.\label{Sol-Pot}
\end{equation}
The $\mathcal{L(F)}$ and $V(y)$ attain finite values at the black hole center and die out sharply at the large $y$. This is consistent with the asymptotic flatness of the black hole metric. It is clear from Eq.~(\ref{Sol-Sc}) that polymerized black hole parameter $k$ can be interpreted as the scalar field charge. Thus, the LQG motivated polymerized black hole described by the metric (\ref{metric1}) can also be interpreted as an exact solution of the Einstein field equations sourced by the minimally coupled phantom scalar field (\ref{Sol-Sc}) with non-zero potential (\ref{Sol-Pot}) and the NED field associated with the magnetic field (\ref{Sol-NED}).  Following the same technique, one can re-interpret the other quantum gravity black hole models having the bouncing geometry and the non-singular cosmological models in the loop quantum cosmology.

\section{Gravitational deflection of light}\label{Sec-3}
\begin{figure}
		\centering
	\includegraphics[scale=0.63, trim = 10cm 0cm 7cm 0cm, clip]{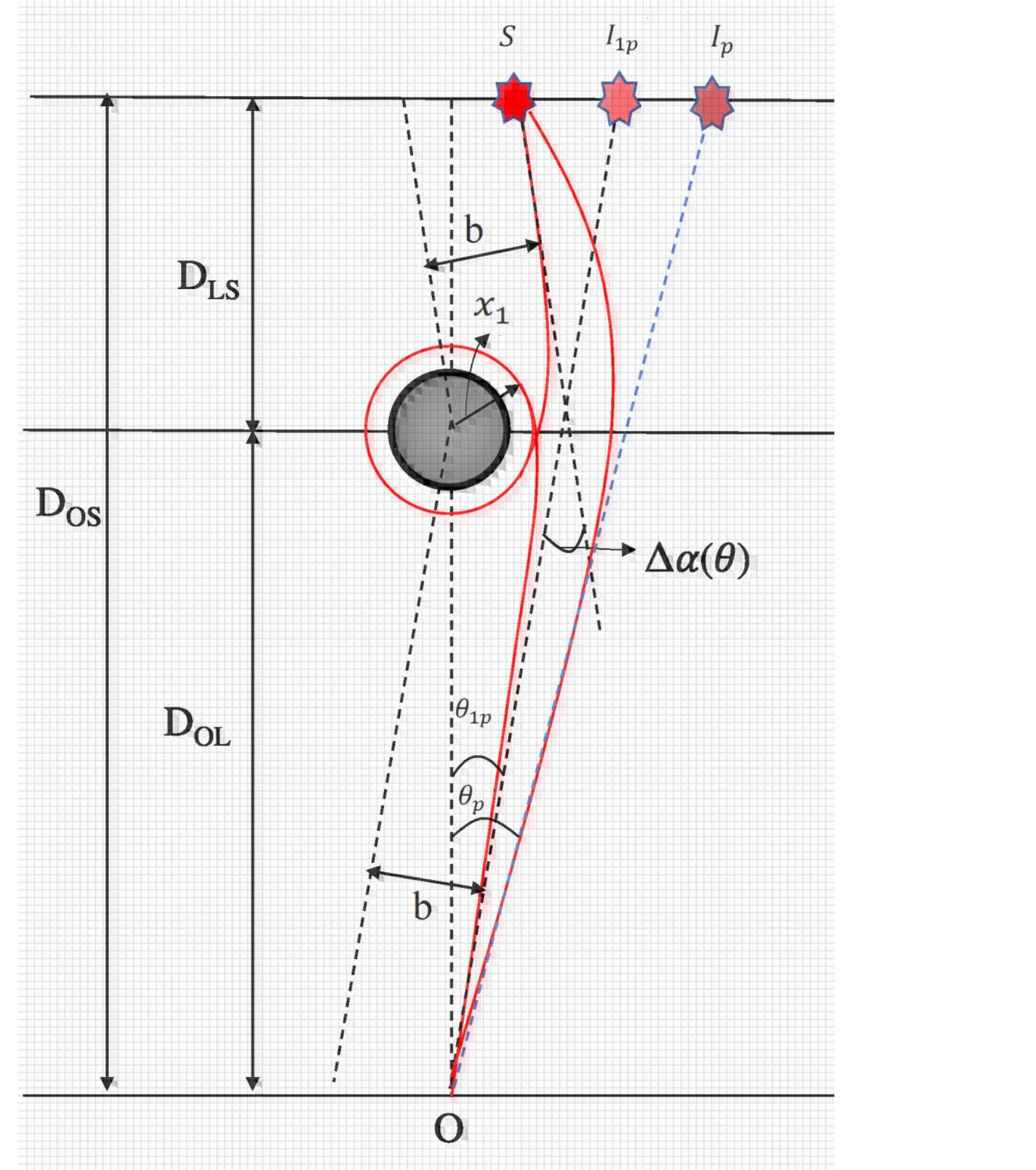}
	\caption{Schematic for the geometrical configuration of gravitational lensing. The source and the observer are on opposite sides of the black hole. The line joining the observer and the black hole is considered as a reference optical axis and all the angles, $\theta$ (image position) and $\beta$ (source position), are measured with respect to this axis at the observer position. $S$, $I_p$, and $I_{1p}$, respectively, are the source position, direct primary image position, and the first relativistic primary image position. Primary images form on the same side of the black hole. }\label{fig:schematic}		
\end{figure}

To study the lensing effects, we adopt the configuration where the black hole $L$ is situated between a point source of light $S$ and an observer $O$. Both the source and the observer are at distances much larger than the horizon radius, $r_+$, from the black hole in the asymptotically flat region. A black hole, by virtue of its strong gravitational field, acts as a \textit{gravitational convex} lens and bends the path of light rays coming from the source \cite{Einstein:1936llh} (cf. Fig.~\ref{fig:schematic}). However, in contrast to an optical convex lens, the amount of deflection experienced by light is inversely proportional to the impact parameter and the distance of the closest approach to the lens center \cite{Weinberg:1972kfs}. 
The presence of the spherical symmetry allows us to fix the plane of photons  motion, $\theta =\pi/2$. The two commuting spacetime isometries along Killing vectors $\xi^{\mu}_{(t)}=\delta^{\mu}_{t}$ and $\xi^{\mu}_{(\phi)}=\delta^{\mu}_{\phi}$ lead to two independent constant of motion for photon geodesics, viz., the energy $E=-p_{\mu}\xi^{\mu}_{(t)}$ and angular momentum $L=p_{\mu}\xi^{\mu}_{(\phi)}$; $p^{\mu}$ is the photon four-momentum \cite{Chandrasekhar:1985kt}.

Here onward, we will be using the metric (\ref{metric1}) form of the polymerized black hole, this is because in this form the areal radius is same as the radial coordinate. We normalized the coordinates by $2M$, viz., $x\equiv r/2M$, $t=t/2M$ and parameter $k\to k/2M$. Solving the geodesic equation of motion for test particles ($\delta=0$ for photon and $\delta=-1$ for massive test particle) yields
\begin{equation}
	\dot{t}=\frac{E}{A(x)},\;\; \dot{\phi}=\frac{L}{C(x)},
\end{equation}
with 
\begin{eqnarray}
	&&-A(x)\dot{t}^2+B(x)\dot{x}^2+C(x)\dot{\phi}^2=\delta\\
	\Rightarrow&&\dot{x}^2+ V_{\text{eff}}=E^2,\label{Rpot}
\end{eqnarray}
where $\dot{x^{\mu}}=\frac{d x^{\mu}}{d\tau}$ and $\tau$ is the affine parameter along the geodesics. $V_{\text{eff}}$ is the radial effective potential. Thus, the photon four-momentum components read as
\begin{equation}
	p^{\mu}=\Big(\frac{-E}{A(x)},\; \pm\sqrt{\frac{E^2}{A(x)B(x)}-\frac{L^2}{B(x)C(x)}},\;0,\; \frac{L}{C(x)}\Big).
\end{equation}
For the sake of the geodesic motion, the relevant quantity is the ratio of $L$ and $E$, which is called the impact parameter 
\begin{equation}\label{eq:impact_parameter1}
	b\equiv \frac{L}{E} 
	=\frac{C(x)\dot{\phi}}{A(x)\dot{t}},
\end{equation}
where $b$ is a dimensionless parameter. The radial effective potential and its derivatives with respect to the radial coordinate $x$ for the photon read as 
\begin{eqnarray}
	V_{\text{eff}}&=&E^2-\left(\frac{E^2}{A(x)B(x)}-\frac{L^2}{B(x)C(x)}\right)\label{v1}\\
	&=& \frac{E^2k^2}{x^2}+\frac{L^2}{x^5} \left(x \sqrt{1-\frac{k^2}{x^2}}-1\right) \left(x^2-k^2\right), \label{v}\\
	V_{\text{eff}}'&=&\frac{k^2 \left(L^2 \left(5 x \sqrt{1-\frac{k^2}{x^2}}-5\right)-2 E^2 x^3\right)}{x^6}+\frac{L^2\left(3-2 x \sqrt{1-\frac{k^2}{x^2}}\right)}{x^4}.
\end{eqnarray}
\begin{figure*}
	\centering
	\includegraphics[scale=0.85]{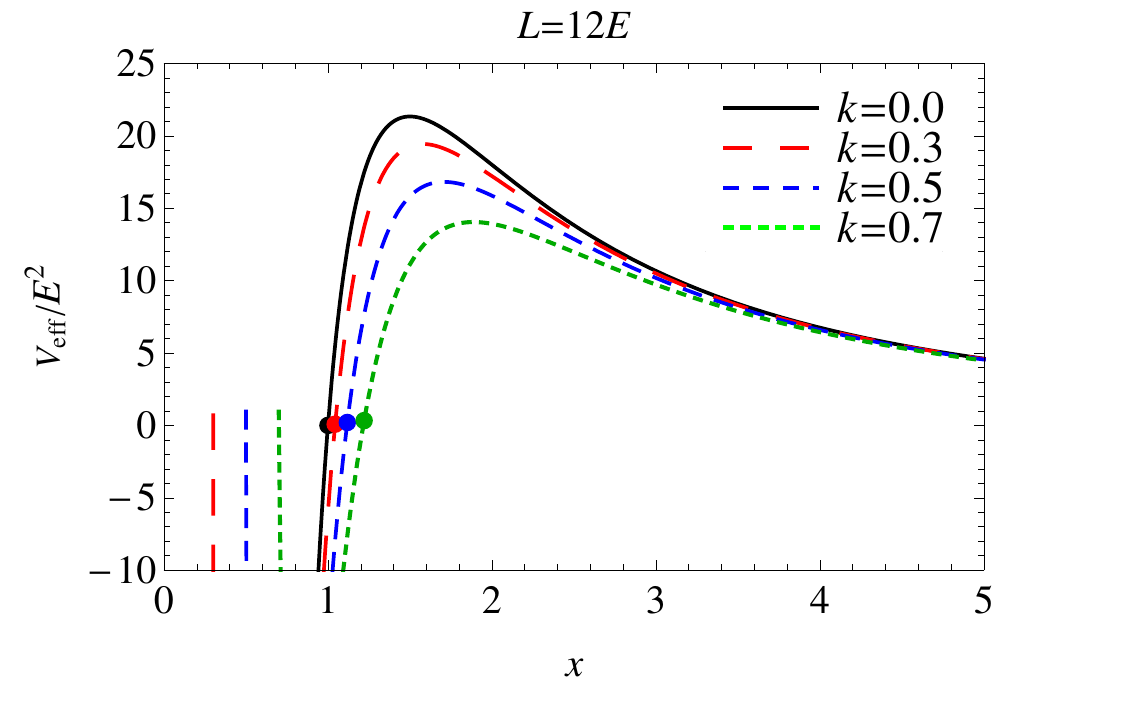}
	\caption{Comparing the radial effective potential of photons with various $k$ values to that of a Schwarzschild black hole (solid black line). $V_{\text{eff}}$ vanishes at asymptotically large distances. For $k\neq0$, $V_{\text{eff}}$ has local minima at $x<x_+$ and maxima at $x>x_+$. The colored points depict the horizon location. For $k\neq 0$, $V_{\text{eff}}$ does not vanish at the horizon but rather $V_{\text{eff}}(x_+)<E^2$. }\label{fig:potential}
\end{figure*}
From Eq.~(\ref{Rpot}), it is clear that light rays can propagate only in the region with $E^2\geq V_{\text{eff}}$. Furthermore, being an  asymptotically-flat spacetime, we obtain $\lim\limits_{x \rightarrow \infty}V_{\text{eff}}=0$, thus, the photon can exist at infinity $x\rightarrow \infty$ with $\dot{x}^2=E^2\geq 0$.

We assume that the light ray coming from a far distant source gets deflected at the closest distance $x = x_0$ from a black hole, and it goes to the observer. At the distance of the closest approach, sometimes called as radial turning point radius $x=x_0$, the energy of photons matches with the radial effective potential, such that $V_{\text{eff}}=E^2$ giving  $\dot{x}=0$. 
Geodesics equations can be recast as follows
\begin{eqnarray}
	\frac{d\phi}{dx}&=&\frac{\dot{\phi}}{\dot{x}} = \pm \frac{1}{E\sqrt{\frac{C(x)}{B(x)}\left(\frac{C(x)}{A(x)b^2}-1\right)}}.\label{phiEq}
\end{eqnarray}
At the distance of closest approach, $\frac{dx}{d\phi}=0$ or $V_{\text{eff}}=E^2$  in Eq.~(\ref{v1}), this yields
\begin{align}
	b=\sqrt{\frac{C_0}{A_0}},\label{Imp}
\end{align}
providing that $B(x)$ and $C(x)$ are nonzero at $x\sim x_0>x_{+}$. Here and hereafter, all the quantities with subscript ``0" are evaluated at $x=x_0$. Because the impact parameter is constant along the geodesics, with the knowledge of $b$, we can determine the distance of closest approach $x_0$ using Eq.~(\ref{Imp}). 
In the close vicinity of the black hole horizon, light rays experience strong gravitational deflection and can make close circular orbits, such that $\dot{x}=\ddot{x}=0$ at $x=x_c>x_+$, which happens for photons having the same energy as the radial potential maximum. $x=x_c$ is the unstable circular photon orbit radius, which can be determined by 
$$V_{\text{eff}}=E^2,\;\; V_{\text{eff}}'=0\; \text{and}\; \lim\limits_{x_0\to x_c} V_{\text{eff}}(x_0)''\geq 0,$$ 
leading to
\begin{equation}
	D(x) \equiv \frac{C'(x)}{C(x)}-\frac{A'(x)}{A(x)}, \label{D1}
\end{equation}
$x_c$ is the largest positive root of $D(x_c)=0$, obtained as
\begin{equation}
	x_c= \frac{1}{2} \sqrt{\frac{3}{2}} \sqrt{4 k^2+3+\sqrt{8 k^2+9}}.
\end{equation}
Light rays with the critical value of impact parameter $b_c=\lim\limits_{x_0\to x_c}b(x_0)$ make an infinite number of loops along the unstable circular orbit of critical radius $x_c$. These orbits are radially unstable, as small radial perturbations drive these photons into the black hole or toward spatial infinity \cite{Chandrasekhar:1985kt}. Indeed, these are the photons that can go closest to the black hole and still escape the black hole and manage to reach a distant observer, therefore, the distance of closest approach is $x_0\geq x_c$. These orbits eventually construct a timelike spherical surface known as the photon sphere, which appears as a \emph{critical curve} on the image plane \cite{Bozza:2001xd}. 

The circular photon orbit radius increases with the polymer parameter $k$, and for $k=0$, the orbit radius is $x_c=3/2$. Therefore, $b_c$ is the minimum value of the impact parameter for continuum turning of light geodesics, such that light rays with the impact parameter $b < b_c$ are captured by the black hole, while those with $b > b_c$ are deflected. These captured light rays eventually fall into a black hole and cast the black hole's \textit{shadow}.

Fig.~\ref{fig:potential} shows the dimensionless radial effective potential for the light rays around the polymerized black hole. For comparison $V_{\text{eff}}$ for the Schwarzschild black hole is also shown. It is evident that $k\neq 0$ causes significant changes in the effective potential. Indeed, the radial effective potential possesses both local maxima and minima, which, respectively, lead to the formation of unstable and stable  photon circular orbits. Therefore, in contrast to the Schwarzschild black hole, around a polymerized black hole anti-photon sphere (stable bound orbits) is also possible. These stable photon orbits, however, invariably occur inside the event horizon and hence have no observational value. In what follows, we will discuss gravitational lensing only around the outer photon sphere.

The light rays traveling from the source to a observer suffer the total deflection angle $\alpha_D$, that is given by \cite{Virbhadra:1999nm,Bozza:2002zj}
\begin{equation}
	\alpha_D (x_0)= I(x_0)-\pi,\label{DefAng1}
\end{equation}
with 
\begin{eqnarray}\label{DefAng2}
	I(x_0)&=&\int_{x_0}^{D_{LS}} \frac{d\phi}{dx}dx+ \int_{x_0}^{D_{OL}} \frac{d\phi}{dx}dx\nonumber\\
	&=&2\int_{x_0}^\infty \frac{d\phi}{dx}dx=\int_{x_0}^\infty \frac{2\, dx}{\sqrt{\frac{C(x)}{B(x)}\left(\frac{C(x)A(x_0)}{C(x_0)A(x)}-1\right)}}.
\end{eqnarray}
Here, $D_{LS}$ and $D_{OL}$, respectively, are the source and observer distances from the black hole, which in this case can be taken as $D_{LS}\to \infty$ and $D_{OL}\to \infty$. It is clear that in the absence of the black hole's gravitational field ($A(x)=B(x)=1, C(x)=x^2$), the $I(x_0)=\pi$ and deflection angle vanish. We name the limits $x_0\gtrsim x_c$ and $x_0\gg x_c$, respectively, as the strong deflection limit and the weak deflection limit.

\subsection{Weak gravitational lensing}
We begin by analyzing the weak gravitational lensing for light rays with a large impact parameter $b\gg b_c$, such that the closest approach distance $x_0$ is very large compared to the photon orbit radius $x_c$. In this case, light rays starting from the source reach the observer without winding around the black hole as the deflection angle is smaller than $2\pi$. Using Eq.~(\ref{DefAng2})
\begin{equation}
	I(x_0)=\int_{x_0}^{\infty}\frac{2\; dx}{\sqrt{\frac{\left(x^2-k^2\right) \left(x^3 \left(\sqrt{x_0^2 -k^2}-1\right)-x_0^3 \left(\sqrt{x^2-k^2}-1\right)\right)}{x x_0^3}}}.\label{DefAng3}
\end{equation}
Let us define $w=\frac{x_0}{x}$ and make Taylor series expansion of the integrand in Eq.~(\ref{DefAng3})
\begin{equation}
	I(x_0)=\int_{0}^{1} 2 f(w)\;dw,\label{DefAng4}
\end{equation}
with 
\begin{eqnarray}
	f(w)&\sim & \frac{1}{ \sqrt{1-w^2}} -\frac{\left(w^3-1\right) }{2 x_0 \left(1-w^2\right)^{3/2}}+ \frac{ \left(2 k^2 \left(3 w^2+1\right) \left(w^2-1\right)^2+3 \left(w^3-1\right)^2\right)}{8 x_0^2 \left(1-w^2\right)^{5/2}} \nonumber\\
	&&- \frac{\left(w^3-1\right) \left(2 k^2 \left(5 w^2+3\right) \left(w^2-1\right)^2+5 \left(w^3-1\right)^2\right)}{16 x_0^3 \left(1-w^2\right)^{7/2}} + \mathcal{O}\left(\frac{1}{x_0^4}\right).\label{fy}
\end{eqnarray}
Integrating $f(w)$ as in Eq.~(\ref{DefAng4}) and using Eq.~(\ref{DefAng1}), the deflection angle reads as
\begin{eqnarray}
	\alpha(x_0)&=& \frac{2}{x_0} + \frac{1}{x_0^2}\left(-1+\frac{15 \pi }{16}+ \frac{5 \pi  k^2}{8}\right)- \frac{1}{x_0^3}\left(\frac{15 \pi }{16} -\frac{61}{12}  \frac{29 k^2}{6}+\frac{5 \pi  k^2}{8}\right) +\mathcal{O}\left(\frac{1}{x_0^4}\right).\label{DefAng5}
\end{eqnarray}
Clearly, the deflection angle inversely depends on the distance of  closest approach to the black hole $x_0$. Using Eq.~(\ref{Imp}), we expand the impact parameter in the series of $1/x_0$ and this gives
\begin{equation}
	\frac{1}{x_0}=\frac{1}{b}+ \frac{1}{2 b^2} +\frac{5+2 k^2}{8 b^3} +\mathcal{O}\left(\frac{1}{b^4}\right).
\end{equation}
Inserting the above expression in Eq.~(\ref{DefAng5}), we obtain
\begin{eqnarray}
	\alpha_D(u)&=&\frac{4 M}{u}+ \frac{4M^2}{u^2} \left(\frac{15 \pi }{16}+\frac{5 \pi  k^2}{8}\right) + \frac{128 M^3 \left(1+k^2\right)}{3 u^3}+\mathcal{O}\left(\frac{1}{u^4}\right),\label{DefAng6}
\end{eqnarray}
where $u= 2M b$ is the re-scaled impact parameter with dimension of length. For the limit $k \to 0$, the Eq.~(\ref{DefAng6}) recovers the deflection angle for the Schwarzschild black hole, which reads as \cite{Virbhadra:1999nm,Bozza:2001xd}
\begin{equation}
	\alpha_D(u)|_{\text{Sch}}=\frac{4 M}{u}+ \frac{15\pi M^2}{4u^2} + \frac{128 M^3}{3 u^3} +\mathcal{O}\left(\frac{1}{u^4}\right).\label{DefAngSch}
\end{equation}
\begin{figure}
		\centering
	\includegraphics[scale=0.74]{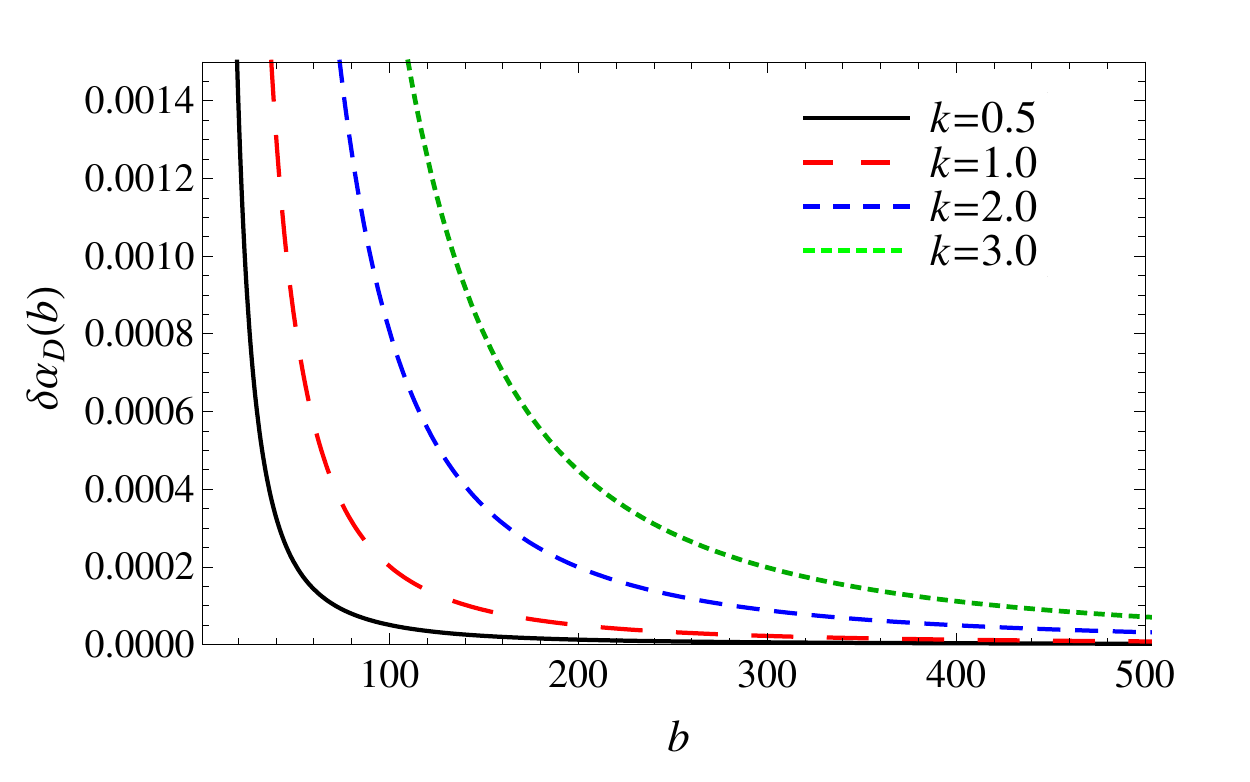}
	\caption{The correction in the light deflection angle $\delta\alpha_D=\alpha_D-\alpha_D|_{Sch}$ for the polymerized black hole from the Schwarzschild black hole. $\delta\alpha_D$ is in units of radian and $b$ is a dimensionless impact parameter. }\label{fig:WeakDef}
\end{figure}
It is clear that quantum effects, by virtue of polymer quantization, contribute positively to the deflection angle, viz., polymerized black hole produces a larger deflection angle than the Schwarzschild black hole even in the weak-deflection limit. The primary quantum correction to the weak deflection angle is of the order of $u^{-2}$. Fig.~\ref{fig:WeakDef} depicts this correction in the deflection angle; $\delta\alpha_D$ is more significant for smaller values of $b$. However, Fig.~\ref{fig:WeakDef} gives the correct description only for $b\gg b_c$. For $b=500$, the correction in polymerized black hole deflection angle with $k=0.5$ is $0.4072$ milli-arc-seconds.

\subsection{Strong gravitational lensing}
As the distance of minimum approach $x_0$ decreases, the deflection angle increases and eventually diverges (cf. from Eq.~(\ref{DefAng5})). However, to get the correct order of divergence for the strong gravitational lensing, we shall expand the deflection angle near the photon sphere $x_0\sim x_c$. For this purpose, we define a new variable $z$ as in Ref.~ \cite{Tsukamoto:2016jzh}
\begin{eqnarray}
	z &=& 1-\frac{x_0}{x}.\label{z}
\end{eqnarray} 
It is worth noting here that an alternate choice of $z$ can also be made, such as $z=\frac{A(x)-A(x_0)}{1-A(x_0)}$, as discussed by Bozza in Ref.~\cite{Bozza:2002zj}. However, the advantage of adopting the choice in Eq.~(\ref{z}) is that it is equally valid for ultra static spacetime (such as wormhole) with constant $A(x)$. Secondly, for other choices of $z$, it is not always possible to obtain the analytic expressions of deflection angle in the strong deflection limit \cite{Tsukamoto:2016oca}.

Using Eq.~(\ref{z}), the integral in Eqs.~(\ref{DefAng2}) and (\ref{DefAng3}) can be re-written as 
\begin{align}
	I(x_0) &= \int_{0}^{1}f(z,x_0) dz,\label{SDefAng1}
\end{align}
with the function
\begin{align}
	f(z,x_0)&=  \frac{2 x_0}{(1-z)^2}\frac{ 1}{\sqrt{\frac{C(x)}{B(x)}\left(\frac{C(x)A(x_0)}{C(x_0)A(x)}-1\right)}}.
\end{align}
Inserting metric functions for the polymerized black hole and $x=\frac{x_0}{1-z}$, this leads to
\begin{eqnarray}
	f(z,x_0)&=&  2x_0^{3/2} \Big\{ (k^2 (z-1)^2-x_0^2)\Big(-\sqrt{x_0^2-k^2}+ \sqrt{x_0^2-k^2 (z-1)^2} \nonumber\\
	&& + z\Big(3-2 \sqrt{x_0^2-k^2 (z-1)^2} + z (z-3+\sqrt{x_0^2-k^2 (z-1)^2})         \Big)\Big)\Big\}^{-1/2}\nonumber\\
	&\equiv&   \frac{2 x_0 }{\sqrt{G(z,x_0)}}.\label{Sf}
\end{eqnarray}
Making the Taylor series expansion of the integrand $G(z,x_0)$ yields
\begin{equation}
	G(z,x_0)=\sqrt{\phi_1 z+ \phi_2 z^2+\mathcal{O}(z^3)},\label{G3}
\end{equation}
with 
\begin{align}
	\phi_1=&  \frac{x_0 \Big(C'(x_0) A(x_0)-A'(x_0) C(x_0)\Big)}{A(x_0) B(x_0)},\nonumber\\
	\phi_2=&\frac{-x_0}{B(x_0)}\Big(3C'(x_0)-\frac{x_0C''(x_0)}{2}+\frac{x_0 C'(x_0)B'(x_0)}{B(x_0)}\Big)+ \frac{x_0^2 C'^2(x_0)}{B(x_0)C(x_0)}\nonumber\\
	&+\frac{3x_0C(x_0)}{A(x_0)B(x_0)}\Big( A'(x_0) -\frac{x_0 A''(x_0)}{6} \Big) - \frac{x_0^2 C(x_0)A'(x_0)}{A(x_0)B(x_0)}\Big(\frac{2 C'(x_0)}{C(x_0)} - \frac{ A'(x_0)}{A(x_0)} - \frac{B'(x_0)}{B(x_0)} \Big).\label{Taylor1}
\end{align}
For the polymerized black hole, these coefficients take the form
\begin{eqnarray}
	\phi_1&=& -3 x_0 +2 x_0 \sqrt{x_0^2-k^2} +\frac{3 k^2}{x_0} (1-\sqrt{x_0^2-k^2}),\nonumber\\
	\phi_2&=&  3 x_0 -\frac{9 k^2}{x_0} +\frac{\Big(18k^4+17 k^2 x_0^2 -2 x_0^4\Big)}{2 x_0 \sqrt{x_0^2-k^2}}.
\end{eqnarray}
In the strong deflection limit $x_0\to x_c$, from Eqs.~(\ref{D1}) and (\ref{Taylor1})
\begin{align}
	&\phi_1\to 0,\nonumber\\
	&\phi_2\to \frac{x_c^2 C(x_c)}{2 B(x_c)}\Big(\frac{C''(x_c)}{C(x_c)} -\frac{A''(x_c)}{A(x_c)} \Big).\label{phi2}
\end{align} 
Therefore, in the limit $x_0\to x_c$, the leading order of the divergence of the integrand term  $f(z,x_0)$ is $z^{-1}$ and that the integral $I(x_0)$ and deflection angle diverge logarithmically, i.e., $\alpha_D\propto \log(z)\propto \log\left(1-\frac{x_c}{x}\right)$  (cf. Eq.~(\ref{Sf})). It is the reason why we truncate the Taylor series expansion in Eq.~(\ref{G3}) at $z^2$. The sole idea is to identify the diverging term and order of divergence, so that the diverging term can be subtracted from $I(x_0)$ to get the regular term $I_R(x_0)$. Dividing the integral $I(x_0)$ into two parts, diverging $I_D(x_0)$ and regular $I_R(x_0)$, the $I_D(x_0)$  reads as
\begin{eqnarray}
	I(x_0)&=&I_D(x_0)+ I_R(x_0),\label{def}\\
	f(z,x_0)&=& f_D(z,x_0) + f_R(z,x_0),\\
	I_D(x_0)&=&\int_{0}^{1}\frac{2 x_0\, dz}{\sqrt{\phi_1 z+ \phi _2 z^2}}.
\end{eqnarray}
Upon integration, the diverging term contribution reads as
\begin{equation}
	I_D(x_0)=\frac{4 x_0}{\sqrt{\phi_2}}\log\Big( \frac{\sqrt{\phi_2}+ \sqrt{\phi_1+ \phi_2}}{\sqrt{\phi_1}}\Big).\label{SD}
\end{equation}
Expanding $\phi_1$ in the close vicinity of $x_c$
\begin{eqnarray}
	\phi_1&=&\frac{x_c C(x_c)}{B(x_c)} \Big(\frac{C''(x_c)}{C(x_c)} -\frac{A''(x_c)}{A(x_c)} \Big)(x_0-x_c) + \mathcal{O}((x_0-x_c)^2).\label{phi1}
\end{eqnarray}
To get the coordinate independent expression of the deflection angle, we express it in terms of the impact parameter. For this purpose, using Eq.~(\ref{Imp}) and making a Taylor expansion around $x_0\sim x_c$, we get
\begin{eqnarray}
	b(x_0)&=&\sqrt{\frac{C(x_c)+(x_0-x_c)C'(x_c) +\frac{1}{2}(x_0-x_c)^2 C''(x_c)}{A(x_c)+(x_0-x_c)A'(x_c) +\frac{1}{2}(x_0-x_c)^2 A''(x_c)}},\nonumber\\
	b(x_0)&=& b(x_c)+ \frac{(x_0-x_c)^2}{4}\sqrt{\frac{C(x_c)}{A(x_c)}} \Big(\frac{C''(x_c)}{C(x_c)} -\frac{A''(x_c)}{A(x_c)} \Big)+ \mathcal{O}((x_0-x_c)^2).\label{SImp}
\end{eqnarray}
Upon eliminating $(x_0-x_c)$ from Eq.~(\ref{phi1}) and using Eq.~(\ref{SImp}), the coefficient $\phi_1$ in the $x_{0}\to x_{c}$ reads as
\begin{equation}
	\phi_{1} =  \frac{2C(x_c)x_c}{B(x_c)} \Big(\frac{C''(x_c)}{C(x_c)} -\frac{A''(x_c)}{A(x_c)} \Big)^{1/2} \left( \frac{b}{b_{c}}-1 \right)^{\frac{1}{2}}.\label{phi1-1}
\end{equation}
Substituting Eqs.~(\ref{phi2}) and (\ref{phi1-1}) into (\ref{SD}), we obtain the divergent part $I_{D}(b)$ of the deflection angle in the strong deflection limit \cite{Tsukamoto:2016jzh}
\begin{align}\label{eq:IDm1}
	I_{D}(b)
	&= {\frac{x_c}{\sqrt{\phi_2}}} \log x^{2}_{m}\Big(\frac{C''(x_c)}{C(x_c)} -\frac{A''(x_c)}{A(x_c)} \Big) -{\frac{x_c}{\sqrt{\phi_2}}} \log \left( \frac{b}{b_{c}}-1 \right)+\mathcal{O}((b-b_{c})\log (b-b_{c})).
\end{align}

To determine the regular part in the deflection angle Eq.~(\ref{def}), we first make the Taylor series expansion of $f_R(z,x_0)$ around $x_0\sim x_c$ and then  integrate. 
\begin{align}\label{SR}
	I_{R}(x_c)&
	\equiv \int^{1}_{0} f_{R}(z,x_c)dz.
\end{align}
This integration can be done analytically, however, it results in a long and complicated expression for the polymerized black hole. Here, we completed our deflection angle calculation, which can be rewritten in a compact form as follows \cite{Tsukamoto:2016jzh}
\begin{equation}\label{DefAng0}
	\alpha_D(u)=-\bar{p}\log \left( \frac{u}{u_{c}}-1 \right) +\bar{q} +\mathcal{O}((u-u_{c})\log(u-u_{c})),
\end{equation}
where $\bar{p}$ and $\bar{q}$ are given by
\begin{equation}\label{eq:abar1}
	\bar{p}=\sqrt{\frac{2B(x_c)A(x_c)}{C^{''}(x_c)A(x_c)-C(x_c)A^{''}(x_c)}}
\end{equation}
and
\begin{equation}\label{eq:bbar1}
	\bar{q}=\bar{p}\log \left[x_c^{2}\left(\frac{C^{''}(x_c)}{C(x_c)}-\frac{A^{''}(x_c)}{A(x_c)}\right)\right] +I_{R}(x_{c})-\pi,
\end{equation}
respectively, where $I_R(x_c)$ is given by the Eq.~(\ref{SR}). Here, $\bar{p}$ and $\bar{q}$ are called the strong deflection limit coefficients. Contribution from the regular part is in the $\bar{q}$ expression. Therefore, in the strong deflection limit $x\to x_c$ or $b\to b_c$, the deflection angle diverges logarithmically (cf. Eq.~(\ref{DefAng0})).  

\begin{figure}
	\centering
	\includegraphics[scale=0.78]{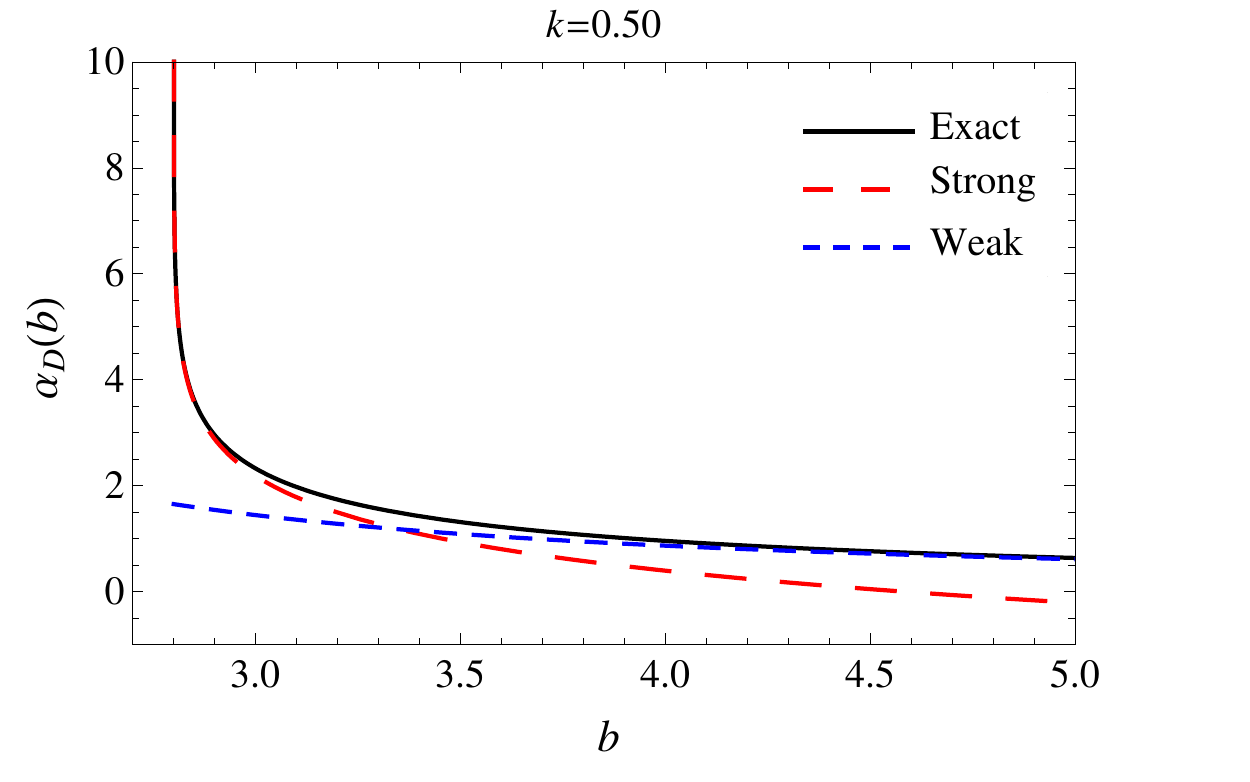}
	\caption{Comparison of the exact numerical deflection angle (black solid line) with the approximate deflection angle in the strong (red dashed line) and weak (blue small dashed line) field limits. For this case, the critical impact parameter is $b_c=2.8012$. $\alpha_D$ is in units of radians.}\label{fig:DefAng1}
\end{figure}

It is interesting to find the value of the impact parameter with which light rays make $n$ complete loops around the black hole with a deflection angle  $\alpha_D=2\pi n$. Inverting the relation in Eq.~(\ref{DefAng0}), the corresponding value of impact parameter reads 
\begin{equation}
	u_n=u_c\Big(1+e^{(\frac{\bar{q}-2\pi n}{\bar{p}})}\Big).
\end{equation}
It is clear that for large $n$, impact parameters exponentially approach the critical value with which light rays make an infinite winding around the black hole. For instance, light rays moving around a polymerized black hole with $k=0.10$ make the first three complete loops for impact parameters $u_n$ and orbit radii $x_n$ as $u_1= 6.61319,\; u_2=6.60633,\; u_3=6.60632071$ and $x_1=2.09412, \, x_2=2.04447,\; x_3=2.042342927$, whereas  $u_c=6.60632069$ and $x_c=2.0423429216$.
\begin{center}
	\begin{figure*}
		\begin{centering}
			\begin{tabular}{c c c c}
				\includegraphics[scale=0.7]{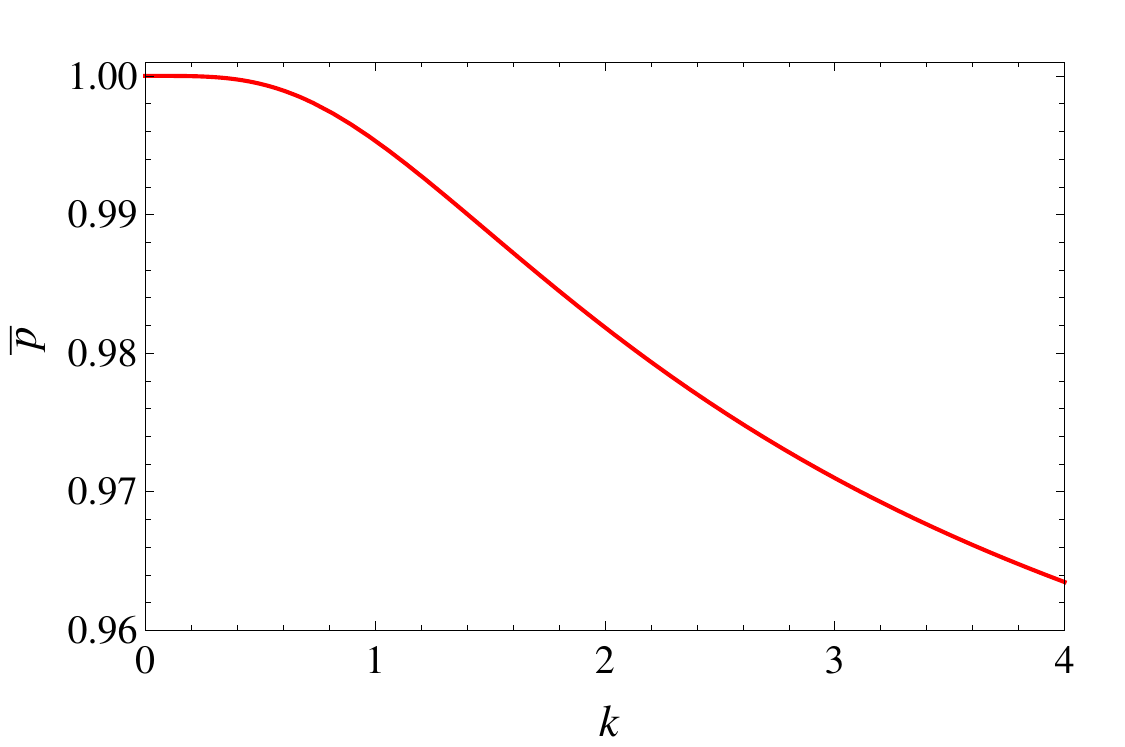}\hspace*{-0.4cm}&
				\includegraphics[scale=0.7]{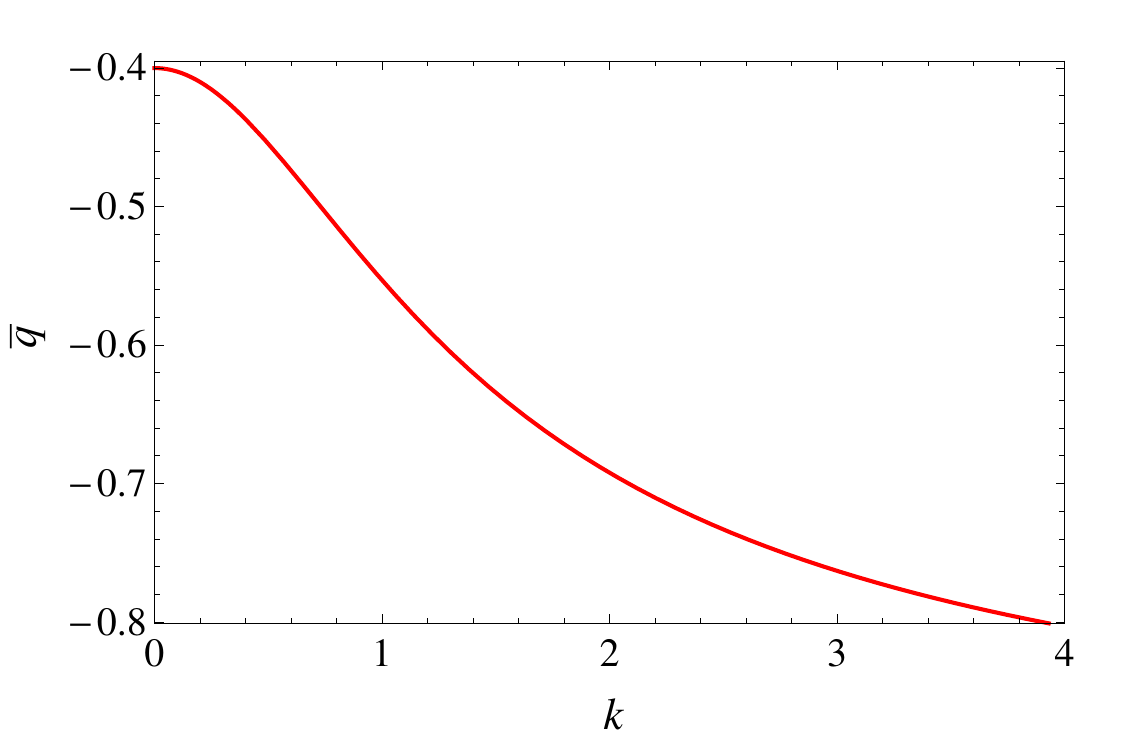}
			\end{tabular}
		\end{centering}
		\caption{The strong lensing coefficients $\bar{p}$ and $\bar{q}$ as a function of $k$.}\label{fig:LensCoeff}		
	\end{figure*}
	\begin{figure*}
		\begin{center}	
			\begin{tabular}{c c}
				\includegraphics[scale=0.68]{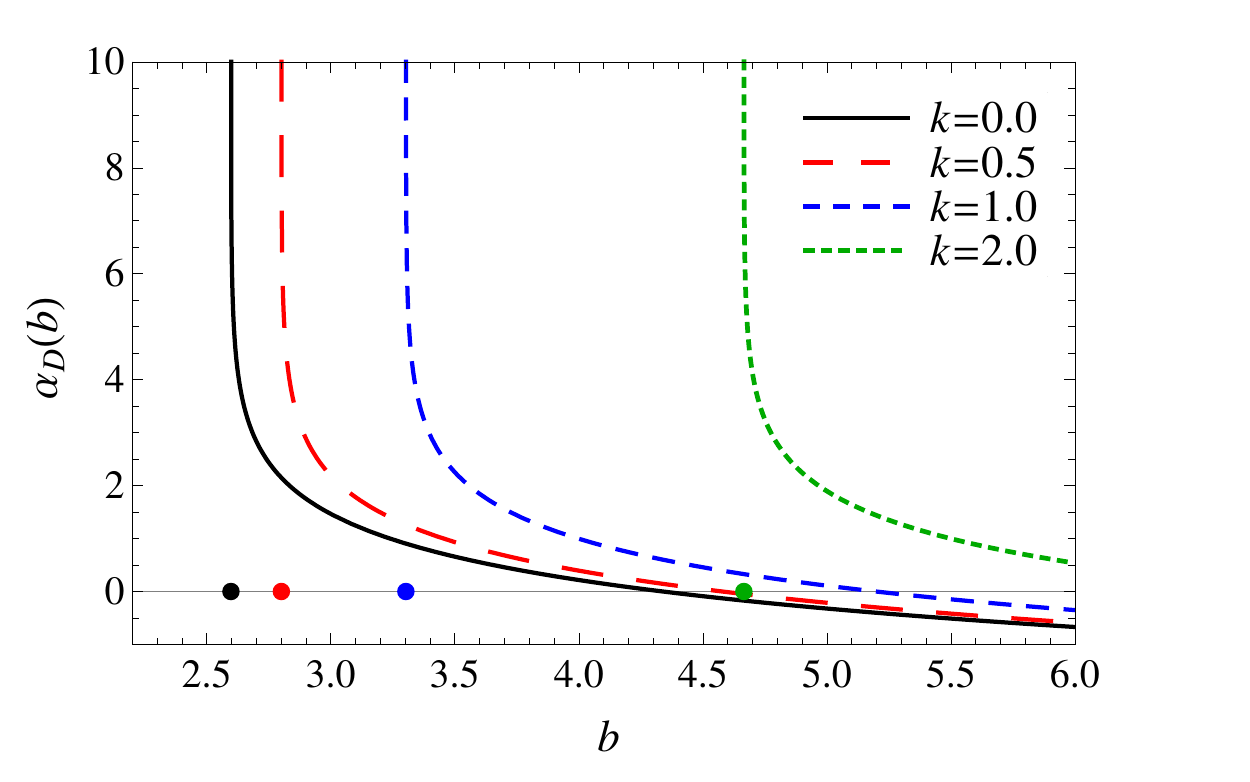}\hspace*{-0.6cm}&
				\includegraphics[scale=0.68]{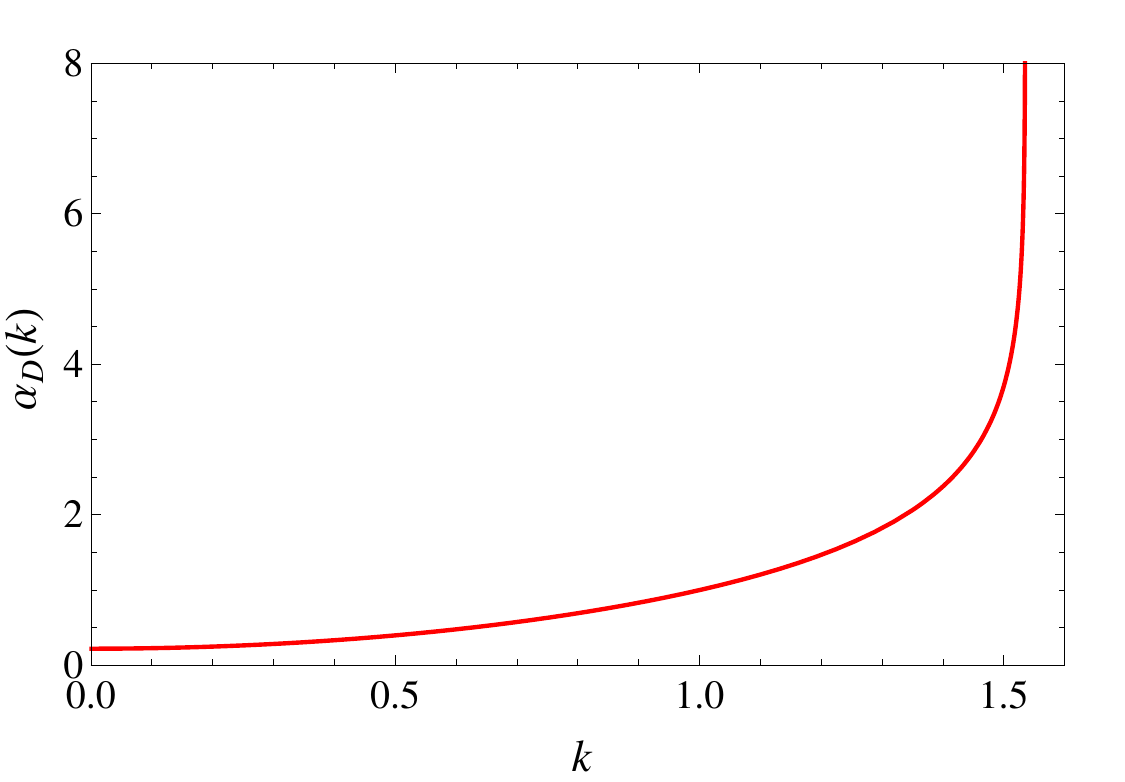}
			\end{tabular}
			\caption{(\textit{Left:}) Deflection angle as a function of impact parameter $b$ for different values of $k$.  Colored points on the horizontal axis correspond to the impact parameter $b=b_c$ at which the deflection angle diverges logarithmically. $\alpha_D$ is in units of radians. (\textit{Right:}) Deflection angle variation with  $k$ for $b=4$. }\label{fig:DefAng}	
		\end{center}	
	\end{figure*}
\end{center}
Fig.~\ref{fig:LensCoeff} depicts the lensing coefficients as a function of $k$. Both $\bar{p}$ and $\bar{q}$ decrease with $k$, and for $k=0$ they smoothly retain the values for the Schwarzschild black hole, viz., $\bar{p}=1$ and $\bar{q}=-0.4002$ \cite{Bozza:2001xd,Bozza:2002zj}. In Fig.~\ref{fig:DefAng} (left figure), the deflection angle $\alpha_D(u)$ is shown as a function of impact parameter $u$ for various values of $k$. Deflection angle decreases with impact parameter and only for $u=u_c$ deflection angle shows divergence. The direct effect of $k$ on the deflection angle $\alpha_D(u)$ is shown in Fig.~\ref{fig:DefAng} (right figure), where we fixed the light impact parameter and plotted $\alpha_D(u)$ as a function of $k$. $\alpha_D(u)$ increases with $k$. Therefore, the deflection angle is higher for the polymerized black holes than for the Schwarzschild black holes. Because the amount of deflection of light rays varies with the $k$, the light density received by the distant observers will be different, which naturally leads to the different observation intensity caused by the black hole shadow, as we will see in the next section.
In Fig.~\ref{fig:DefAng1}, we plotted the deflection angle analytically calculated in the weak (\ref{DefAng6}) and strong (\ref{DefAng0}) field limits as a function of $u$ for $k=0.50$ and compared it with the exact deflection angle calculated numerically using (\ref{DefAng1}). The strong deflection angle is an excellent approximation for photons passing close to the photon sphere.

\section{Lensing observables}\label{Sec-4}
Light rays traveling from the source to the observer are deflected, near a black hole, from their original path by an angle $\alpha_D$ and thus the observer sees the image of the source at the angular position $\theta$, which is different from the source position $\beta$. The lens equation establishes the relation between the image position and the source position in terms of the deflection angle by the following relation \cite{Virbhadra:1999nm,Bozza:2002zj,Bozza:2007gt}
\begin{eqnarray}
	\epsilon D_{OS}\tan\beta &=& \frac{D_{OL}\sin\theta -\epsilon  D_{LS}\sin(\alpha_D-\theta)}{\cos(\alpha_D-\theta)}.\label{lensEq}
\end{eqnarray}
This equation is known as the Virbhadra--Ellis lens equation \cite{Virbhadra:1999nm}. As shown in the previous section, in the strong deflection limit, light can make $n\in N$ numbers of loops around the black hole before escaping to the observer, and the effective deflection angle can be defined as the $\bar{\alpha}_D=\alpha_D-2\pi n$, such that $2\pi n$ is the deflection angle for the $n$ complete loops. All the angles in Eq.~(\ref{lensEq}) should be within $(0, 2\pi)$ and thus we shall use the deflection angle $\bar{\alpha}_D$ rather $\alpha_D$. We assume that the source and images are close to the optical axis, thereby in the small-angle limit, $\beta\ll 1, \theta=b/D_{OL}\ll 1$, expectedly deflection angle is also small $\bar{\alpha}_D\ll 1$ and Eq.~(\ref{lensEq}) reduces to the following form \cite{Bozza:2007gt, Eiroa:2012rw}
\begin{eqnarray}\label{beta}
	\epsilon\,\beta &=& \theta - \epsilon \frac{D_{LS}}{D_{OS}}\bar{\alpha}_D.
\end{eqnarray}
Here, $\epsilon$ takes the values $+1$ or $-1$, respectively, describing the direction of light propagation around the black hole. For $\epsilon =+1(-1)$, light rays cross from the front side (backside) of the black hole, and the resulting images appear on the same side (opposite side) of the optical axis with respect to the source. These images are called as the primary ($\epsilon=1$) and secondary ($\epsilon=-1$) images. For $b\ll b_c$ the deflection angle is always smaller than $2\pi$ and the resulting images are called \textit{direct images}, whereas if the impact parameter is close to its critical value, then photons can make several loops around the black hole before escaping to the observer, such photons make higher-order images, commonly known as \textit{relativistic images}, as introduced by Virbhadra and Ellis \cite{Virbhadra:1999nm,Virbhadra:2008ws}. It is important to note that, while there are multiple relativistic images on one side of the optical axis, there can be only one primary image that forms due to light deflection in the weak-field without looping of the light ray around the lens \cite{Virbhadra:2008ws}.
These direct and relativistic images form on both the same and opposite sides of the source, which are characterized by the value of $\epsilon$. Hereafter, we use subscripts $p$ and $s$, respectively, for primary and secondary images. Similarly, $\theta_{p}$ and $\theta_{s}$ defines, respectively, \textit{direct} primary and secondary images with zero winding around the black hole, whereas, $\theta_{np}$ and $\theta_{ns}$ stand, respectively, for the \textit{relativistic} primary and secondary images of order $n$, such that $n=1$ corresponds to the outermost relativistic image.

To obtain the image position, the scheme is to first calculate the deflection angle $\alpha_D$ in terms of the impact parameter and then express it in terms of $\theta$ using $\theta=b/D_{OL}$ and substitute in Eq.~(\ref{beta}). Then, for a given source position $\beta$, we solve the lens Eq.~(\ref{beta}) for the image position $\theta$.
Expanding the deflection angle about $(\theta_n ^0)$ to the first order \cite{Tsukamoto:2016jzh}
\begin{equation}\label{Taylor}
	\alpha_D(\theta_n) = \alpha_D(\theta_n ^0) +\frac{\partial \alpha_D(\theta_n)}{\partial \theta } \Bigg |_{\theta_n ^0}(\theta_n-\theta_n ^0)+\mathcal{O}(\theta_n-\theta_n ^0)^2,
\end{equation}
where $\theta_n ^0$ and $\alpha_D(\theta_n ^0)=2\pi n$ are the image position, and the deflection angle for the light rays making exactly $n$ complete loops around the black hole. 
Using $\theta=b/D_{OL}$ in the deflection angle Eq.~(\ref{DefAng0}) \cite{Tsukamoto:2016jzh}
\begin{eqnarray}
	\theta_n^0&=& \theta_{\infty}\Big( 1+e^{\frac{\bar{q}-2\pi n}{\bar{p}}}\Big),
\end{eqnarray}
where $\theta_{\infty}=b_c/D_{OL}$ is the critical curve or photon sphere angular radius. Clearly, the relativistic image angular position $\theta_n^{0}$ rapidly decrease with $n$ and eventually approach the photon sphere, $\theta_n^{0}\to \theta_n^{\infty}$, as the $n\to \infty$. Differentiating deflection angle with respect to $\theta$ in Eq.~(\ref{DefAng0}) 
\begin{equation}
	\frac{\partial \alpha_D(\theta_n)}{\partial \theta }\Bigg |_{\theta_n ^0}=-\frac{\bar{p}}{\theta_n^0-\theta_{\infty}},
\end{equation}	
and substituting it in Eq.~(\ref{Taylor}) gives
\begin{eqnarray}
	\bar{\alpha}_D&=&\alpha_D(\theta_n) - \alpha_D(\theta_n ^0)= -\frac{\bar{p}(\theta_n - \theta_n^0)}{\theta_{\infty} e^{\frac{\bar{q}-2\pi n}{\bar{p}}}}.\nonumber
\end{eqnarray}
This gives the excess in deflection angle to the $2\pi n$ for the light rays making $n$ complete loops around the black hole. Angular position of the $n^{th}$ relativistic image can be determined by solving Eq.~(\ref{beta}) for $\theta$ and keeping only lowest order terms in $\theta_{\infty}$ \cite{Bozza:2002zj}
\begin{eqnarray}
	\theta_n&=&\frac{\epsilon\, D_{LS}\bar{p}\theta_{n}^0 + \epsilon\, D_{OS}\beta\theta_{\infty}e^{\frac{\bar{q}-2\pi n}{\bar{p}}}}{\epsilon\, D_{LS}\bar{p}+D_{OS}\theta_{\infty}e^{\frac{\bar{q}-2\pi n}{\bar{p}}}},\nonumber\\
	\theta_n&\approx & \;\epsilon\, \theta_{n}^{0}+(\beta-\epsilon\,\theta_n^0)	\frac{D_{OS}\theta_{\infty}}{D_{LS}\bar{p}}e^{\frac{\bar{q}-2\pi n}{\bar{p}}}.\label{imgposition}
\end{eqnarray}
This defines the first lensing observables. The correction to $\theta_n^0$, given in the second term, falls very rapidly with the winding number. Interestingly, for the $b_c\ll D_{OL}$, $\theta_n$ is not sensitive to the source position and is mainly defined by $\theta_n^0$. 
One particularly interesting case is $\beta=0$, viz., perfect alignment of the source, black hole, and observer. In this case, the light that starts off from the source can reach the observer from all possible directions, and a point-like source appears as a circular ring, known as the Einstein ring \cite{Luminet:1979nyg,Bozza:2004kq}. The angular radii of the $n^{th}$ relativistic Einstein ring can be obtained by substituting $\beta\to 0$ in Eq. ~\ref{imgposition}, given by
\begin{eqnarray}
	\theta_n^E&=&\theta_{\infty}\Big(1  -\frac{D_{OS}\theta_{\infty}}{D_{LS}\bar{p}}e^{\frac{\bar{q}-2\pi n}{\bar{p}}}\Big) \Big( 1+e^{\frac{\bar{q}-2\pi n}{\bar{p}}}\Big).
\end{eqnarray}
Whereas for the non-zero value of $\beta$, the Einstein ring gets broken, and we get multiple images. As $\beta$ increases, primary images move away from the optical axis and always form outside the Einstein ring $\theta_{np}\geq \theta_E$, whereas secondary images move towards the optical axis and always form inside the ring $\theta_{ns}\leq \theta_{E}$. Both sets of images approach the Einstein ring as $\beta\to 0$, merging into a single degenerate ring image of radius $\theta_E$ for $\beta=0$.  Thus, in principle, a black hole produces two infinite sequences of relativistic images whose positions can be calculated numerically using Eq.~(\ref{imgposition}). The deflection angle calculated for the weak deflection limit in Eq.~(\ref{DefAng6}) together with the lens equation (\ref{beta}) is suitable for determining the direct primary and secondary image positions, which reads as \cite{Virbhadra:2008ws}
\begin{equation}
	\theta_{p,s} =\beta + \frac{D_{LS}}{D_{OS}} \left(\frac{2 }{b}+ \frac{1}{b^2} \left(\frac{15 \pi }{16}+\frac{5 \pi  k^2}{8}\right)+\mathcal{O}\left(\frac{1}{b^3}\right) \right),
\end{equation}
where $b=\theta_{p,s} D_{OL} $. Another, important lensing observable is the angular separation between the photon sphere and the outermost relativistic image, which is given by \cite{Bozza:2002zj}
\begin{align}
	s&\equiv \theta_1-\theta_{\infty}\approx \theta_1^0- \theta_{\infty}^0 =\theta_{\infty}e^{\frac{\bar{q}-2\pi }{\bar{p}}}.
\end{align}
The significance of $s$, which is the value to compare with the observation's resolution in order to distinguish amongst a group of relativistic images, is that it is independent of the source position $\beta$. Another effect of gravitational lensing is the so-called magnification effect. The ratio of the image flux (product of its surface brightness and the solid angle it subtends on the sky) to the unlensed source flux is known as the image magnification. However, according to Liouville's theorem, gravitational lensing preserves the surface brightness. Therefore, the image magnification turns out to be the ratio of the solid angles of the image and of the unlensed source made at the observer 
\begin{align}
	\mu&= \frac{\sin\theta}{\sin\beta}\frac{d\theta}{d\beta}\equiv \mu_t\,\mu_r,\\
	\mu_n &= \frac{\theta_{\infty}D_{OS}e^{\frac{\bar{q}-2\pi n}{\bar{p}}}}{\beta \bar{p}D_{LS}}\Big(\theta_n^0+ \frac{D_{OS}\theta_{\infty} e^{\frac{\bar{q}-2\pi n}{\bar{p}}}}{D_{LS}\bar{p}} (\beta-\theta_n^0)   \Big),\\
	\mu_n&\approx\frac{\theta_{\infty}^2D_{OS}e^{\frac{\bar{q}-2\pi n}{\bar{p}}}\Big(1+e^{\frac{\bar{q}-2\pi n}{\bar{p}}}\Big)}{\beta \bar{p}D_{LS}}+\mathcal{O}\Big(\frac{\theta_{\infty}^3}{D_{LS}^2}\Big).\label{mag}
\end{align}
Here, $\mu_t=\frac{\sin\theta}{\sin\beta}$ and $\mu_r=\frac{d\theta}{d\beta}$ are the tangential and radial magnifications, respectively. Divergence of tangential (radial) magnification is called tangential (radial) caustic. Quite evidently, Einstein rings ($\beta=0$) correspond to the tangential caustic. Equation (\ref{mag}) infers that the magnification is very faint unless the lens and the source are highly aligned, and then it linearly diverges for the perfect alignment as in the Einstein ring. Secondly, as expected,  the first relativistic image is the brightest one and $\mu$ falls rapidly for the higher-order relativistic images. Therefore the outermost set of images, one on each side of the optic axis, is observationally the most significant. The sum of magnification of all the images can be calculated as \cite{Bozza:2002zj,Bozza:2008ev}
\begin{equation}
	\mu=\sum_{n=1}^{\infty}\mu_n=\frac{\theta_{\infty}^2D_{OS}\Big( 1+e^{\frac{2\pi}{\bar{p}}}+e^{\frac{\bar{q}}{\bar{p}}}\Big)e^{\frac{\bar{q}}{\bar{p}}}}{\beta \bar{p}D_{LS} (e^{\frac{4\pi}{\bar{p}}-1})}.\label{mag1}
\end{equation}

The lensing observable is the ratio of the magnifications of the outermost image to the sum of the other images, which can be calculated using Eqs.~(\ref{mag}) and (\ref{mag1}) \cite{Bozza:2002zj,Bozza:2008ev}
\begin{eqnarray}
	r_{mag}&=&\frac{\mu_1}{\sum_{n=2}^{\infty}\mu_n}
	\approx \frac{(e^{\frac{4\pi}{\bar{p}}}-1)(e^{\frac{2\pi}{\bar{p}}}+e^{\frac{\bar{q}}{\bar{p}}})}{e^{\frac{2\pi}{\bar{p}}}+ e^{\frac{4\pi}{\bar{p}}} + e^{\frac{\bar{q}}{\bar{p}}}}.
\end{eqnarray}
The magnification observable $r_{mag}$ is independent of the source position rather solely depends  on the metric  parameters through the lensing coefficients $\bar{p}$ and $\bar{q}$.
\begin{figure*}
	\begin{tabular}{c c}
		\includegraphics[scale=0.6]{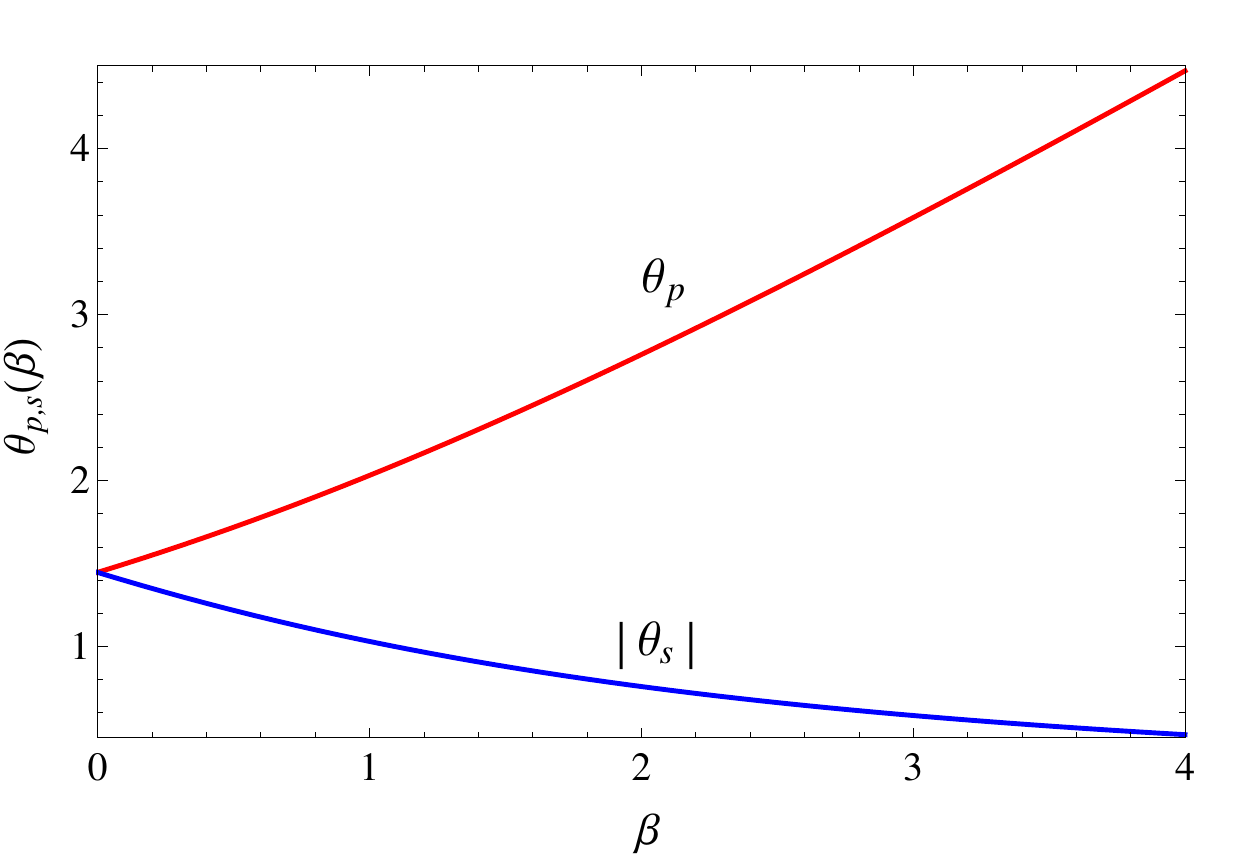}& \includegraphics[scale=0.65]{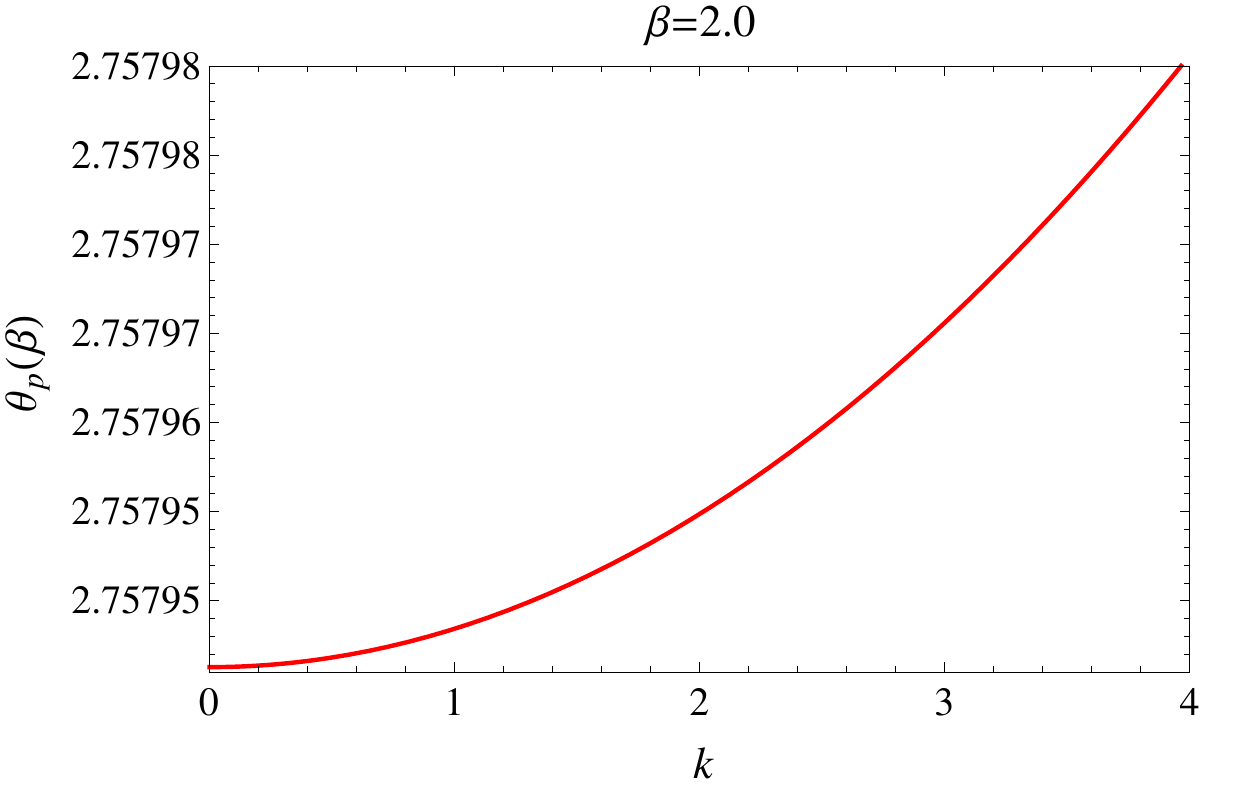}	   
	\end{tabular}
	\caption{(\textit{Left:}) The angular position of the direct primary and secondary images $\theta_{p,s}$ varying with source position $\beta$ and $k=1$. (\textit{Right:}) The angular position of the direct primary image $\theta_p$ varying with $k$ for $\beta=2\, as$. All angles, source and image positions, are in units of $\mathcal{O}(as)$.}\label{fig:Image}
\end{figure*}

\begin{figure*}
	\begin{tabular}{c c}
		\includegraphics[scale=0.7]{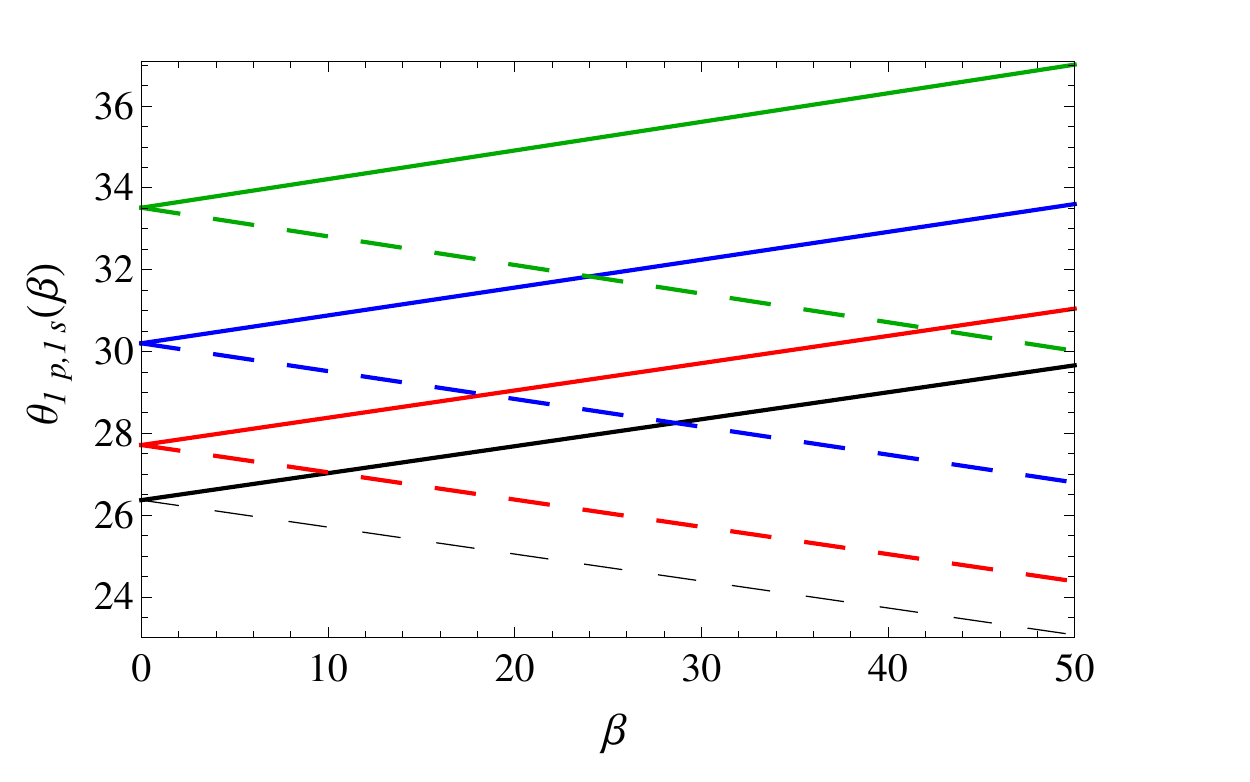}&\hspace{-1.5cm} \includegraphics[scale=0.7]{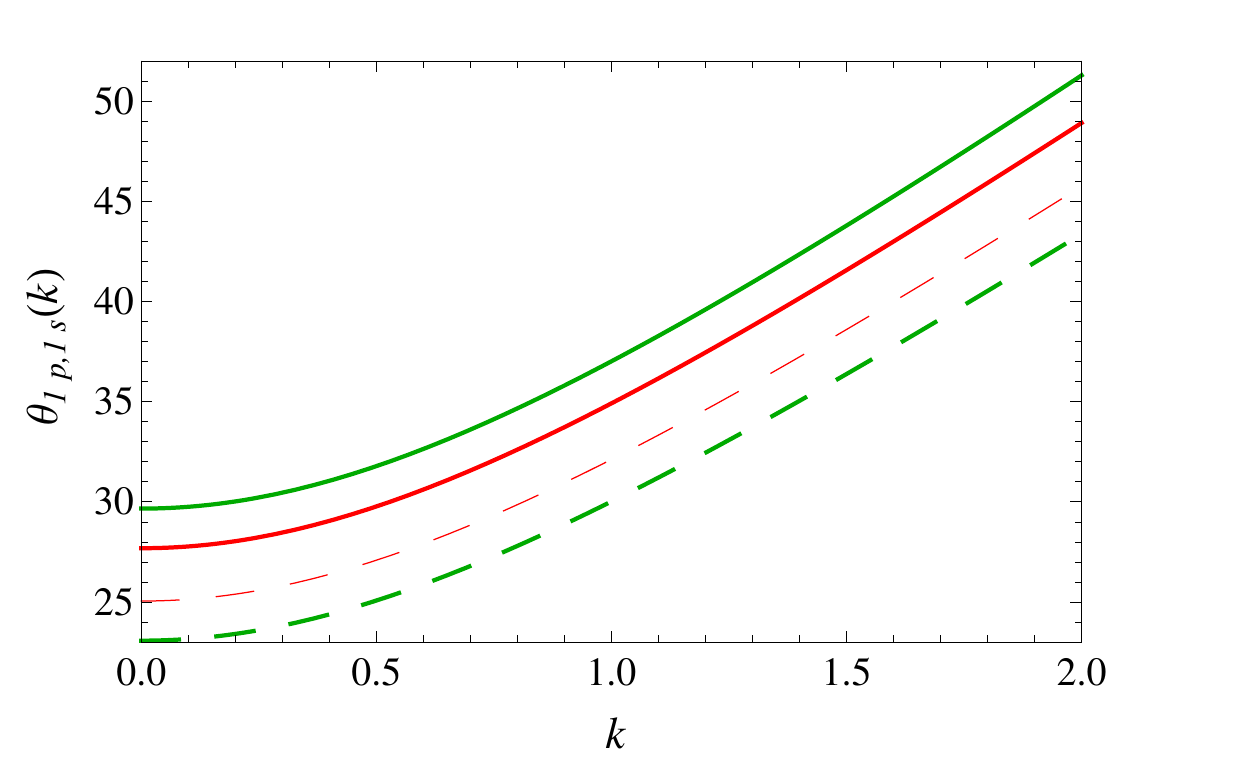}	   
	\end{tabular}
	\caption{(\textit{Left:}) The angular position of the first relativistic primary image $\theta_{1p}$ (solid lines) and the secondary image $|\theta_{1s}|$ (dashed lines) varying with source position $\beta$ for $k=0$ (black lines), $k=0.4$ (red lines), $k=0.7$ (blue lines) and $k=1.0$ (green lines). (\textit{Right:}) Angular positions of the first relativistic primary images $\theta_{1p}$ and secondary images $|\theta_{1s}|$ (dashed lines) varying with $k$ for $\beta=20\,$as (red lines) and for $\beta=50\,$as (green lines). $\beta$ is in units of $\mathcal{O}$(as) and image positions $\theta$ are in units of $\mathcal{O}(\mu\,$as).}\label{fig:Image1}
\end{figure*}

Gravitational lensing also leads to a time delay in image formation. Light rays forming the second relativistic image travel $2\pi u_2$ additional distance through their journey from source to observer. This excess path leads to a finite time lag between the formation of the first and second relativistic images, i.e., the second relativistic image forms $\delta t_{12}=2\pi u_2$ time later than the first image. Similarly, the time difference between the direct secondary and primary images is due to the different paths traveled by photons, it can be calculated as follows \cite{1992grle.book.....S}
\begin{equation}
	\Delta t_{ps}=4\Big(\frac{\theta_s^2-\theta_p^2}{2|\theta_p\theta_s|}+\log|\frac{\theta_s}{\theta_p}| \Big).
\end{equation} 
Direct secondary image forms after $\Delta t_{ps}$ time of the primary image formation. It is important to note that for $\beta=0$, the direct primary and secondary images form at the same angular distance from optical axis, i.e. $\theta_s=\theta_p$, therefore, time delay vanishes. In addition, the time delay between the images is a useful observable quantity only in the case of transient sources. 
\begin{table*}
			\centering
	\resizebox{0.8\textwidth}{!}{
		\begin{centering}	
			\begin{tabular}{l l l l l l l   }
				\hline\hline
				
				$k$  &	$\beta$  &  $\theta_p$(as) & $\theta_s$(as) &  $\mu_p$ & $\mu_s$  & $\Delta t_{ps}$(min)  \\\hline
				\hline
				0	&	0	&	0.1365405 & -0.1365262		& $2.2\times 10^{15}$ & $-2.2\times 10^{15}$ &0\\
				\hline
				0 & $10^{-6}$  & 0.1365410 & -0.13652 & 68267.2 & -68266.2. & $2.9\times 10^{-4}$ \\
				0 & $10^{-3}$  & 0.1370413 & -0.13602 & 68.768 & -67.768 & 0.0194789\\
				0 & $10^{-1}$  & 0.1954053 & -0.09539 & 1.31296& -0.312962 &   1.96298 \\
				0 & 1 		   & 1.018306  & -0.01829 & 1.00032 & -0.0003230 & 41.7508 \\
				0 & 3          & 3.006201  & -0.00618 & 1.00006 & $-4.24 \times 10^{-6}$ & 326.64 \\
				0 & 5          & 5.003726  & -0.00371 & 1.0 & $-5.52\times 10^{-7}$ & 893.275 \\
				\hline
				2 & $10^{-6}$ & 0.1365600	&-0.13652 &	68267.2	&-68266.2	&$1.02\times 10^{-3}$\\
				2 & $10^{-3}$ & 0.1370603&	-0.13600&68.768	&-67.768	&0.020208  \\
				2 & $10^{-1}$ & 0.1954178 &-0.09536 &1.31293 &-0.312933 &1.96397\\
				2 & 1 & 1.018307 &-0.01825 &1.00032 &-0.0003224 &41.8283\\
				2 & 3 & 3.006201	&-0.00614	&1.00005	&$-4.21\times 10^{-6}$	&328.625	 \\
				2 & 5 & 5.003726 &-0.00367 &1. &$-5.46\times 10^{-7}$ & 902.527 \\	
				\hline
				4 & $10^{-6}$ & 0.1366169 & -0.13644 & 68267.2 & -68266.2& 0.0032121 \\
				4 & $10^{-3}$ & 0.1371170 & -0.13595 & 68.7679 & -67.7679 & 0.022398 \\
				4 & $10^{-1}$ & 0.1954552 & -0.09528 & 1.31285 & -0.312847 & 1.96693 \\
				4 & 1 & 1.018309 & -0.01814 & 1.00032 & -0.0003205 & 42.0648 \\
				4 & 3 & 3.006201 & -0.00603 & 1.00004 & $-4.14\times 10^{-6}$ & 334.895  \\
				4 & 5 & 5.003726 & -0.00355 & 1. & $-5.29 \times 10^{-7}$ & 932.837  \\	
				
				\hline\hline
			\end{tabular}		
	\end{centering}	}	
	\caption{Direct image positions, their magnifications, and the time delay between secondary and primary images for Sgr~A* with different values of $k$ and $\beta$ in the weak deflection limit. All angles are in as, and the time delay is in units of minutes. }
	\label{table1}  
\end{table*}  

\begin{table}
			\centering
	\resizebox{0.8\textwidth}{!}{
		\begin{centering}	
			\begin{tabular}{l l l l l l l   }
				\hline\hline
				
				$k$  &	$\beta$  &  $\theta_p$(as) & $\theta_s$(as) &  $\mu_p$ & $\mu_s$  & $\Delta t_{ps}$(hrs)  \\\hline
				\hline
				0	&	0	&	1.2532	& -1.2532		& $2.2\times 10^{15}$ & $-2.2\times 10^{15}$ &0\\
				\hline
				0 & $10^{-6}$ & 1.253211 & -1.253199 & 626603. & -626602. & 0.000691 \\
				0 & $10^{-3}$ & 1.253710 & -1.252699 & 627.102 & -626.102 & 0.057268 \\
				0 & $10^{-1}$ & 1.304207 & -1.204196 & 6.78097 & -5.78097 & 5.66552 \\
				0 & 1 & 1.849271 & -0.849259 & 1.26728 & -0.26727 & 58.1032 \\
				0 & 3 & 3.454617 & -0.454605 & 1.01762 & -0.01762 & 204.469 \\
				0 & 5 & 5.296520 & -0.296509 & 1.00314 & -0.00314 & 418.255 \\
				\hline
				2 & $10^{-6}$  & 1.253226 & -1.253184 & 626603. & -626602. & 0.002385 \\
				2 & $10^{-3}$  & 1.253725 & -1.252684 & 627.102 & -626.102 & 0.058962 \\
				2 & $10^{-1}$  & 1.304221 & -1.204180 & 6.78097 & -5.78097 & 5.66722 \\
				2 & 1 & 1.849280 & -0.849239 & 1.26727 & -0.26727 & 58.1056 \\
				2 & 3 & 3.454620 & -0.454579 & 1.01762 & -0.01762 & 204.479 \\
				2 & 5 & 5.296522 & -0.296480 & 1.00314 & -0.00314 & 418.289 \\
				\hline
				4 & $10^{-6}$ & 1.253271 & -1.253139 & 626603. & -626602. & 0.007465 \\
				4 & $10^{-3}$ & 1.253770 & -1.252639 & 627.102 & -626.102 & 0.064042 \\
				4 & $10^{-1}$ & 1.304264 & -1.204134 & 6.78096 & -5.78096 & 5.67232 \\
				4 & 1 & 1.849308 & -0.849177 & 1.26727 & -0.26726 & 58.1129 \\
				4 & 3 & 3.454631 & -0.454499 & 1.01762 & -0.01761 & 204.510 \\
				4 & 5 & 5.296527 & -0.296395 & 1.00314 & -0.00314 & 418.3910 \\
				\hline\hline
			\end{tabular}		
	\end{centering}	}	
	\caption{Direct image positions, their magnifications, and the time delay between secondary and primary images for M87* for different values of $k$ and $\beta$ in the weak deflection limit. All angles are in as, and the time delay is in units of hrs. }
	\label{table2}  
\end{table}

\begin{table*}
			\centering
	\resizebox{1.1\textwidth}{!}{
		\begin{centering}	
			\begin{tabular}{l l l l l l l l l l l l l }
				\hline\hline
				
				{$k$ }	&  $\theta_{1p}$	&	$\theta_{2p}$	&	$\theta_{1s}$	&	$\theta_{2s}$	&	$\theta_{\infty}$	&	$\theta_{E1}$	&	$\theta_{E2}$	&	$\mu_{1}(10^{-18})$	&	$\mu_{2}(10^{-21})$	&	$s$	&	$r_{\text{mag}}$  &	$\Delta t_{12}$(min)  \\\hline
				\hline
				0. &  31.911&  25.1542& -18.4347& -25.129& 25.1415& 25.1728& 25.1416&  822.336&  1533.75&  0.0314646&  535.159&  10.7011\\
				0.5 & 33.9967 &   27.1209 &   -20.2831 &   -27.0954 &   27.1081 &   27.1399 &   27.1081 &  902.215 &   1676.83 &   0.0320006 &   537.046 &   11.5381\\
				1. & 39.1502 &   31.9777 &   -24.8451 &   -31.9517 &   31.9646 &   31.9976 &   31.9647 &  1109.57 &   2009.78 &   0.0332436 &   551.084 &   13.6052 \\
				1.5 & 45.8413 &   38.2465 &   -30.6946 &   -38.2202 &   38.2333 &   38.2679 &   38.2333 & 1405.08 &   2438.98 &   0.0349631 &   575.094 &   16.2734\\
				2. & 53.274 &   45.1613 &   -37.095 &   -45.1344 &   45.1478 &   45.1845 &   45.1478 &  1772.11 &   2944.17 &   0.0370898 &   600.905 &   19.2164\\
				2.5 & 61.0956 &   52.3977 &   -43.7501 &   -52.37 &   52.3838 &   52.4229 &   52.3839 & 2204.23 &   3522.75 &   0.0395238 &   624.712 &   22.2963\\
				3. & 69.1404 &   59.8115 &   -50.5369 &   -59.7827 &   59.797 &   59.8386 &   59.7971 &  2698.52 &   4173.71 &   0.0421754 &   645.551 &   25.4516 \\
				3.5 & 77.3231 &   67.3316 &   -57.3985 &   -67.3016 &   67.3165 &   67.3608 &   67.3166 &  3253.46 &   4896.11 &   0.0449797 &   663.498 &   28.6522\\
				4. & 85.5965 &   74.92 &   -64.3063 &   -74.8887 &   74.9043 &   74.9514 &   74.9044 &  3868.19 &   5689.15 &   0.0478931 &   678.923 &   31.8818\\		
				\hline\hline
			\end{tabular}
	\end{centering}	}	
	\caption{Relativistic image positions, their magnifications, relativistic Einstein rings, and strong lensing observables for the supermassive black holes Sgr A* for different values of $k$ and $\beta=1$as. All angles are in $\mu$as.}
	\label{table3}  
\end{table*}  

\begin{table*}
	\resizebox{1.1\textwidth}{!}{
		\begin{centering}	
			\begin{tabular}{l l l l l l l l l l l l l }
				\hline\hline
				
				{$k$ }	&  $\theta_{1p}$	&	$\theta_{2p}$	&	$\theta_{1s}$	&	$\theta_{2s}$	&	$\theta_{\infty}$	&	$\theta_{E1}$	&	$\theta_{E2}$	&	$\mu_{1}(10^{-18})$	&	$\mu_{2}(10^{-21})$	&	$s$	&	$r_{\text{mag}}$  &	$\Delta t_{12}$(hrs)  \\\hline
				\hline
				0. & 19.8563 & 19.7822 & -19.7573 & -19.782 & 19.782 & 19.8068 & 19.7821 & 4.75466 & 8.86797 & 0.02475 & 535.159 & 289.648 \\
				0.5 & 21.4049 & 21.3295 & -21.3041 & -21.3293 & 21.3293 & 21.3545 & 21.3294 & 5.21651 & 9.69525 & 0.02517 & 537.046 & 312.304 \\
				1. & 25.2293 & 25.1508 & -25.1242 & -25.1506 & 25.1506 & 25.1768 & 25.1507 & 6.41541 & 11.6203 & 0.02615 & 551.084 & 368.254 \\
				1.5 & 30.1661 & 30.0831 & -30.0548 & -30.0829 & 30.0829 & 30.1104 & 30.083 & 8.12403 & 14.1019 & 0.02750 & 575.094 & 440.473 \\
				2. & 35.6121 & 35.5236 & -35.4932 & -35.5234 & 35.5235 & 35.5526 & 35.5235 & 10.2462 & 17.0228 & 0.02918 & 600.905 & 520.133 \\
				2.5 & 41.3118 & 41.2171 & -41.1843 & -41.2169 & 41.217 & 41.2481 & 41.217 & 12.7446 & 20.3681 & 0.03109 & 624.712 & 603.497 \\
				3. & 47.1514 & 47.05 & -47.0147 & -47.0498 & 47.0498 & 47.083 & 47.0499 & 15.6025 & 24.1319 & 0.03318 & 645.551 & 688.902 \\
				3.5 & 53.075 & 52.9666 & -52.9286 & -52.9663 & 52.9664 & 53.0018 & 52.9664 & 18.8111 & 28.3088 & 0.03539 & 663.498 & 775.532 \\
				4. & 59.0526 & 58.9368 & -58.8961 & -58.9366 & 58.9367 & 58.9743 & 58.9367 & 22.3654 & 32.894 & 0.03768 & 678.923 & 862.948 \\
				\hline\hline
			\end{tabular}
	\end{centering}	}	
	\caption{Relativistic image positions, their magnifications, relativistic Einstein rings, and strong lensing observables for the supermassive black holes M87* for different values of $k$ and $\beta=1$as. All angles  are in $\mu$as.}
	\label{table4}  
\end{table*}  

\section{Lensing by supermassive black holes}\label{Sec-5}	
We model two astrophysically important supermassive black holes Sgr~A* and M87*, residing at the center of Milky-Way and nearby M87 galaxy, as the polymerized black hole and calculate the image positions and lensing observables. We consider, for Sgr~A* black hole $M=4.3 \times 10^6 M_{\odot}$ and $D_{OL}=8.35\times 10^3$ pc  \cite{Do:2019txf}, and for M87* black hole $M=6.5\times 10^9 M_{\odot}$ and $D_{OL}=16.8$ Mpc \cite{Akiyama:2019cqa}. 
Fig.~\ref{fig:Image} depicts the variation in direct primary and secondary images positions with $\beta$ and $k$, which is valid in the weak deflection angle limit. The primary (secondary) image always forms outside (inside) the Einstein ring. Although, the primary image position moves farther and farther away from the optical axis with increasing $\beta$, it weakly depends on $k$. The secondary image moves toward the optical axis with increasing $\beta$, such that the angular separation between the primary and secondary image increases with $\beta$. 

For Sgr~A* black hole, we consider the S2 star as a source -- one of the best known candidate for gravitational lensing in strong fields. Taking the source distance $D_{LS}=10^4M$, we calculate the lensing observables.
Table.~\ref{table1} and~\ref{table2} summarizes the angular position and magnification of direct image $(\theta_{p,s}, \mu_{p,s})$ and the time delay $\Delta t_{ps}$ between the secondary and primary images for the Sgr~A* and M87* black holes. Numerical comparison with the $k=0$ case suggests that for the polymerized black hole the direct images form farther away compared to the Schwarzschild black hole. Direct images magnifications weakly depend on $k$ but sharply fall with $\beta$. The angular position for direct images is of $\mathcal{O}(as)$ whereas the time delay is in minutes and hours for Sgr~A* and M87* black holes, respectively. Time delay has weak dependence over $k$ and increases slowly with it.

For the strong gravitational lensing, the first relativistic primary and secondary image positions are shown and compared with those for the Schwarzschild black hole in Fig.~\ref{fig:Image1}. Angular position of relativistic images is of $\mathcal{O}(\mu\,as)$ and increases with $k$. It might be easy to distinguish the direct image from the rest of the relativistic images, which are rapidly converging to the $\theta_{\infty}$. The relativistic image position, magnification, Einstein ring size, and lensing observables are calculated for Sgr~A* and M87* black holes and shown in tables.~\ref{table3} and \ref{table4}, respectively. As expected, the relativistic images are much fainter compared to the direct images, and magnification decreases with the increasing source position. In addition, secondary images are demagnified than the primary images. The Einstein ring size is of $\mathcal{O}(\mu\, as)$ and it grows with increasing values of $k$. The separation observable $s$ also increases with $k$, suggesting that for polymerized black hole with large values of $k$ the outermost relativistic image can be distinguished from the pack of the higher order relativistic images. Similarly, the time delay $\Delta t_{12}$ between the first two relativistic images is of the order of minutes for Sgr~A* black hole, and of the order of hours for the M87* black hole and increases with $k$. Interestingly, for a given source position, the time delay between the first and second relativistic images is larger than the time delay between the direct primary and secondary images, i.e, $\Delta t_{ps}< \Delta t_{12}$.

\section{Polymerized black hole shadows under different accretion flows }\label{Sec-6}
The black hole's gravitationally lensed light rays, emanating from all the sources in the sky, construct the black hole image on the observer's screen. To mathematically generate this image it is practical to employ the relativistic backward ray-tracing method \cite{Zhu:2015zca,Narayan:2015cua}. In this method, we trace the light rays backward in time from the observer's screen using Eq.~(\ref{phiEq}). Depending on the impact parameter $b\geq b_c$, the backtraced light rays from the observer can reach the source either directly or after completing any numbers of loops around the black hole. We assign some brightness to these light rays' directions. In contrary, backtraced light rays with $b<b_c$ get trapped in the black hole gravitational field and eventually spiral down to the black hole horizon and fall into it, accounting for the dark region on the observer's screen. This brightness depression on the observer's plane, enclosed by the bright circular ring at $b_c$, is known as the ``black hole shadow" \cite{Bardeen:1973tla}. 
On the other hand, for light rays with $b\to b_c$, the number of winding $n$ around the black hole rises exponentially such that for $b=b_c$, $n\to \infty$ and light rays converge to the critical curve. The path integral of such strongly lensed photons, arising from near-critical light rays, through the emission region diverge logarithmically, thus the image intensity, resulting in a bright ring \cite{Chael:2021rjo, Gralla:2019xty}. These lensed light rays appear within a narrow angular band, called the \emph{photon ring}, at the shadow boundary on the observer's sky. The photon ring, as defined here, was called the ``shadow apparent boundary" by Bardeen \cite{Bardeen:1973tla} and the ``critical curve" by Gralla et al. \cite{Gralla:2019xty}. While the photon rings or critical curve solely depends on the spacetime geometry, the shadow intensity distribution depends additionally on the details of the accretion models and emission processes. The possibilities of estimating the black hole parameters from the photon ring structure are discussed in detail in refs.~\cite{Broderick:2021ohx,Kumar:2018ple}. 

For an optically thin isotropic emitting region transparent to its own emission around a black hole, the image plane region $b<b_c$ would be dark while $b>b_c$ would be uniformly bright with a narrow bright ring with diverging intensity at $b\to b_c$. Interestingly enough, the shadow boundary location is independent of the inner radius at which the accreting gas stops radiating \cite{Narayan:2019imo}. For the Schwarzschild black hole, the shadow boundary appears for $b_c=3\sqrt{3}/2$. On the other hand, if the black hole is backlit by a distant planar and sufficiently large source screen with isotropic and uniform brightness, the black hole will cast a slightly larger shadow, extending out the critical curve; for Schwarzschild black hole, the shadow boundary appears at $b=6.17/2$ still with a photon sphere at $b=b_c=3\sqrt{3}/2$. This additional dark region $6.17/2\leq b\leq 3\sqrt{3}/2$ accounts for the light rays having deflection angle $\pi/2<\alpha_D < 3\pi/2$, whereas the main bright region $b>6.17/2$ corresponds to the light having deflection angle smaller than the $\pi/2$. Close to the critical curve, a series of converging and demagnified photon rings exist that account for the multiple winding around the black hole \cite{Gralla:2019xty}.

Whereas for planar and optically thin accretion disks, whose emission is confined to the equatorial plane, the central brightness depression extends simply to the lensed position of the inner edge of the disk, and is bounded by a bright photon ring with diverging intensity. However, in contrast to the spherical accretion model, this photon ring can typically be decomposed into a series of infinitely many concentric discrete photon rings, which are organized self-similarly and depend on the number of loops executed around the black holes. However, only a few subrings can be resolved as the higher-order rings are highly closely packed and demagnified \cite{Luminet:1979nyg,Beckwith:2004ae}. In particular, for the accretion disk extending up to the black hole's event horizon, the shadow boundary does not coincide with the critical curve but rather is restricted to a much smaller area-\textit{an inner shadow}--whose edge lies near the direct lensed image of the equatorial horizon \cite{Beckwith:2004ae,Chael:2021rjo}; for Schwarzschild black hole ($x_+=1$) it appears at $x=1.45$. The inner shadow forms by the light rays, which fall into a black hole without crossing the equatorial plane once, i.e., $n<1/4$. Furthermore, these models still feature a photon ring. For these models, the darkest region in the observed image will correspond to the inner shadow. However, due to the increasingly large gravitational redshift at the event horizon, the image brightness falls sharply around the inner shadow edge. In contrast, the spherical accretion models do not give rise to the inner shadow. Due to this reason, the EHT images released thus far do not resolve the inner shadow of M87, as they also lack the requisite resolution. 

The black hole shadow features depend not only on the spacetime geometry but also on the surrounding accretion details. This section analyzes the effect of the polymerized black hole accretion flow on the shadow images by considering three different scenarios namely, a black hole under the static spherical accretion, a black hole surrounded by an optically thin accretion disk, and radially infalling accretion onto the black hole.

\begin{figure*}
	\begin{tabular}{ c c}
		\includegraphics[scale=0.63]{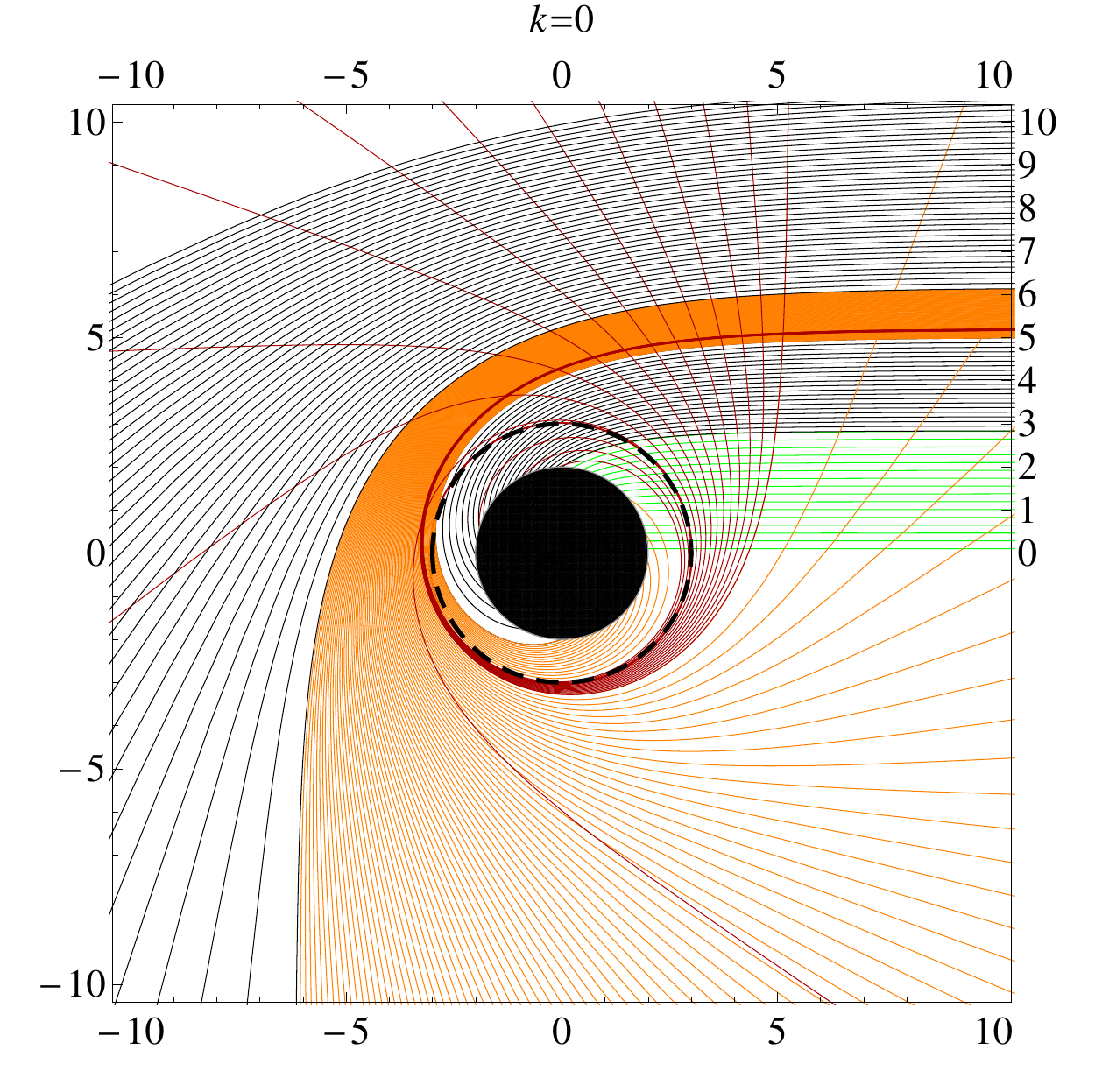}&
		\includegraphics[scale=0.63]{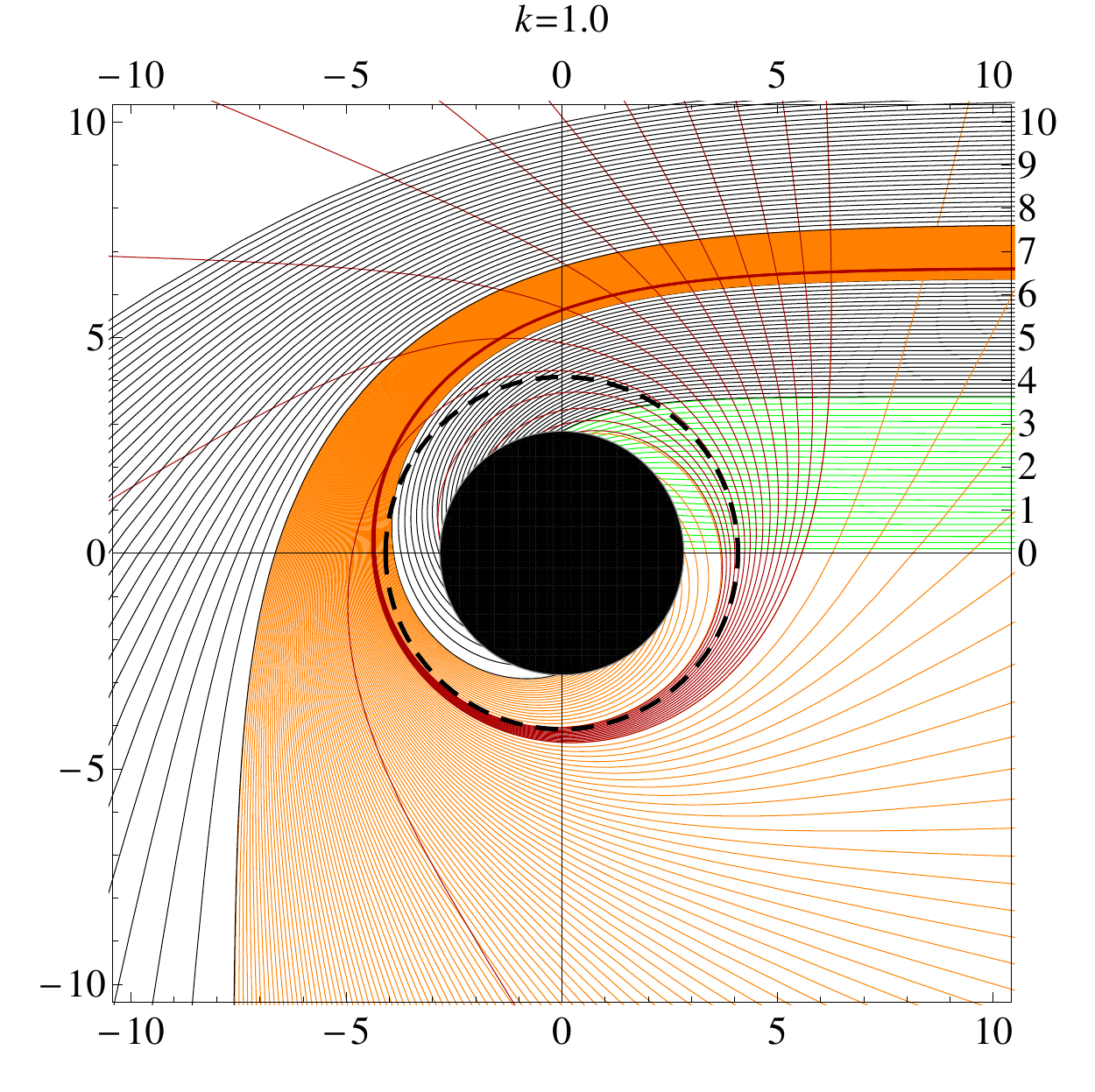}
	\end{tabular}
	\caption{Photon geodesics as a function of impact parameter $u$. Colors classify geodesics based on their number of equatorial plane crossing $n=\frac{\phi}{2\pi}$ (see text for details). The black hole is shown as a solid disk and the critical curve as a closed black dashed circle.}\label{fig:Orbit}
\end{figure*}

\subsection{Static spherical accretion flow}
We begin by considering that the polymerized black hole is surrounded by an optically thin, spherically symmetric, and isotropic radiating gas, which is at rest and extends up to the horizon. In the rest-frame of the gas, the emitted specific intensity is $\mathcal{I}_{\nu}^{em}$ at photon frequency $\nu_{em}$. We consider an observer at a far distance from the black hole $x_{obs}\to \infty$; the observed specific intensity $\mathcal{I}_{\nu }^{obs}$ at the photon frequency $\nu_{obs}$ can be obtained by integrating the emissivity along the photon path $\gamma$ as follows \cite{Jaroszynski:1997bw,Bambi:2013nla}
\begin{equation}
	\mathcal{I}_{\nu }^{obs}=\int_{\gamma}^{}z^3 j(\nu_e)\,d\ell,\label{emission}
\end{equation}
where the redshift factor $z$ quantifies the change in a photon's frequency as it traverses through spacetime from the point of emission $x_{emit}$ to the point of detection $x_{obs}$ defined as
\begin{equation}
	z=\frac{\nu_{obs}}{\nu_{e}}= A(x)^{1/2},\label{redshift}
\end{equation}
and $j(\nu_{em})$ is the emissivity per-unit volume (specific emissivity) measured in the rest-frame of the gas. Here, we assume that the emission is monochromatic and the emission radial profile is $1/x^2$, that is \cite{Falcke:1999pj,Bambi:2013nla}
\begin{equation}
	j(\nu_{em})=\frac{\delta (\nu_{em}-\nu_{obs})}{x^2},\label{emissitivity}
\end{equation}
where $\delta$ is the Dirac delta function. The infinitesimal proper length along the photon path is 
\begin{eqnarray}
	d\ell&=&\sqrt{B(x)\, dx^2+C(x)\, d\phi^2}= \sqrt{\frac{B(x)C(x)}{C(x)-A(x)b^2}}\,dx.
\end{eqnarray}  
Therefore, the observed total photon intensity can be obtained by integrating Eq.~(\ref{emission}) for all observed frequencies, and it takes the following form \cite{Narayan:2019imo,Bambi:2013nla}
\begin{equation}
	\mathcal{I}_{obs}=\int_{\gamma}\frac{A(x)^{3/2}}{x^2} \,\sqrt{\frac{B(x)C(x)}{C(x)-A(x)b^2}}\,dx.\label{emission1}
\end{equation}
\begin{figure}
		\centering
	\includegraphics[scale=0.8]{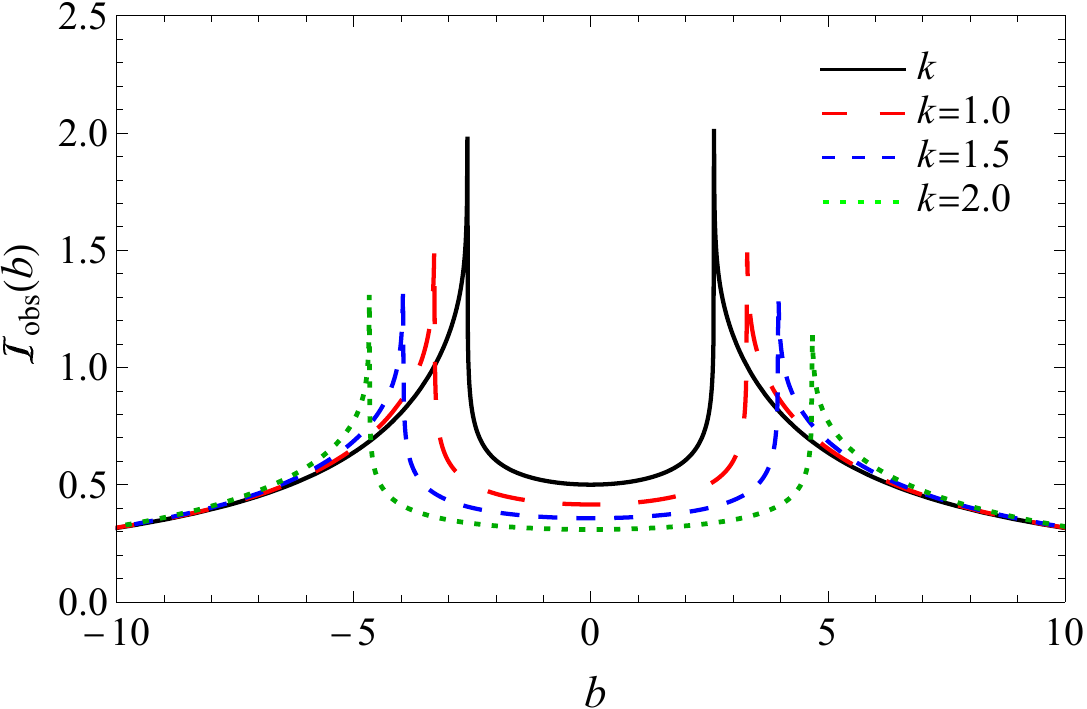}
	\caption{The observed intensity profile for a static spherically symmetric accretion around polymerized black hole as a function of impact parameter $b$. For comparison the intensity distribution for the Schwarzschild black hole is also shown with black line.}\label{fig:StaticIntensity}
\end{figure}
\begin{figure*}
	\begin{tabular}{c c c}
		\includegraphics[scale=0.5]{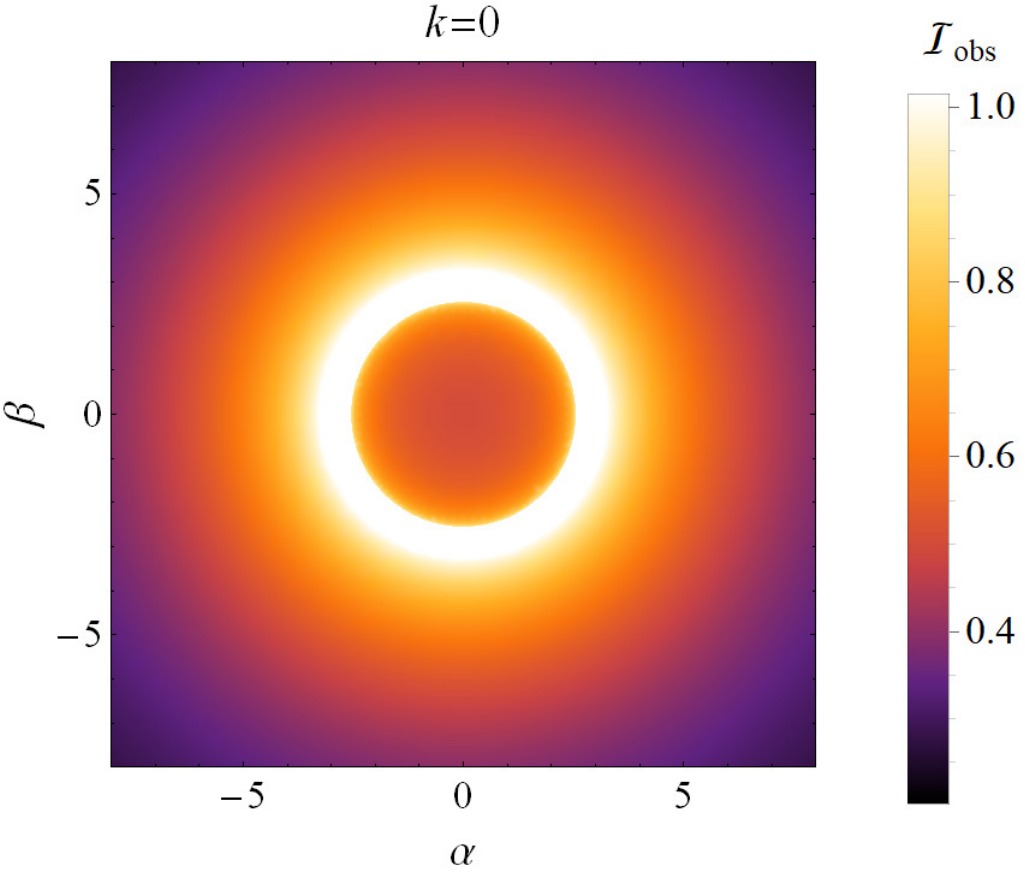}\hspace*{-0.2cm}&
		\includegraphics[scale=0.5]{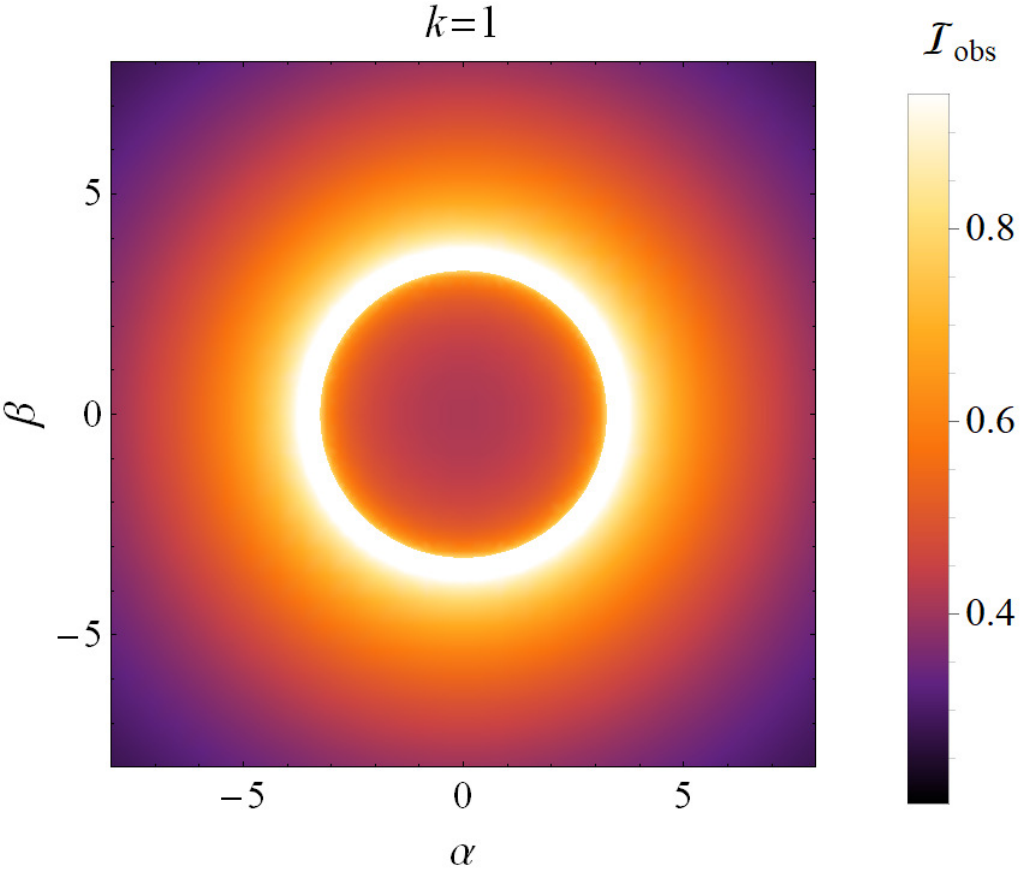}\hspace*{-0.2cm}&
		\includegraphics[scale=0.5]{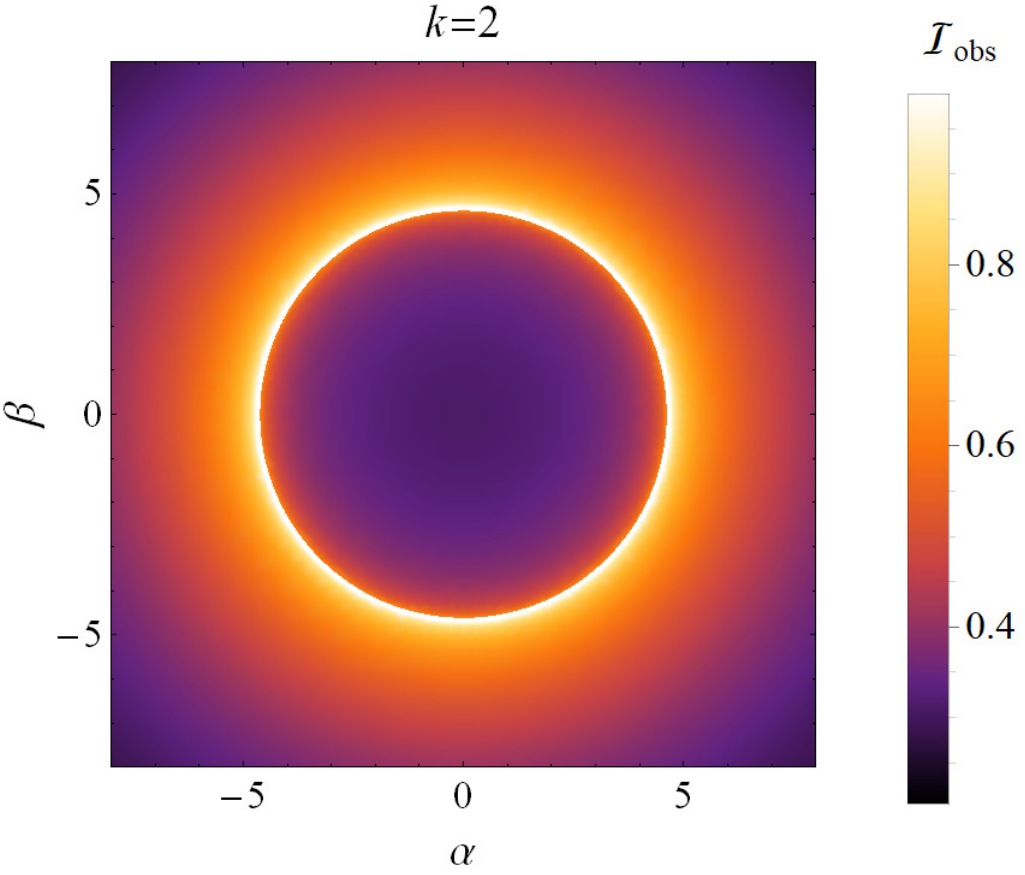}
	\end{tabular}
	\caption{Polymerized black hole shadows with the static spherical accretion for the different values of $k$ as seen by a distant observer. }\label{fig:StaticShadow}
\end{figure*}
The $\gamma$ in Eq.~(\ref{emission1}) signifies that the integral has to be evaluated along the path of the photon. It is interesting to note that, unlike the photon circular orbit radius and the shadow size, which are determined only by $A(x)$, the intensity distribution additionally depends on the metric function $B(x)$. Because the gas is uniformly distributed, and optically thin, emitted light can propagate in all directions and travel arbitrarily large distances without being absorbed or scattered. For $b\leq b_c$, the light rays are backtraced from the observer to the horizon, whereas for $b>b_c$, light rays are backtraced from the observer to some turning point and then to the emitter position. Using Eq.~(\ref{emission1}), we calculated the observed total photon intensity and showed it in Fig.~\ref{fig:StaticIntensity}. With the decreasing impact parameter ($b>b_c$), the $\mathcal{I}_{obs}$ increases rapidly and reaches a peak at $b=b_c$ and then sharply falls to a lower value. This expected intensity depression at the center is the black hole shadow signature. These shadows are shown in Fig.~\ref{fig:StaticShadow}, where we present the two-dimensional intensity map in celestial coordinates ($\alpha,\beta$). Different colors correspond to different values of the observed intensity, and we use one color function for all shadow plots, where the greater (smaller) intensity means the brighter (darker) color. The salient feature of the shadow is that the intensity is circularly symmetric and it possesses a circular bright ring at $b=b_c$ with the strongest intensity, which in principle is the position of the photon sphere with diverging intensity. However, due to the limitation of calculation accuracy and the logarithmic form of the divergence (integrand in the Eq.~(\ref{emission1}) diverges for the limit $x\to x_c$), the calculated intensity will never reach infinity (cf. Fig.~\ref{fig:StaticIntensity}). It is worth noting here that the peak is not at the position of the inner edge of the emitting gas, $x=1$, rather it is at the lensed position of the photon ring $b_{c}$. From Fig.~\ref{fig:StaticShadow}, we can directly compare the intensity magnitude inside and outside the photon ring, and it is clear that the inner region of the photon ring is not completely dark with zero intensity, such as would be observed if the radiating gas were entirely behind the black hole. The non-zero intensity at $b=0$ arises because the radiating gas is also present along the lines of sight that intersect the surface of the black hole, and tiny fraction of that radiation inside the photon ring can always escape to an observer at infinity. In particular, for $x > x_c$, the solid angle of the escaping rays is $2\pi (1+\cos\theta)$ and for $x<x_c$ the solid angle of escaping rays is $2\pi(1-\cos\theta)$ with $\theta$ as
\begin{equation}
	\theta=\arcsin\Big(\frac{x_c^{3/2}}{x}\sqrt{A(x)} \Big).
\end{equation}
The net luminosity observed at infinity
\begin{eqnarray}
	L_{\infty}&=&\int_{x_+}^{x_c}4\pi x^2j(\nu_e)2\pi (1-\cos\theta)dx+ \int_{x_c}^{\infty}4\pi x^2j(\nu_e)2\pi (1-\cos\theta)dx.
\end{eqnarray}

Figure~\ref{fig:StaticShadow} implies that the polymerized black holes have larger shadows with darker interiors than Schwarzschild black holes. In addition, polymerized black holes have a smaller brightness near the photon ring than those of Schwarzschild black holes. The fraction intensity depression $f_c$, defined as the ratio of intensity at the center $b=0$ and just outside the photon ring $b=b_c+0.1$, for $k=0, 1$ and 2, respectively, is $f_c=0.30, 0.288$, and $0.272$. $f_c$ decreases with $k$ and this can be seen in Fig.~\ref{fig:StaticShadow}. 
The EHT telescopes observed shadow images with a finite angular resolution, which is equivalent to the blurring of the theoretical image (see appendix A for details).

\subsection{Accretion disk flow}\label{accretiondisk}
In our next example, we analyze a simple case of the emission from an optically and geometrically thin disk-shaped accretion flow at the equatorial plane outside the polymerized black hole. We further assume that the disk emits isotropically in the rest frame of the observer located at a far distance from the black hole in the north pole direction. The light trajectories are shown in Fig.~\ref{fig:Orbit}. We conveniently orient our setup, such that the vertical black line represents the black hole's accretion disk at $\theta=\pi/2$ and the observer is on the right hand side of the black hole ($\theta=0$) that is shown as the black disk. In order to implement our numerical simulation of the black hole shadow, we first summarize the important light trajectories.

\subsubsection{Direct emission, lensed ring and photon ring}
In earlier subsections, we classified the light geodesics based on their impact parameter, whether they plunge into the black hole ($b\leq b_c$) or scatter and escape to the observer ($b>b_c$). Following Gralla \textit{et al.}~\cite{Gralla:2019xty}, we further characterized light ray trajectories by the number of the crossing of the black hole's equatorial plane $n=\frac{\phi}{2\pi}$ outside the horizon. Here, $\phi$ is the total shift in the azimuthal angle for a given light ray trajectory outside the horizon. 
\begin{itemize}
	\item Direct Rays Type-1: Light rays crossing equatorial plane only once ($\frac{1}{2}\leq n< \frac{3}{4}$) with $\pi\leq \phi<\frac{3\pi}{2}$. While back-tracing these rays meet the background source ($b>b_c$), and make dominant contribution to the black hole image. These rays are deflected by angle less than $\pi/2$. Light rays not being deflected at all by the black hole ($b\gg b_c$) follow straight line motion and correspond to $n=\frac{1}{2}$. These rays are shown as black curves outside the critical curve in Fig.~\ref{fig:Orbit}.
	\item Direct Rays Type-2: Rays crossing equatorial plane only once ($\frac{1}{2}\leq n< \frac{3}{4}$) with $\pi\leq \phi<\frac{3\pi}{2}$. The salient feature of these rays is that while back-tracing these rays \emph{do not} meet the background source but rather fall into the black hole ($b<b_c$), this makes them different from the direct rays type-1, shown as black curves inside the critical curve in Fig.~\ref{fig:Orbit}. These rays are also deflected by an angle less than $\pi/2$. 
	\item Lensing Ring: Rays crossing the equatorial plane twice ($\frac{3}{4}\leq n< \frac{5}{4}$) with $\frac{3\pi}{2}\leq \phi<\frac{5\pi}{2}$. While light rays with $b>b_c$ connect the observer to the source on the same side, those with $b<b_c$ fall into the black hole. These rays are shown as orange curves in Fig.~\ref{fig:Orbit}.\\
	\item Photon Ring: Rays crossing the equatorial plane three or more times ($n> \frac{5}{4}$) with $\phi>\frac{5\pi}{2}$ and $b\gtrsim b_c$. These light rays follow multiple winding around the black hole. These rays are shown as red curves in Fig.~\ref{fig:Orbit}.\\
	\item Critical Curve: Rays following infinite winding around the black hole $n\to\infty$ with $b=b_c$. Higher-order photon rings rapidly converge to the critical curve shown as a black dashed circle in Fig.~\ref{fig:Orbit}.\\
	\item Inner Shadow: Rays that do not cross the equatorial plane before intersecting the event horizon ($b<b_c$). These rays are shown as green curves inside the critical curve in Fig.~\ref{fig:Orbit}.
\end{itemize}

The impact parameter window allowing a certain number of half-orbits $n>\frac{5}{4}$ is quickly diminished and corresponding rings are highly demagnified \cite{Gralla:2019drh}.  Each winding of light rays around a black hole constructs a new closed photon ring that is indexed by the equatorial plane crossing number. For instance, light rays with $\frac{5\pi}{2}+2m\pi<\phi\leq \frac{7\pi}{2}+2m\pi$ with $m$ as an integer, construct a $m$th order photon ring around the black hole. These higher-order photon rings asymptote to the critical curve. The contribution of these higher-order photon rings to the black hole's optical appearance is exponentially suppressed as compared to that of the direct emission. Interestingly, light rays with $b<b_c$ will also perform a number of half orbits on their trip down to the event horizon. These orbits are indeed crucial for the accretion disk models where the inner edge of the disk is allowed to extend inside the outer photon sphere. 

\begin{figure}
		\centering
			\hspace{-0.6cm}
	\includegraphics[scale=0.39]{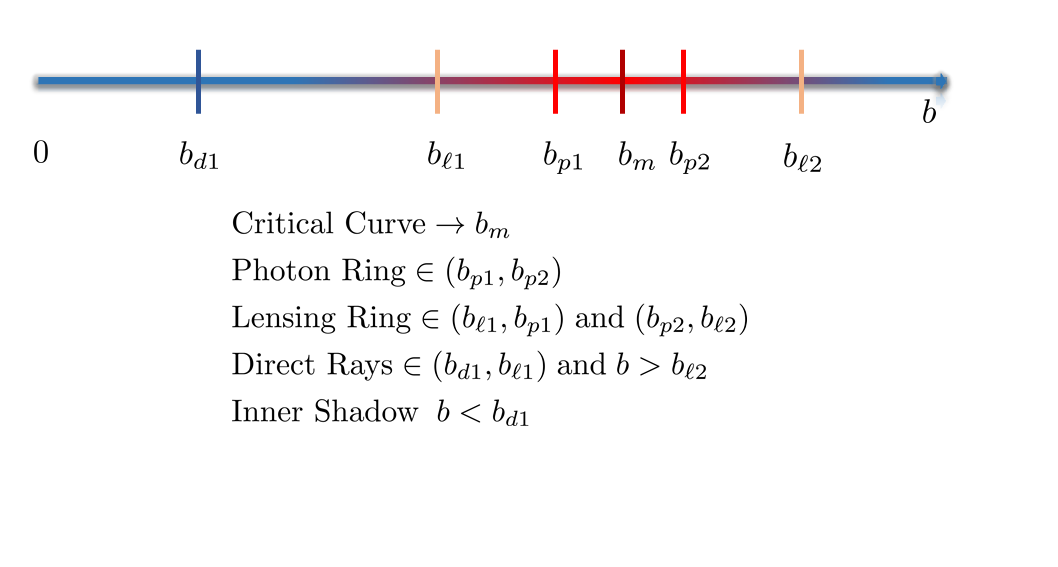}
	\vspace{-1.2cm}	
	\caption{Various light ray trajectories classified by their impact parameter.}\label{fig:Orbit1}
\end{figure}
\begin{table*}
	\begin{tabular}{|c||c|c|c|}
		\hline 
		Classes	 & $k=0$ & $k=0.5$ & $k=1.0$\\
		\hline\hline
		Direct Rays Type-1 & $u\notin (5.0152, 6.1669)$ &  $u\notin (5.40767,6.58743)$ &  $u\notin (6.3787, 7.6383)$\\
		& $u> 6.1669$ & $u>6.58743$ & $u> 7.6383$\\
		\hline
		Direct Rays Type-2 & $2.8477<u< 5.0152$ & $3.071<u<5.40761$& $3.626<u< 6.3787$\\
		\hline 
		Lensing Ring& $u\in (5.0152, 5.18781)$ & $u\in (5.40767, 5.59362)$  & $u\in (6.3787, 6.59598)$ \\ 
		&   $u\in (5.22793, 6.1669)$ & $u\in(5.63493, 6.58743)$ & $u\in (6.640098, 7.6383)$  \\
		\hline 
		Photon Ring	& $u\in (5.18781, 5.22793)$ &  $u\in ( 5.59362, 5.63493)$& $u\in(6.59598, 6.640098 )$ \\ 
		\hline 
		Critical Curve	& $u\equiv u_c=5.1962$ & $u\equiv u_c=5.60259$ & $u\equiv u_c=6.60632$  \\ 
		\hline 	Inner Shadow &  $u<2.8477$&  $u<3.071$& $u< 3.626$ \\ 
		\hline 
	\end{tabular}	\caption{In this table we depict the various lensing features for different values of $k$. The impact parameter for the direct, lensing, photon rings amd inner shadows are shown and compared with the Schwarzschild black hole values (see Sec. III for details).}\label{T1}
\end{table*} 
\begin{figure}
		\centering	
	\includegraphics[scale=0.8]{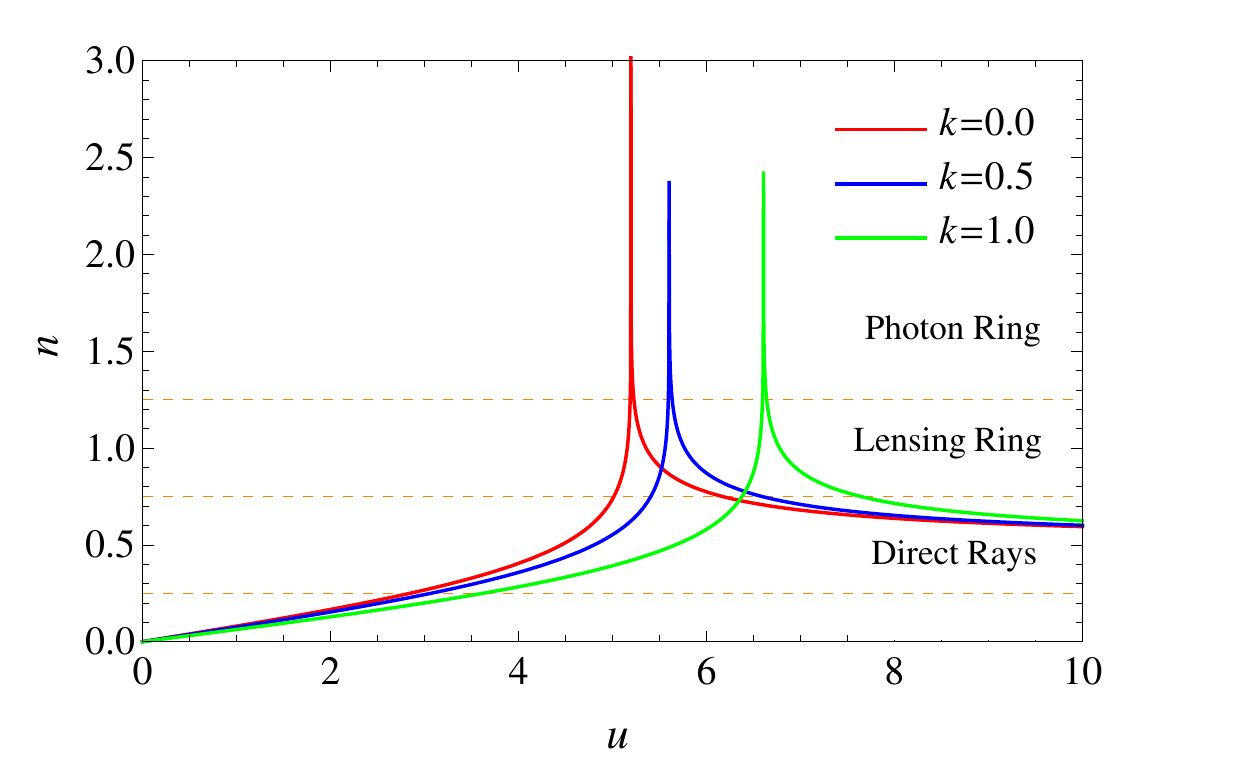}
	\caption{Number of half-orbits $n$ as a function of impact parameter $b$ for different values of $k$. Three orange colored dashed horizontal lines are for photon ring, lensed, and direct emission regions, respectively, with $n=5/4, \,3/4,\, 1/4$ (from top to bottom). }\label{fig:Orbit2}
\end{figure}

With this classification, as summarized in Fig.~\ref{fig:Orbit1}, we first back-trace the light rays with $b>b_c$ from the observer at $x_{obs}$ till the distance of closest approach $x_0$ and then to the emitter. While for the light ray with $b<b_c$, the tracing ends as the ray reaches the horizon at a finite value of
$\phi$
\begin{align}
	& b>b_c,\qquad n=\frac{1}{2\pi}\Big(\int_{x_{emit}}^{x_{0}}\frac{d\phi}{dx}dx+\int_{x_0}^{x_{obs}}\frac{d\phi}{dx}dx\Big),\\
	&b<b_c,\qquad n=\frac{1}{2\pi}\int_{x_+}^{x_{obs}}\frac{d\phi}{dx}dx.
\end{align}
We numerically computed the impact parameter ranges for the lensing ring, photon ring, and direct emission and summarized them in the table.~\ref{T1}. Figure~\ref{fig:Orbit2} shows the total number of orbits
as a function of the $b$ and different values of $k$. With the increase of the value of $k$, the range of $b$ occupied by the lensing ring and the photon ring becomes large and the corresponding impact parameters also increase. This implies that the polymerized black holes have thicker lensing and photon rings compared to those for the Schwarzschild black hole. Using the strong lensing deflection angle Eq.~(\ref{DefAng0}), one can calculate the impact parameter $b$ corresponding to light rays with number of crossings of the equatorial plane $n$, 
\begin{equation}
	b=b_c \left(1+e^{-\frac{(2n-1)\pi -\bar{q}}{\bar{p}}}\right),
\end{equation}
which implies that as $n\to \infty$, $b\to b_c$.

\subsubsection{Shadow images}
\begin{figure*}
		\centering
	\begin{tabular}{c c c}
		\includegraphics[scale=0.55]{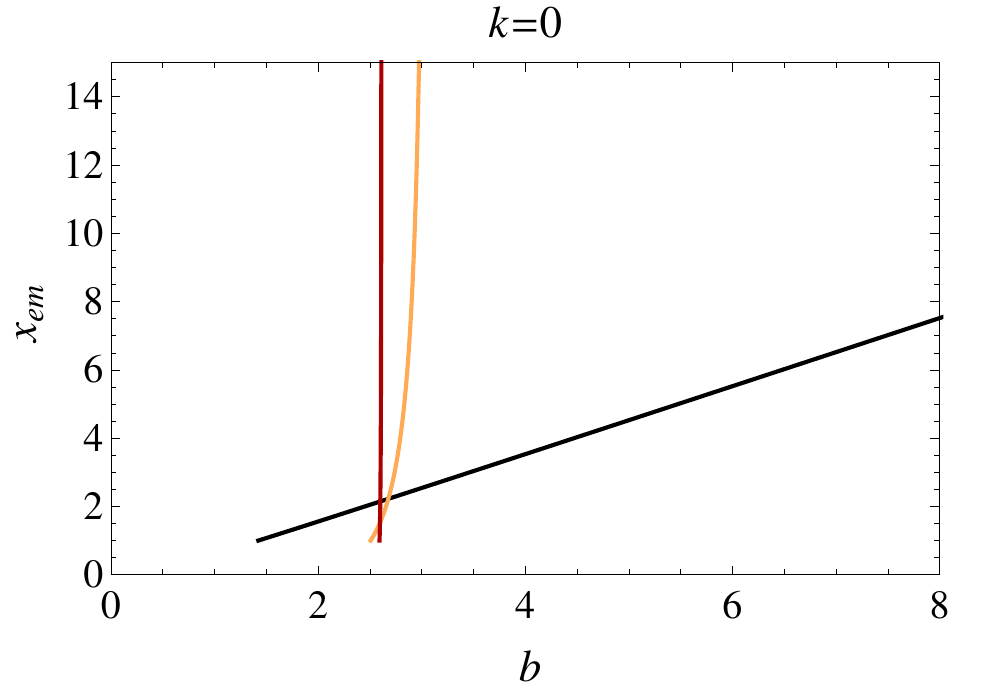}\hspace*{-0.5cm}&
		\includegraphics[scale=0.55]{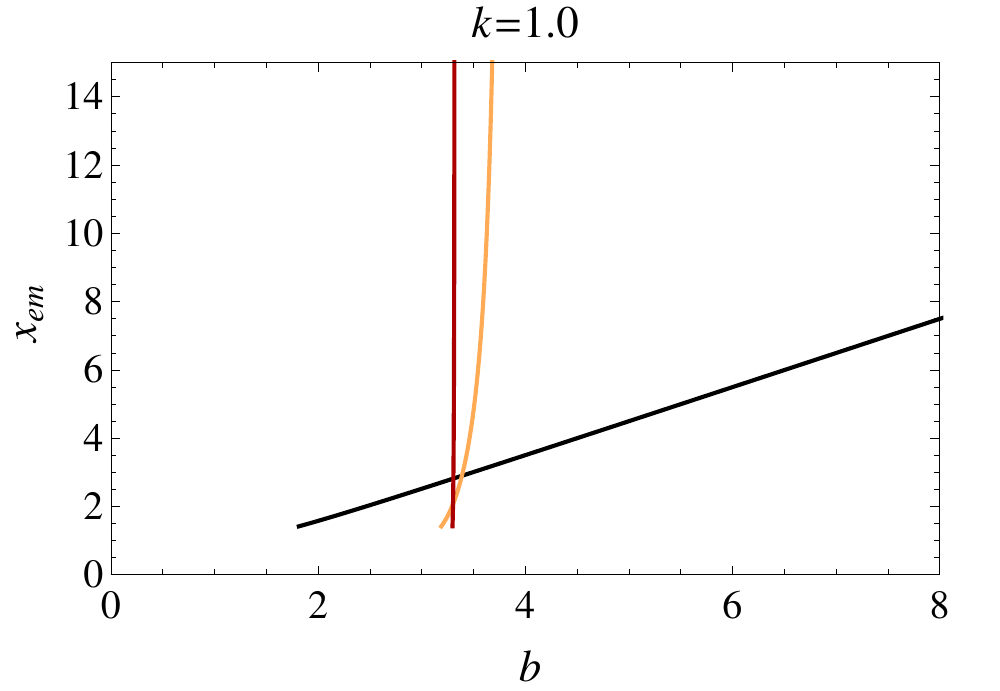}\hspace*{-0.5cm}&
		\includegraphics[scale=0.55]{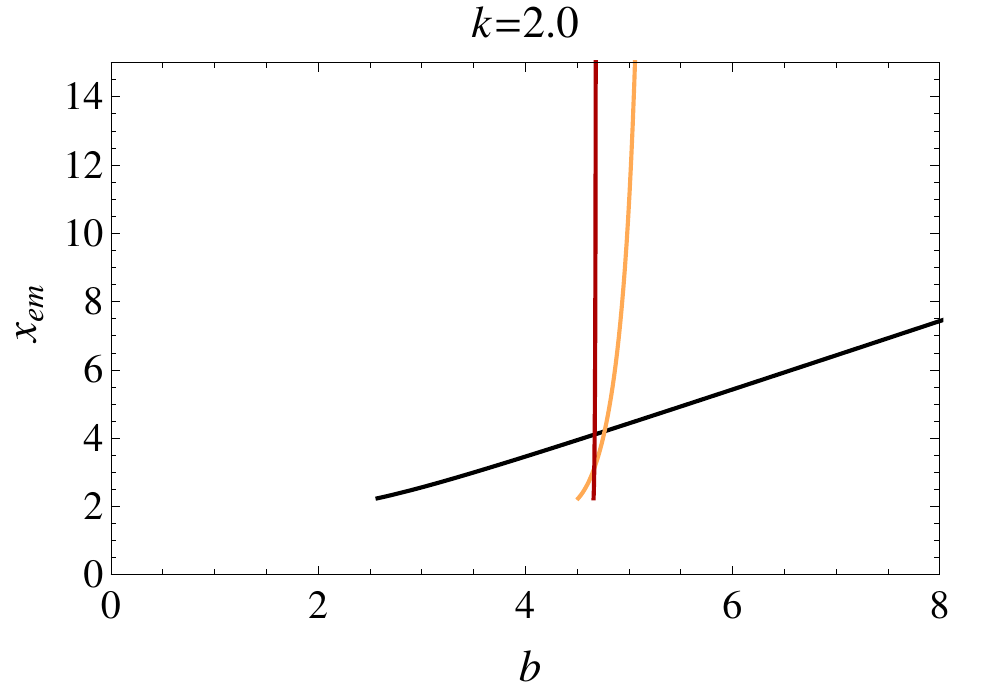}
	\end{tabular}
	\caption{First, second, and third transfer functions are shown with varying impact parameter $b$, respectively, with black, orange, and red colors lines. They also correspond to the direct, lensing rings and photon rings. The slope of each curve is interpreted as the demagnification factor of the corresponding emission (see text for details).}\label{fig:transfer}
\end{figure*}

To construct the shadow images, we are presuming that the light is only emitted from the accretion disk and that other effects, such as light absorption or reflection, are insignificant. The observed photon specific intensity at frequency $\nu_{obs}$ is $\mathcal{I}_{\nu }^{obs}=z^3 \mathcal{I}_{\nu }^{em}$, which can be integrated for the full frequency spectrum to get the total observed photon intensity 
\begin{equation}
	\mathcal{I}_{obs}=z^4 \mathcal{I}_{em},
\end{equation}
where $z$ is the redshift factor (\ref{redshift}). As shown in Fig.~\ref{fig:Orbit}, for the direct emission (black trajectories), the backtraced light from the observer falls on the front side of the accretion disk, whereas for the lensed emission (orange trajectories), the light bent around the black hole crosses the equatorial plane once and falls on the back side of the accretion disk. The light is even directed to make a complete loop around black hole and return to the front side of the accretion disk for the photon ring emission (red trajectories). Nevertheless, as earlier discussed, depending on the impact parameter $b$, light rays pass through the accretion disk for $n(b)$ times. With each crossing the light rays pick up a certain intensity  and transmit it to the observer. Hence, the total observed intensity at the observer's screen is the sum of the intensity from each intersection
\begin{equation}
	\mathcal{I}_{obs}=\sum_{n} A(x)^2 \mathcal{I}_{em}|_{x=x_n(b)},\label{diskintensity}
\end{equation}
where $x_n(b)$, known as the transfer function, is the radial position of the $n_{th}$ crossing of the accretion disk. In this case, careful consideration is necessary to find higher-order $x_n(b)$ as these are highly sensitive to $b$. Gralla \textit{et al.} \cite{Gralla:2019xty} identified the slope $dx/db$ as the demagnification factor of the corresponding light rays. For the purpose of this work, $n = 1; 2; 3$ denotes the direct, lensed and photon ring emission, neglecting additional intersections with the disk since they will presumably contribute much less to the total intensity \cite{Gralla:2019drh}. We calculated $x_n(b)$ and showed their behavior with $b$ in Fig.~\ref{fig:transfer}. The first three transfer function are shown with the black, orange and red lines. The black line, $x_1(b)$, accounting for the first transfer function and direct image of the disk, has a slope approximately equal to 1, therefore, it makes the major contribution to the disk image intensity. Orange $x_2(b)$ and red $x_3(b)$ lines have slope much larger than 1, and accounts for the demagnified lensed images of the back side and front side of the disk, respectively. Here, we consider a toy accretion disk model, previously investigated by \cite{Guo:2021bhr,Guerrero:2021ues,Uniyal:2022vdu,Li:2021riw,Okyay:2021nnh}.
The inner edge of the accretion disk matches with the innermost stable circular orbit (ISCO) location and the emission exists only for $x>x_{isco}$
\begin{equation}
	\mathcal{I}_{em}(x) = \begin{cases} 
		\Big(\frac{1}{x - (x_{isco} - 1)}\Big)^2 & x\geq x_{isco}\\
		0 & x< x_{isco}.
	\end{cases}\label{model1}
\end{equation}

\begin{figure*}
	\begin{tabular}{c c c}
		\includegraphics[scale=0.5]{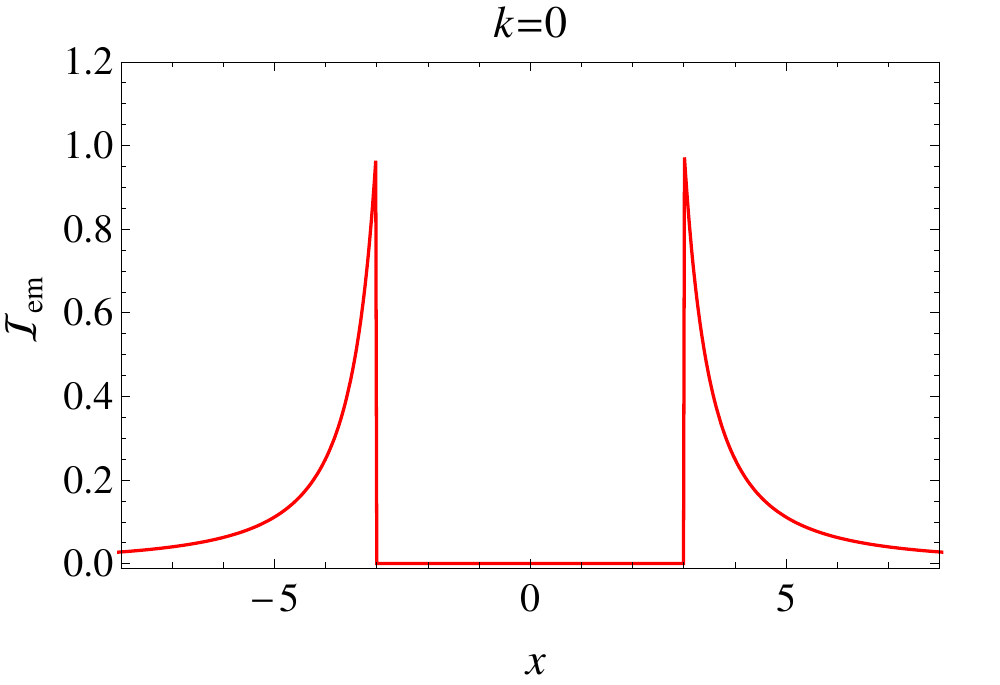}\hspace*{-0.2cm}&
		\includegraphics[scale=0.5]{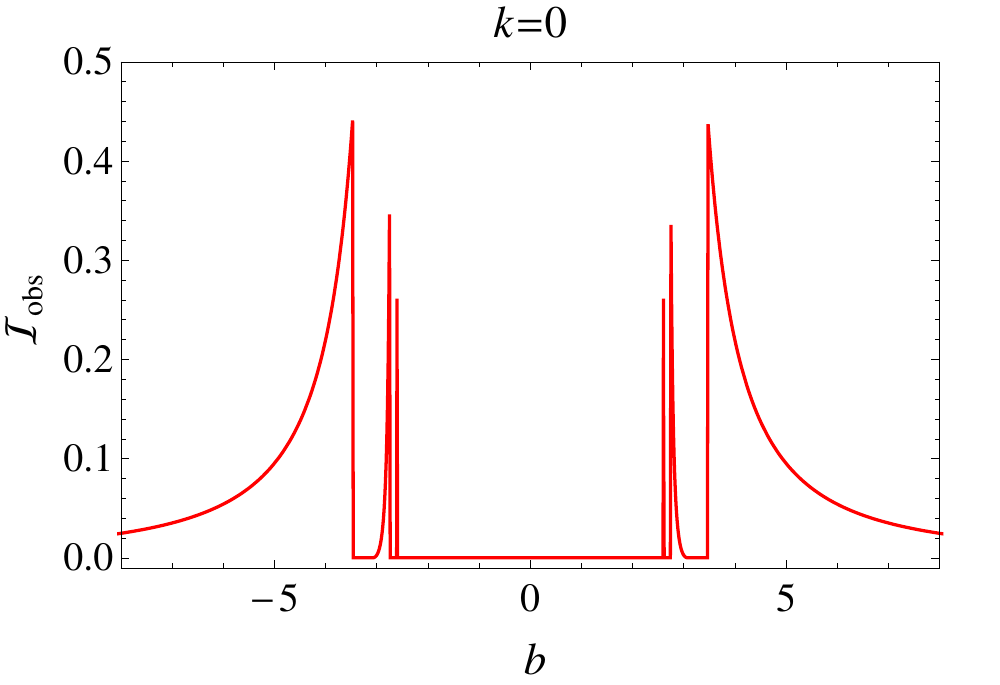}\hspace*{-0.2cm}&
		\includegraphics[scale=0.5]{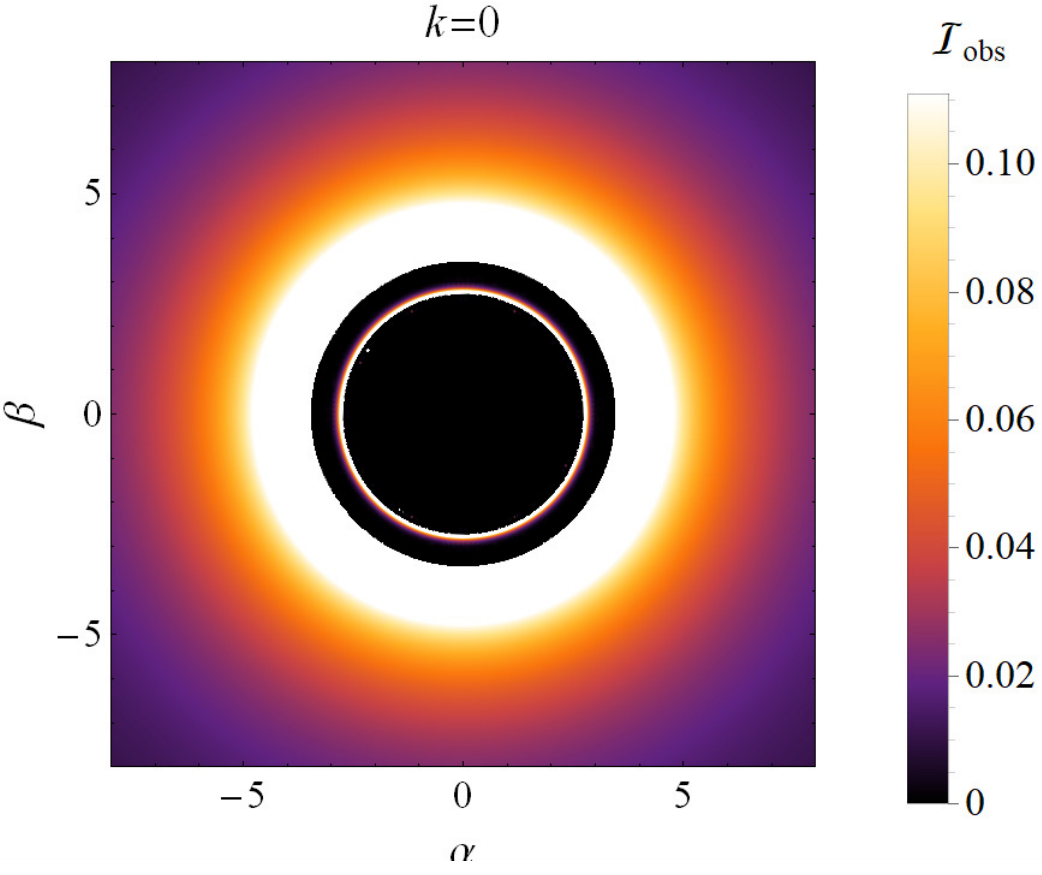}\\
		\includegraphics[scale=0.5]{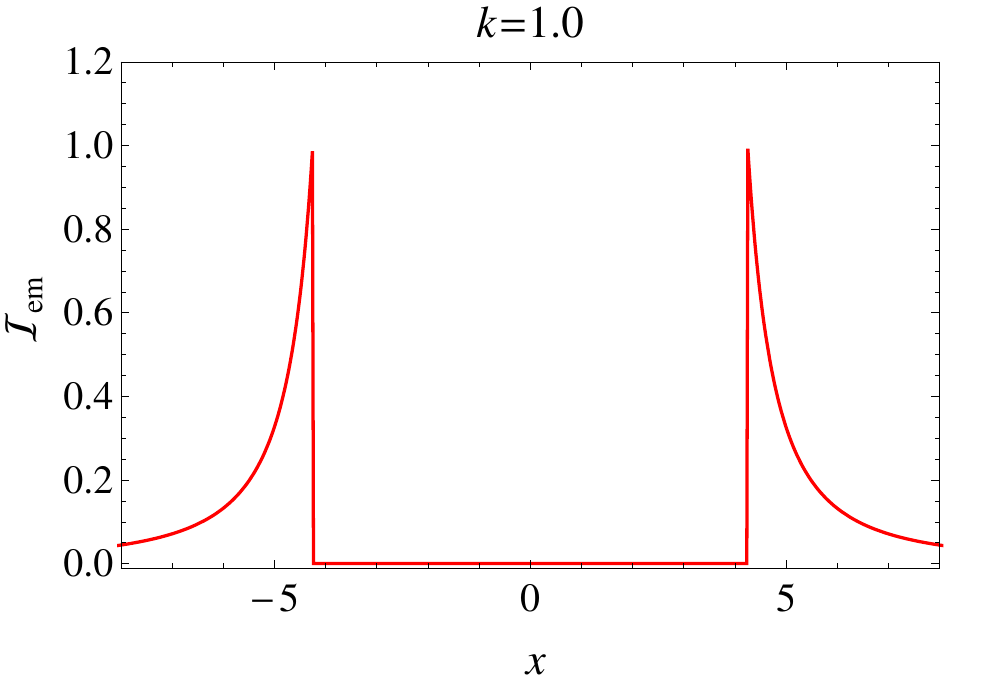}\hspace*{-0.2cm}&
		\includegraphics[scale=0.5]{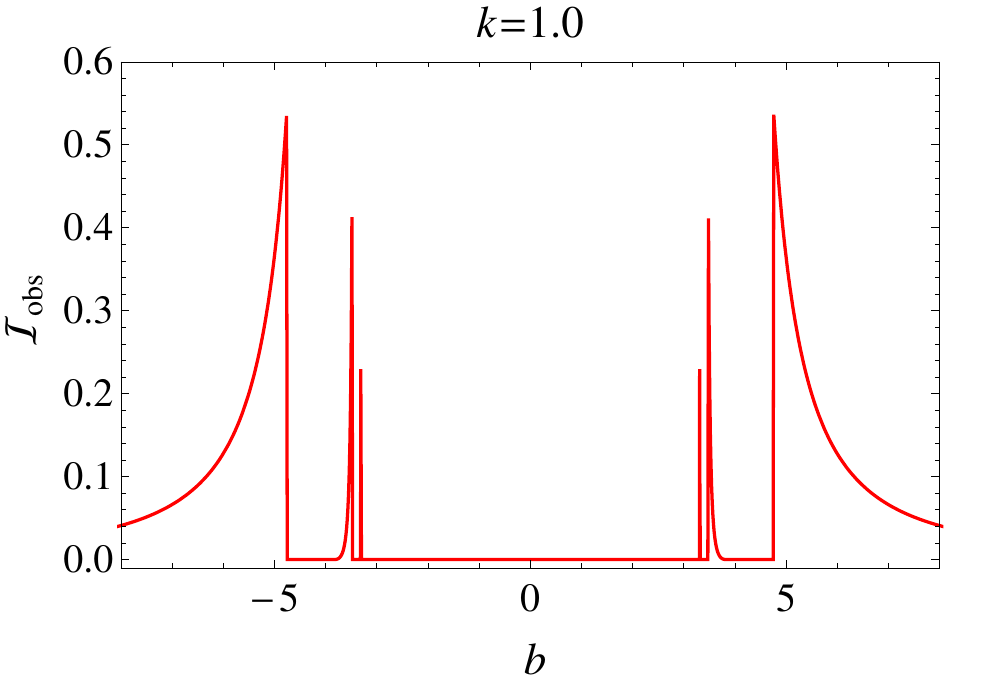}\hspace*{-0.2cm}&
		\includegraphics[scale=0.5]{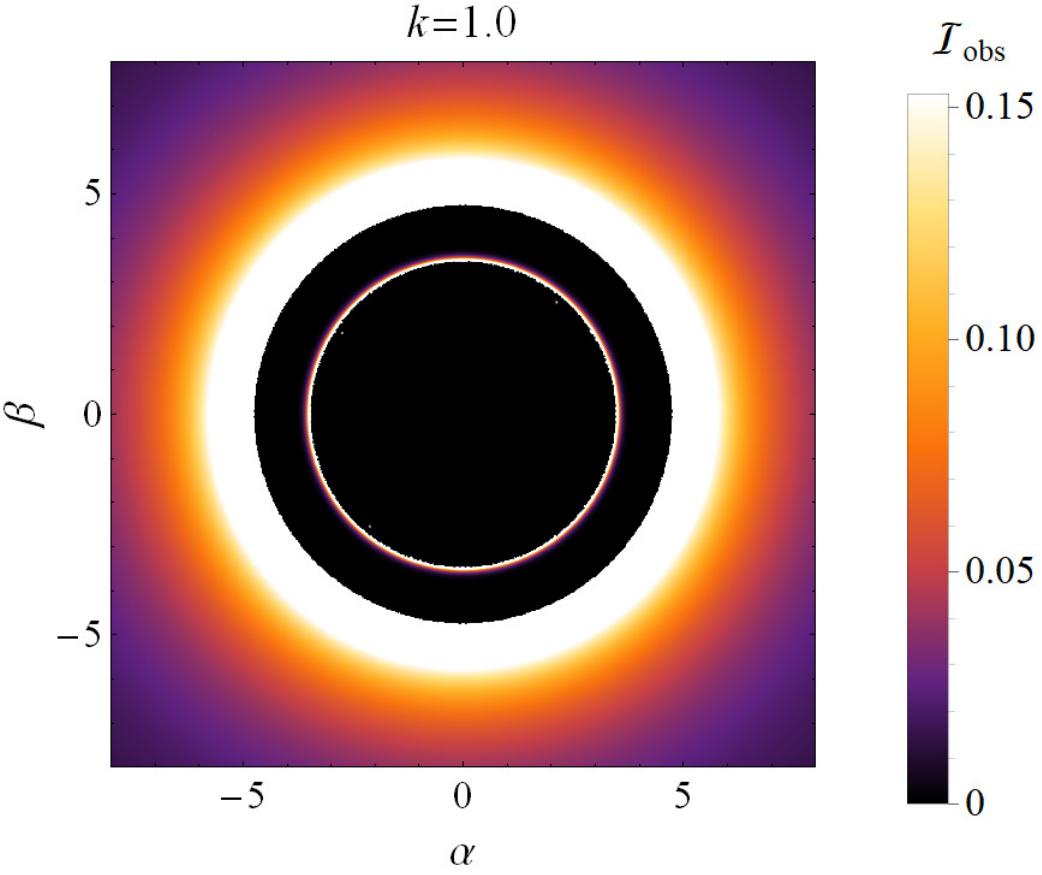}
	\end{tabular}
	\caption{Total emission intensity as a function of $x$ (left), the observed  intensity as a function the $b$	(middle), and the shadow images (right) for the polymerized black hole surrounded by an accretion disk. } \label{fig:diskShadow}
\end{figure*}

The black hole's ISCO radius can be determined by $V_{\text{eff}}=E^2, V_{\text{eff}}'=0,$ and $V_{\text{eff}}'' =0$. Using Eqs.~(\ref{diskintensity})-(\ref{model1}), the total emitted intensity $\mathcal{I}_{em}(x)$ as a function of $x$, the total observed intensity $\mathcal{I}_{obs}(b)$ as a function of $b$, and the two-dimensional image in celestial coordinates are plotted in the Fig.~\ref{fig:diskShadow}. The observed intensity plot in Fig.~\ref{fig:diskShadow} allows for a clear view of the three intensity peaks of light rings associated to direct, lensed, and photon ring emissions. For $k=0$, the emission function peaks at the ISCO, $x=3$, while, the observed direct emission peaks at $b=3.49$ and has an additional lensed image emission at $2.775\leq b\leq 3.025$ and a photon ring at $b=2.628$. Similarly, for $k=1.0$, the emission function peaks at the ISCO, $x=4.249$, while, the observed direct emission peaks at $b=4.759$ and has an addition lensed image emission at $3.492\leq b\leq 3.734$ and a photon ring at $b=3.321$. The contribution from the photon rings and the lensing rings emission is small compared to the direct emission. Because the accretion disk is orthogonal to the line of sight of the observer to the black hole, the lensed image of the disk is circularly symmetric as shown in the right panel of Fig.~\ref{fig:diskShadow}. 
The outermost boundary of the dark region and the inner bright ring, respectively, are due to the direct emission coming from disk and the lensed emission forming the secondary images of disk. Even though the direct emission dominates the total intensity map, a bright extended lump of radiation enclosing a thinner and dimmer ring and an even thinner photon ring (which is barely visible at naked eye) are the prominent features in the image. As a result, when compared to the spherical model, the geometrically thin disks model has different shadow features. An interesting and important analytical study of higher-order ring images of accretion disk around black hole is presented in Refs.~\cite{Bisnovatyi-Kogan:2022ujt,Wielgus:2021peu}.

\subsection{Infalling spherical accretion flow}
Let us consider a more realistic scenario, where the surrounding optically-thin radiating gas undergoes radially free-fall motion onto the black hole but emitting isotropically. The observed specific intensity for the photon frequency $\nu_{obs}$ at the point $(\alpha, \beta)$ on the observer's screen is still defined by Eq.~(\ref{emission}). However, because of the relative motion between the infalling gas and the static observer, the redshift factor is different from the static accretion case. Indeed, the redshift factor is evaluated as \cite{Narayan:2019imo,Bambi:2013nla}
\begin{equation}
	z=\frac{p_{\rho} u _o^{\rho}}{p_{\sigma } u _e^{\sigma }}. \label{EQ3.6}
\end{equation}
Here,  $p^{\mu }=\dot{x}_{\mu }$ is the photon four-momentum, and $u_o^{\mu }=(1,0,0,0)$ is the static observer's four-velocity at far distance from the black hole, and $u_e^{\mu }$ is the four-velocity of the accreting gas emitting the radiation under radial free fall, which is
\begin{equation}
	u_e^{\mu}=\Bigg(\frac{1}{A (x)},\, -\sqrt{\frac{1-A (x)}{A (x) B (x)}},\, 0,\,0\Bigg),
\end{equation}
such that $u_e^{\mu}u_{e\mu}=-1$. Hence, the redshift factor of the infalling accretion is calculated as
\begin{equation}
	z=\left(\frac{1}{A (x)}-\frac{p_r }{p_t}\sqrt{\frac{1-A (x)}{A (x) B (x)}} \right) ^{-1}, 
\end{equation}
which, as expected, is a function of $x$ and $b$. Whereas the photon four-momentum satisfies $p_{\mu}p^{\mu}=0$ and
\begin{equation}
	p_r=\pm p_t \sqrt{ \frac{B(x)}{A(x)}-\frac{b^2 B(x)}{C(x)}}, 
\end{equation}
where, the sign $+$ ($-$) corresponds to the photon moving toward (moving away from) black hole. The infinitesimal proper length as measured in the rest-frame of the accreting gas can be defined as \cite{Shaikh:2018lcc}
\begin{equation}
	d l_{prop}=-p_\mu u_e ^\mu d\lambda=\frac{p_t}{z p^x} d x,
\end{equation}
and not $\frac{p_t}{z p_x} d x$ as wrongly reported in many papers.
Integrating the observed specific intensity over all the observed photon frequencies, we get the total observed photon intensity \cite{Narayan:2019imo,Bambi:2013nla,Shaikh:2018lcc}
\begin{equation}
	\mathcal{I}_{{obs}}\propto -\int _{\gamma}\frac{z^4 }{x^2 }\, d l_{prop} \propto -\int _{\gamma}\frac{z^3  p_t }{x^2 p^x}\, dx.
\end{equation}
\begin{figure}
		\centering	
	\includegraphics[scale=0.8]{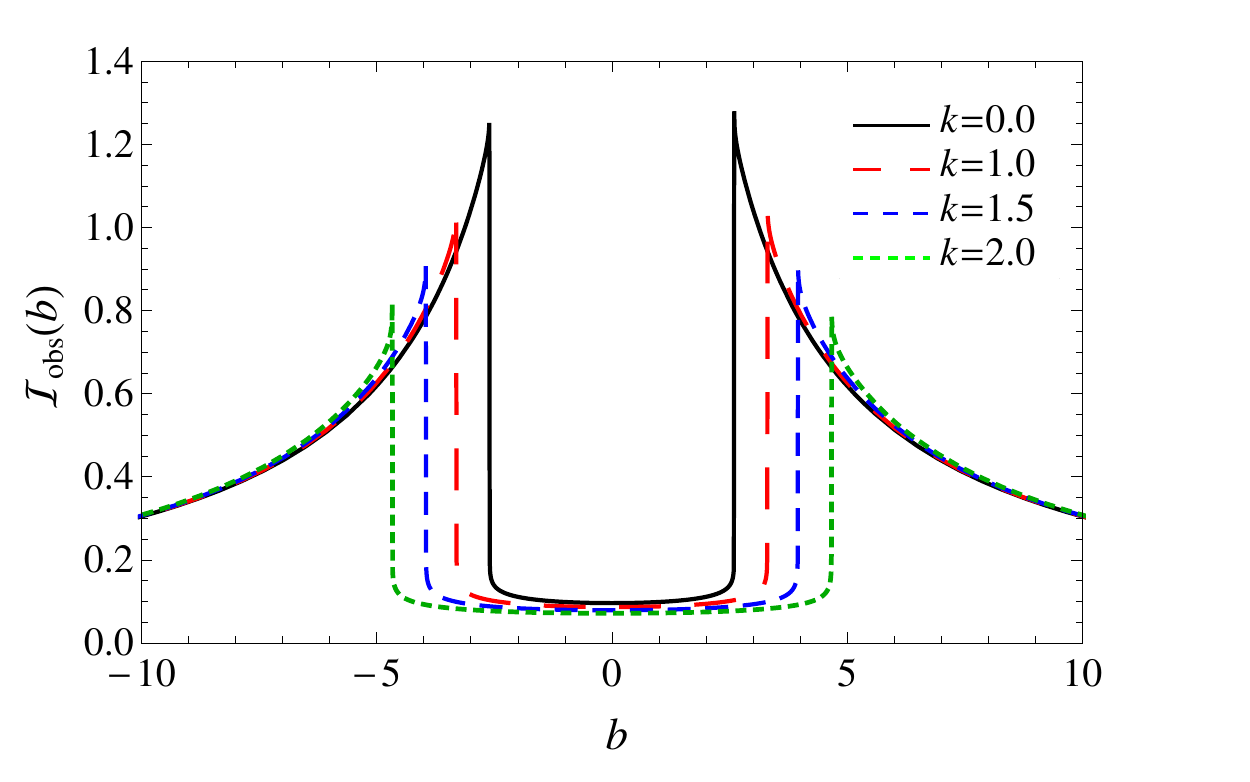}
	\caption{The observed intensity profile for a radially infalling accretion around a polymerized black hole as a function of the impact parameter $b$. For comparison, the intensity distribution for the Schwarzschild black hole is also shown with a black line. } \label{fig:InfallIntensity}
\end{figure}	
\begin{figure*}
	\begin{tabular}{c c c}
		\includegraphics[scale=0.5]{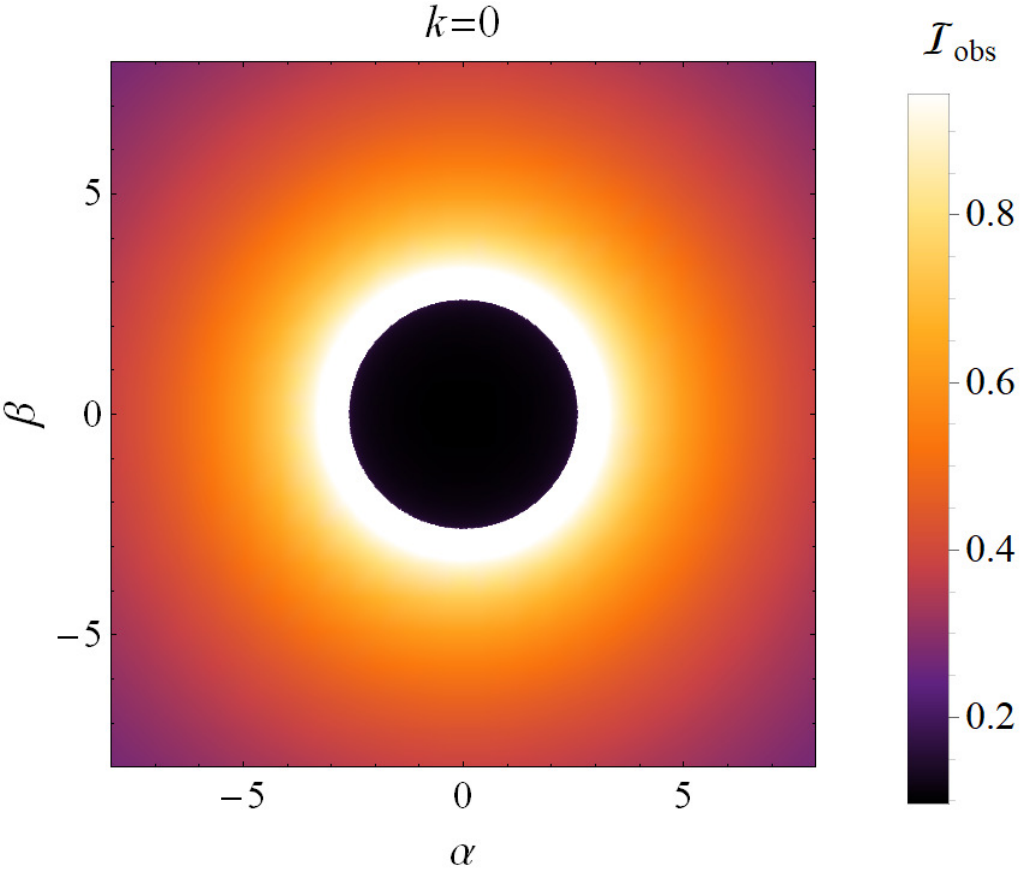}\hspace*{-0.2cm}&
		\includegraphics[scale=0.5]{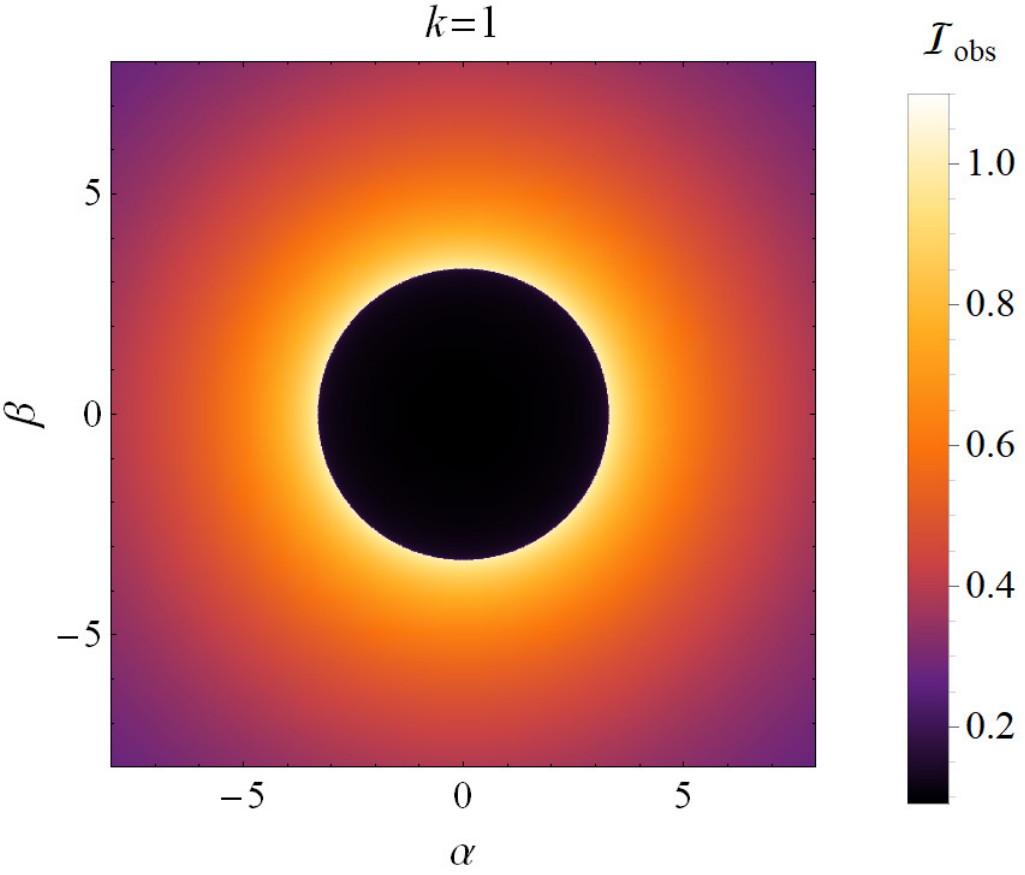}\hspace*{-0.2cm}&
		\includegraphics[scale=0.5]{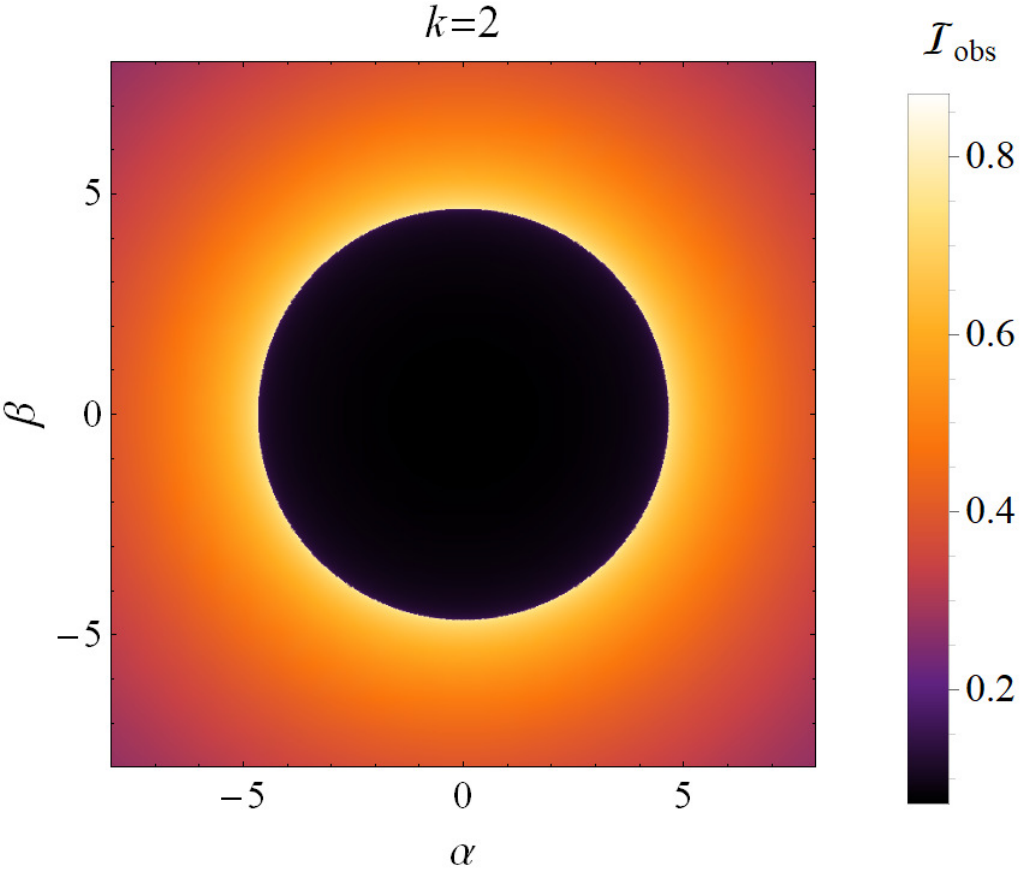}
	\end{tabular}
	\caption{Polymerized black hole shadows with the radially infalling spherical accretion for the different values of $k$ as seen by a distant observer. } \label{fig:InfallShadow}
\end{figure*}

Because the observer is on one side of the black hole, for the radiating matter at the opposite side of the black hole, the infalling matter and the emitted light rays that reach the observer have motion in the same direction. In contrast, for the radiating matter on the same side of the black hole, the emitted light rays move against the infalling matter direction to reach the observer. As a result, both light rays experience distinct redshifts-- Doppler beaming. For $b\leq b_c$, all light rays are backtraced from the observer to the horizon and are highly redshifted and contribute to the dark shadow interior. However, light rays with $b\geq b_c$, are redshifted from the observer $x_{obs}$ to the turning point $x_{tp}$ and blueshifted from the turning point $x_{tp}$ to the emitter position $x_{emit}$. The blueshifted photon illuminates the black hole image. The observed intensity takes the form
\begin{equation}
	\mathcal{I}_{{obs}}({\nu_{o}})\propto -\int _{x_{emit}}^{x_{tp}}\frac{z_-^3\,  p_t }{x^2 |p^x|}\, dx  +\int _{x_{tp}}^{x_{obs}}\frac{z_+^3\, p_t }{x^2 |p^x|}\, dx,
\end{equation}
where $z_+$ and $z_-$, respectively, are the redshift and blueshift of the light rays.

We calculated and depicted the observed intensity as a function of $b$ for different values of $k$ in Fig.~\ref{fig:InfallIntensity}. The two-dimensional images of the shadow with an infalling spherical accretion seen by a distant observer are shown in Fig.~\ref{fig:InfallShadow}. The intensity distribution qualitatively resembles that for the static accretion model, with intensity rising sharply with decreasing $b$, reaching a peak at $b=b_c$, and then dropping to significantly lower values inside the peak. However, there are important differences. The central region inside the photon ring has a severely reduced brightness. In fact, it is clear that the shadow interior with infalling accretion in Fig.~\ref{fig:InfallShadow} is darker than the corresponding region for the static accretion as shown in Fig.~\ref{fig:StaticShadow}. The clear contrast between these static and infalling models is due to the Doppler effect because of the infalling gas, which is more noticeable near the black hole's event horizon. In addition, the photon orbit radii and the shadow radii remain unchanged with accretion model. This confirms that the shadow is an inherent property of spacetime and that the behavior of the accretion flow surrounding the black hole only effects the intensity of the images. The most striking feature of these images is the very high sensitivity of the shadow size to the value of $k$. Such a large difference in the image size would easily be detected with current observational techniques. 

Shaikh \textit{et al.} \cite{Shaikh:2018lcc,Shaikh:2019hbm}, and Joshi \textit{et al.} \cite{Joshi:2020tlq} have reported that some naked singularities may also show black hole like shadow features. An interesting comparison between Schwarzschild black hole shadows produced under spherical and optically thin accretion in Newtonian model, static accretion model, and radially accretion model is presented in \cite{Narayan:2019imo}. The bottom line of the discussion above is that under spherical accretion polymerized black holes cast larger shadows with lower brightness contrast compared to that for the Schwarzschild black hole. The fraction intensity depression $f_c$, for $k=0, 1$ and 2, respectively, is $f_c=0.077, 0.0844$, and $0.092$.

\section{Constraints from the EHT 2017 observations}\label{Sec-7}
The structure of Sgr~A* and M87* black holes images gets completely washed out by interstellar scattering at cm and higher wavelengths. However, at the mm wavelength, it is possible to image the emission region of these supermassive black holes due to three favorable reasons: (i) neglecting interstellar scattering, (ii) better angular resolution, and (iii) the compact synchrotron emitting region becomes optically thin. Indeed, the advent of the EHT, a global very long baseline interferometry (VLBI) network of radio telescopes observing at a frequency of 230GHz, made it possible to make horizon-scale observations of the supermassive black holes at the centers of galaxies. Recently, EHT unraveled the first shadow images of Sgr~A* black hole \cite{Akiyama:2019cqa,Akiyama:2019eap,EventHorizonTelescope:2019pgp}, which complement the observed shadow of M87* black hole in 2019 \cite{EventHorizonTelescope:2022urf,EventHorizonTelescope:2022wok,EventHorizonTelescope:2022xnr,EventHorizonTelescope:2022xqj}. The observed shadows of both the Sgr~A* and M87* black holes have a common feature that show a bright ring of emission surrounding a brightness depression. The size of the emission ring has been measured with unprecedented accuracy. The observed shadows of the Sgr~A* and the M87* black holes have been extensively used to test alternative to Kerr black hole \cite{EventHorizonTelescope:2021dqv,Vagnozzi:2022moj,Bambi:2019tjh,EventHorizonTelescope:2020qrl,Younsi:2021dxe,Afrin:2021wlj,Gralla:2020pra,Ghosh:2022kit,Vagnozzi:2019apd,Kumar:2019ohr,Allahyari:2019jqz}. Because the black hole spin introduces minor corrections to the size of the shadow \cite{Kumar:2018ple,EventHorizonTelescope:2021dqv,EventHorizonTelescope:2022xqj}, we can safely focus on nonspinning spacetimes. In addition, the spherically symmetric black hole shadow size only involves the information of the $tt-$component of the metric. Fortunately, the polymerized parameter $k$ appears in the $tt-$component of metric (\ref{metric1}), and as we have seen in the earlier sections, the shadow size strongly depends on the value of the $k$. Our goal in this section is to compare the LQG-motivated polymerized black hole shadows with those observed for the Sgr~A* and M87* and to use the EHT bounds to place constraints on the polymerized black hole.

\subsection{Constraints from M87* shadow}
The angular diameter of the observed emission ring for the M87* black hole image is $\theta_d=(42\pm 3)\mu$as, with the estimated mass $M=(6.5\pm 0.9)\times 10^9 M_{o}$ and distance $D_{OL}=16.8\pm 0.7$ Mpc. For a realistic source model, the emission is not restricted to lie exactly at the photon ring rather it preferentially falls outside the photon ring (as shown in subsection~\ref{accretiondisk}). Furthermore, it is important to note that the EHT observation of the M87* black hole did not directly measure the shadow angular diameter rather measure the size of the bright ring of emission, which is appropriately calibrate to determine the shadow diameter \cite{Akiyama:2019cqa}. For this purpose, the EHT first computed the fractional difference between the inferred angular gravitational radius of the M87* black hole based on the EHT data and the values measured using stellar orbit observations \cite{Akiyama:2019eap}. Then this value yields a prediction for the M87* shadow angular diameter $\theta_{sh}$:
\begin{equation}
	\delta=\frac{\theta_{sh}}{\theta_{sh, Sch}}-1,
\end{equation}
where $\theta_{sh, Sch}=6\sqrt{3}\theta_g$ is the Schwarzschild black hole shadow angular diameter with $\theta_g=M/D_{OL}$. For the M87* black hole, $\delta_{ M87*}=-0.01\pm 0.17$ \cite{EventHorizonTelescope:2020qrl,EventHorizonTelescope:2021dqv}; the M87* shadow is consistent to within $17\% $ for a $68\%$ confidence interval of the size predicted for a Schwarzschild black hole $\theta_{sh}=6\sqrt{3}(1-0.01\pm 0.17)\theta_g$. 

\begin{figure}
		\centering	
	\includegraphics[scale=0.53]{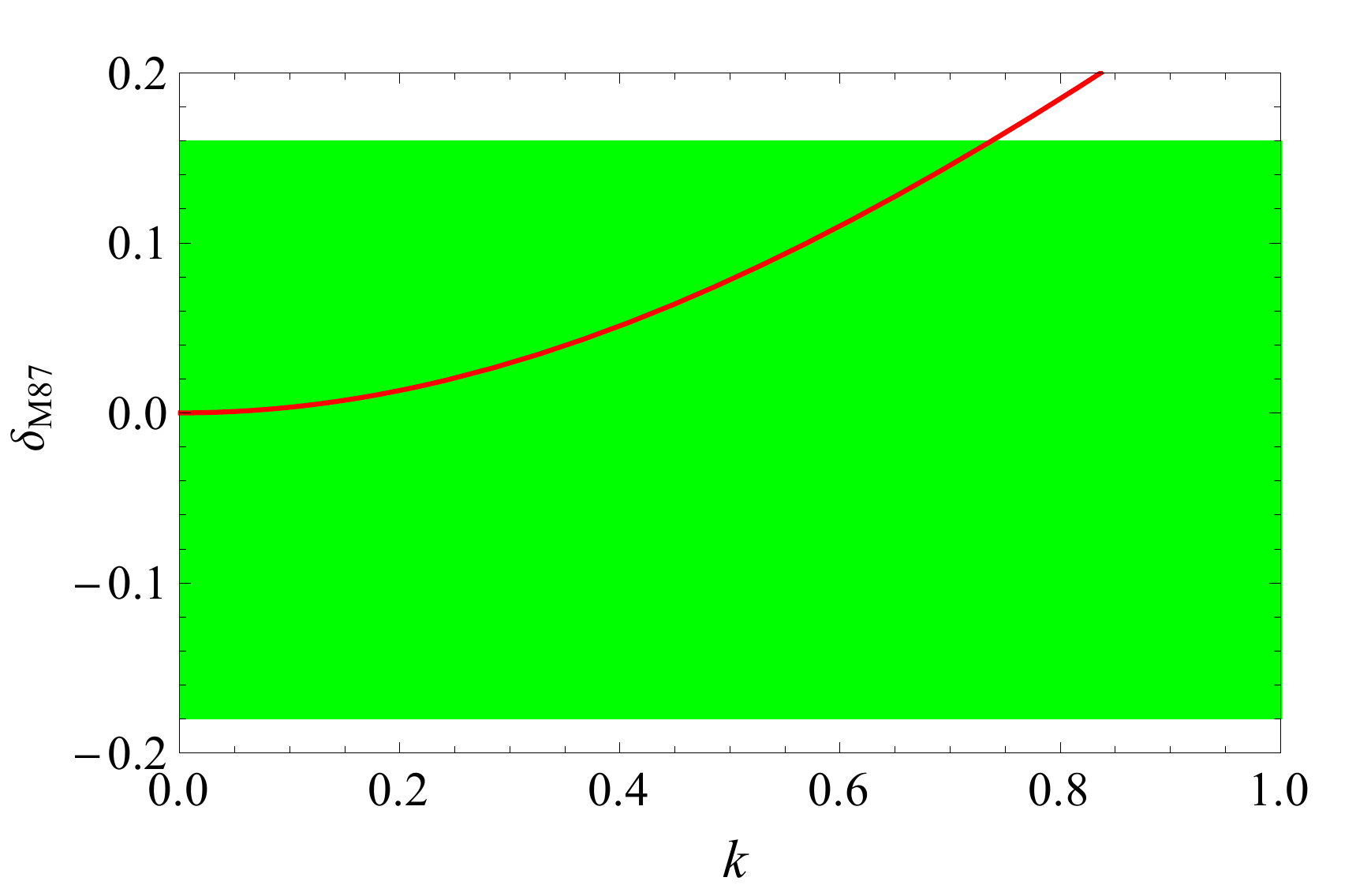}
	\caption{Fractional difference $\delta_{M87}$ varying with $k$. The green shaded region gives the $1\sigma$ confidence region $\delta_{M87}=-0.1\pm 0.17$.} \label{fig:m87}
\end{figure}

We consider the polymerized black hole model for the M87* and calculate the $\delta_{ M87*}$ and depict it in Fig.~\ref{fig:m87}.  Because the polymerized black hole shadow size increases with $k$, the lower bound of $\delta_{ M87*}$ is irrelevant for our purpose of constraining polymerized black hole parameters. 
Polymerized black hole with $k\leq 0.742$ satisfies the $1\sigma$ bound for M87* shadow size. 

\subsection{Constraints from Sgr~A* shadow}
Sgr~A* black hole shadow images have advantages to test the nature of astrophysical black hole (i) Sgr~A* black hole mass bridges the gap between the stellar black holes observed by the LIGO and Virgo, and the M87* black hole, and thus probes a significantly distinct curvature scale ($10^{6}$ order of higher curvature than the M87*) (ii) independent prior estimates for mass to distance ratio are used for Sgr~A*. Most prominently, for the Sgr~A* black hole, the EHT not only measured the emission ring angular diameter $\theta_d=(51.8\pm 2.3)\mu$as but also estimated the shadow diameter $\theta_{sh}=(48.7\pm 7)\mu$as with the priors $M=4.0^{+1.1}_{-0.6} \times 10^6 M_{o}$ and $D_{OL}=8.15\pm 0.15$ kpc \cite{EventHorizonTelescope:2022xnr}. For instance, EHT used three independent imaging algorithms, namely, \texttt{ eht-imaging}, \texttt{SIMLI}, \texttt{DIFMAP}, to determine the Sgr~A* shadow morphology. The most likely averaged measured value of the shadow angular diameter from these three algorithms is in the range $\theta_{sh}\in (46.9-50)\,\mu$as, and the $1\sigma$ credible interval is $41.7-55.6\,\mu$as \cite{EventHorizonTelescope:2022xnr,EventHorizonTelescope:2022urf}. 
EHT used the two separate priors for the Sgr~A* angular size from the Keck and Very Large Telescope Interferometer (VLTI) observations and the three independent imaging models to estimate the bounds on the fraction deviation observable $\delta$ \cite{EventHorizonTelescope:2022xnr,EventHorizonTelescope:2022urf}
\begin{align}
	\delta_{Sgr}= \begin{dcases*} -0.08^{+0.09}_{-0.09}\;\;\;\;\; & \text{VLTI}\\
		-0.04^{+0.09}_{-0.10}\;\;\;\;\; &\text{Keck}	 
	\end{dcases*} 
\end{align}
\begin{figure}
		\centering	
	\includegraphics[scale=0.53]{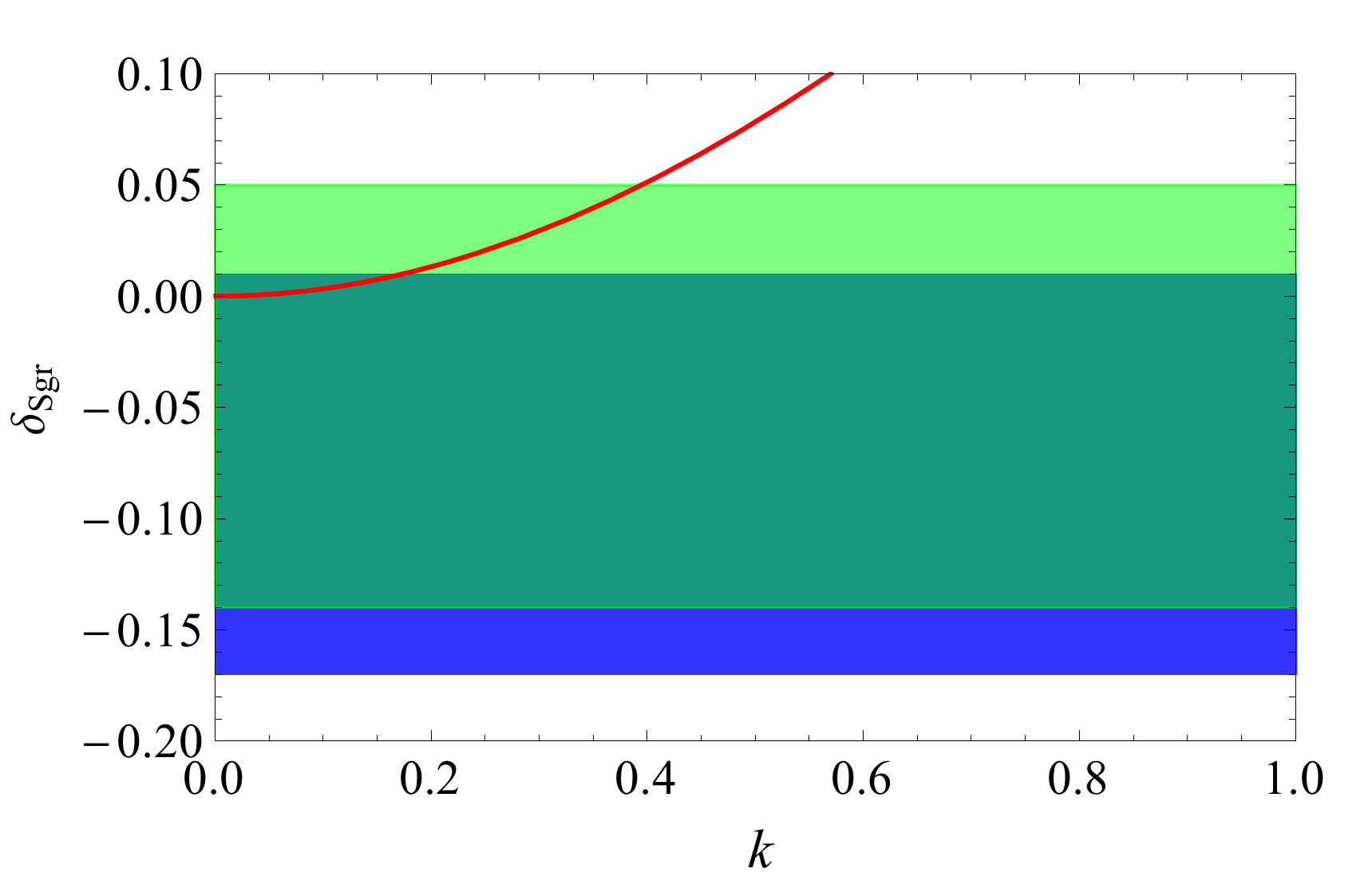}
	\caption{Fractional diameter deviation observable $\delta_{Sgr}$ varying with $k$. The green and blue shaded regions, respectively, give the $1\sigma$ confidence regions for the \texttt{Keck} estimate and \texttt{VLTI} estimate.} \label{fig:sgr2}
\end{figure}

In Fig.~\ref{fig:sgr2}, we depict the $\delta_{Sgr}$. Following \cite{EventHorizonTelescope:2022xnr}, we choose the \texttt{ ehtimaging +Keck+GRMHD} and \texttt{eht-imaging +VLTI+GRMHD} combinations as the two fiducial cases to calculate constraints on the polymerized black hole parameter. In particular, Keck and VLTI measurements within the $1\sigma$ confidence interval, respectively, constraint $k\leq 0.396$ and $k\leq 0.180$.

\section{Conclusion}\label{Sec-8}
In this paper, we have carried out a comprehensive study of the static and spherically symmetric polymerized black holes. These black holes are motivated by the LQG principles and semi-polymerization technique. In addition, these black holes are not only free from the curvature singularity at the center, geodescially comeplete and are globally regular, but also free from the blue-shift mass instability as they possess only a single horizon and are globally hyperbolic. These salient features distinguish them from the other regular black holes. In the limit $k\to 0$, the polymerized black hole metric recovers the Schwarzschild metric, and the horizon size increases with $k$.  We addressed the question of whether the polymerization corrections can leave imprints at observationally accessible length scales outside the horizon. To answer this, we computed the light deflection angle in the weak and strong gravitational lensing limits, and calculated the image position magnification and time delay. We compared the obtained results for the polymerized black hole with those corresponding to the Schwarzschild black hole. Using the ray-tracing technique, we identified the light trajectories contributing to the direct, lensed, and photon ring emission of the surrounding accretion disk . Black hole shadows under different accretion models are constructed.

The following are the major outcomes
\begin{itemize}
	\item By treating the quantum geometry corrections as an ``effective" matter contribution, we proved that the polymerized black hole emerges as a solution of Einstein's field equations sourced by the phantom scalar field and NED fields associated with a magnetic charge.
	\item Contrary to Schwarzschild black holes, the polymerized black holes possess both the unstable and stable photon circular orbits. The unstable (stable) orbits are outside (inside) the event horizon.
	\item In the polymerized black hole spacetime, although the images form farther away from the black hole center, they are more magnified compared to the Schwarzschild black hole.
	\item Polymerized black holes have larger shadows with darker interior than those for the Schwarzschild black hole.
	\item Polymerized black holes have thicker lensing and photon rings compared to those for the Schwarzschild black hole, which can be resolved with the next-generation EHT..
	\item For the thin disk accretion, there is not only a dark central area, but also the photon rings and lensing rings outside of the black hole shadow.
	\item Polymerized black holes with $k\leq 0.742$ satisfy the $1\sigma$ bound for the observed M87* shadow angular diameter.
	\item Keck and VLTI measurements of $\delta$ for the Sgr~A* black hole shadow put constraints, respectively, $k\leq 0.396$ and $k\leq 0.180$.
	\item For microscopic values of $k$, the polymerization corrections are still small to be detectable with the present technology using the supermassive black hole observations.
\end{itemize}

While it is unlikely that evidence for a microscopic length scale can be found in the supermassive black hole observations, it should be kept in mind that we do not strictly speaking know that the minimal length scale is identical to the Planck scale and not lower, and scientific care demands that every new range of parameter space be scrutinized.  Nevertheless, it is both necessary and really timely to accurately compute as many quantum gravity concrete predictions as possible not only to make progress in finding the correct theoretical model but also to increase the possibility of finding these potential signals through next-generation EHT (via shadow) or LIGO/LISA (via gravitational wave echo). In this work, we restricted ourselves only to non-rotating spacetime. The inclusion of the rotation anticipate interesting features. The gravitational lensing and shadow of the rotating polymerized black holes are part of future work. However, given the significance of accretion upon black holes, a proper understanding necessitates relativistic magnetohydrodynamic simulations of the hot plasma flow with effects of magnetic fields. These are beyond the scope of the present paper. 


\section{Acknowledgments} 
It is a pleasure to thank Prof. Kirill Bronnikov, Prof. Luciano Rezzolla, Dr. Rajibul Shaik and Dr. Prashant Kocherlakota for insightful discussions, and Prof. Sunil Maharaj, Prof. Sushant Ghosh for the comments on the initial draft. R.K.W. would like to thank the 
University of KwaZulu-Natal and the NRF for the postdoctoral research fellowship, and the ITP, Goethe University, Frankfurt for the hospitality.

\appendix
\section{Gaussian filtering of accretion images}
\begin{figure*}
	\centering
	\begin{tabular}{c c c}
		\includegraphics[scale=0.42]{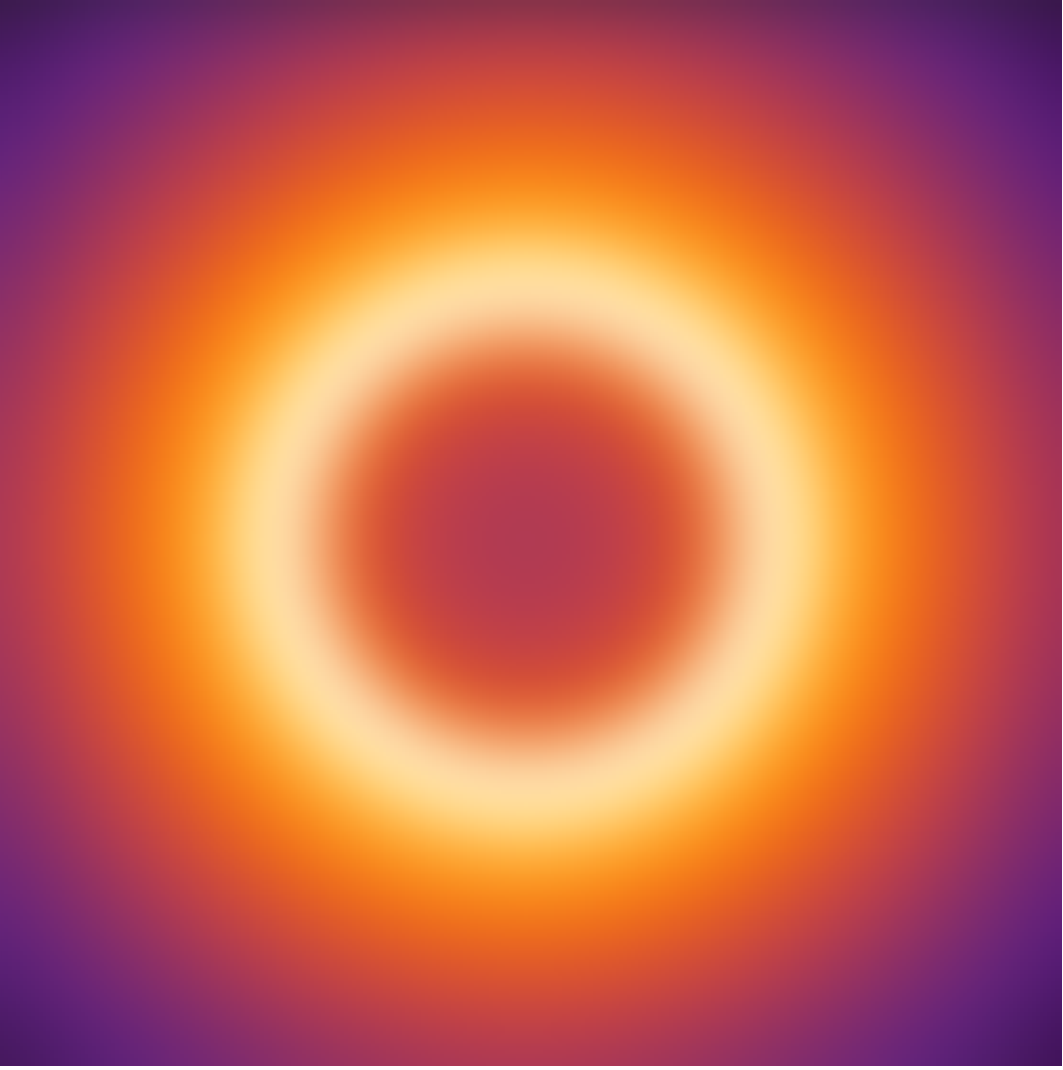}&
		\includegraphics[scale=0.42]{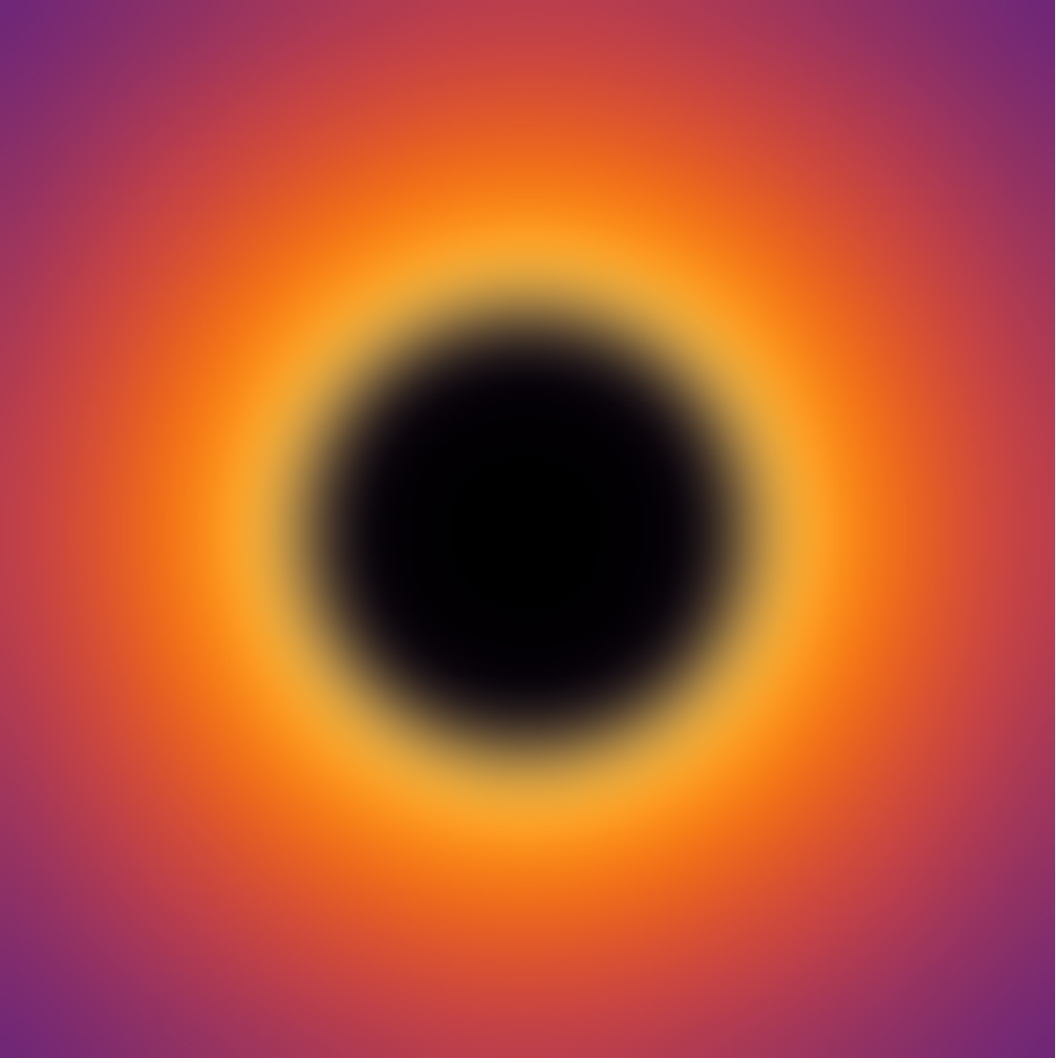}&
		\includegraphics[scale=0.42]{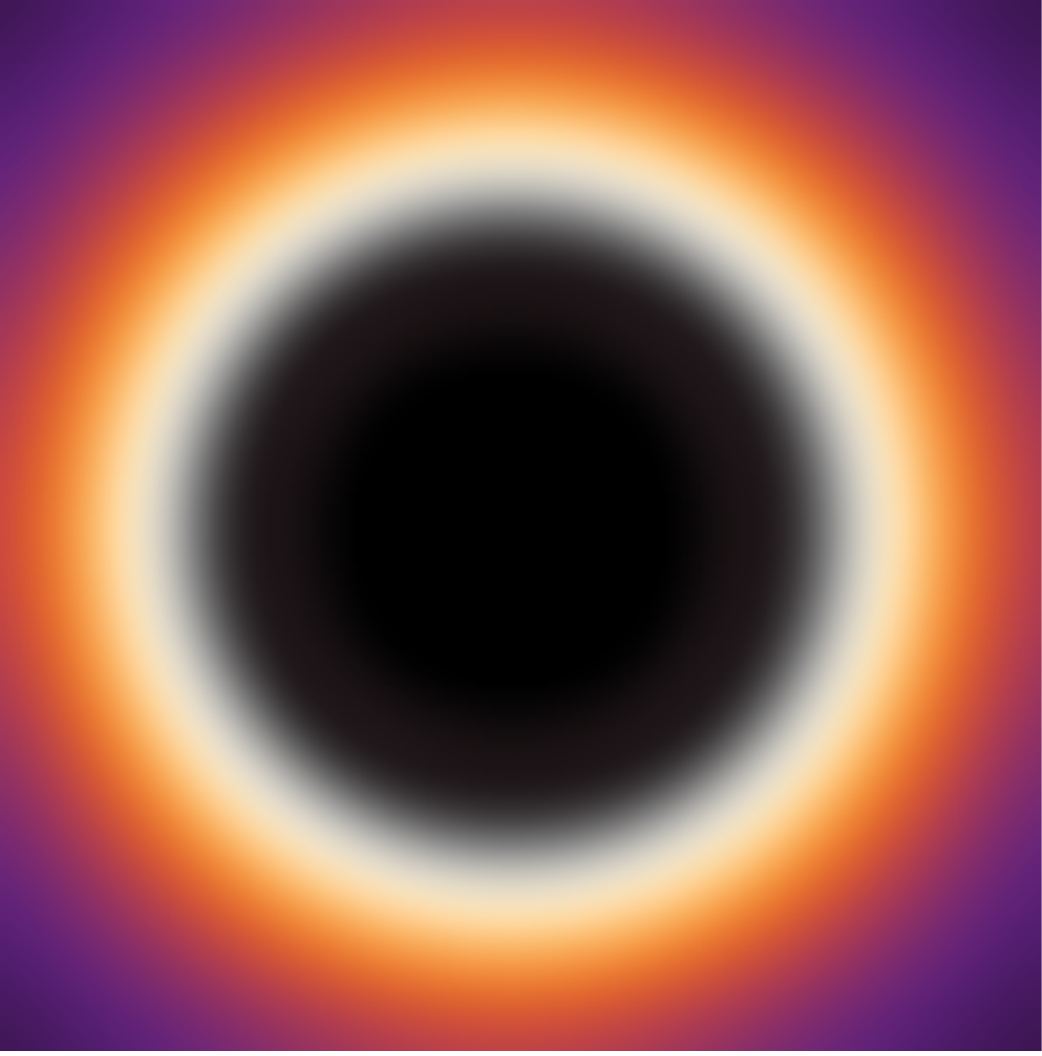}
	\end{tabular}
	\caption{Polymerized black hole shadows with the static (left), infalling (middle), and accretion disk (right) by utilizing a Gaussian filter with a standard deviation of 1/12 the field of view. }\label{fig:Shadow}
\end{figure*}

Because, the EHT does not have the infinite angular resolution, we expect to see a blurred images of astrophysical black holes, as captured for the M87* and Sgr A* black holes. To mimic the finite angular resolution, we have convoluted the original intensity function with a Gaussian distribution and synthetically produced the blurred images. Figure~\ref{fig:Shadow}, shows these blurred shadows under static, infalling and accretion disk flows. Clearly, comparing Fig.~\ref{fig:Shadow} and \ref{fig:diskShadow}, the blurring washes out the lensing ring and photon ring features. Nevertheless, it is difficult to obtain the ring information with the current resolution of EHT.

\bibliography{LQGbib}

\end{document}